\documentclass[aps,prd,twocolumn,eqsecnum,amssymb,amsmath,showpacs,a4paper, superscriptaddress]{revtex4-2}

\usepackage{graphicx}
\usepackage{amsfonts}
\usepackage{amsmath}
\usepackage{hyperref}
\usepackage{units}
\usepackage{color}
\usepackage{dcolumn}
\usepackage{bm}
\usepackage{float}
\usepackage{cleveref}
\usepackage[normalem]{ulem}
\usepackage{appendix}
\usepackage{mathtools}
\usepackage{esvect}

\usepackage[colorinlistoftodos, size=tiny, bordercolor=white]{todonotes}
\usepackage{comment}

\graphicspath{ {images/} }
\usepackage{mathrsfs}
\usepackage{amssymb}
\usepackage{dsfont}
\usepackage{enumitem}
\usepackage{gensymb}
\usepackage{bm}

\usepackage{mathtools}
\usepackage{chngcntr}
\counterwithout{equation}{section}

\makeatletter
\let\cat@comma@active\@empty
\makeatother

\newcommand{\grad}{\bm{\nabla}}
\newcommand{\ex}{\mathrm{ex}}

\newcommand{\tp}{{tp}}
\newcommand{\lr}{{lr}}
\newcommand{\Ri}{\mathrm{R}}
\newcommand{\Le}{\mathrm{L}}
\newcommand{\as}{\mathrm{as}}

\begin{document}

\title{A primer on the analogue black hole bomb with capillary-gravity waves}

\author{Sam~Patrick}
\email{sampatrick31@googlemail.com}
\affiliation{Department of Physics, King’s College London, University of London, Strand, London, WC2R 2LS, UK}

\author{Th\'eo Torres}
\affiliation{Aix Marseille Universit\'e, CNRS, Centrale Marseille, LMA UMR 7031, Marseille, France}

\date{\today}

\begin{abstract}
\noindent
Draining vortices with a free surface are frequently employed as rotating black hole simulators, both in theory and experiments. However, most theoretical work is restricted to the idealised regime, where wave dispersion and dissipation are neglected. We investigate the role of these effects on the analogue black hole bomb, an instability resulting from rotational superradiant amplification in confined systems.
We reveal that the dispersion of deep water capillary-gravity waves significantly modifies the unstable mode eigenfrequencies, whereas viscosity only affects those with high frequencies.
Furthermore, if the circulation is less than an order 1 multiple of the drain rate, superradiance does not occur and the vortex is stable.
The instability is maximised in small systems with high flow velocities, provided there is sufficient space between the vortex and the outer boundary for the first excited state to lie inside the superradiant bandwidth.
Implications for experiments on analogue black holes and free surface vortices are discussed.
\end{abstract}
\maketitle

\section{Introduction}

Draining vortices are the archetypal example of an analogue rotating black hole \cite{schutzhold2002gravity}.
The basic idea is that the draining fluid speeds up as it impinges on the central outlet, and if the fluid velocity manages to surpass the speed of long wavelength excitations of the fluid, the system possess an analogue horizon.
The presence of a horizon leads to the possibility of simulating Hawking radiation from a black hole \cite{hawking1974black,unruh1981experimental}.
Extensive work on the analogue Hawking effect has been performed in one dimensional (1D) flows where the fluid accelerates linearly and the analysis is simpler.
Detailed numerical \cite{unruh1995sonic,corley1996hawking} and analytic \cite{macher2009black,coutant2012black,coutant2016imprint} analyses have revealed the persistence of the effect in fluid-like systems and are supported by experimental measurements in both classical \cite{weinfurtner2011measurement,euve2016observation} and quantum systems \cite{steinhauer2016observation,munoz2019observation}.
These results are generally taken as an indication of the universality of the Hawking effect, that is, a robustness against modifications to the short wavelength part of the governing theory \cite{unruh2005universality}.

In two dimensions (2D), the analogy with black holes becomes much richer.
In particular, rotating systems like the draining vortex can also possess an analogue ergosphere, the region of spacetime around a black hole where an observer would need to travel faster than the speed of light to remain at rest \cite{wald2010general}.
The existence of such a region implies the amplification of waves in a certain frequency band, i.e. \textit{rotational superradiance} \cite{penrose1971extraction,starobinskii1973amplification}, which causes an evaporating black hole to shed its angular momentum more rapidly than its mass \cite{page1976particle,gregory2021black}.
Many theoretical studies demonstrate the occurrence of analogue superradiance using a simple, shallow water model of a draining vortex \cite{basak2003reflection,basak2003superresonance,berti2004quasinormal,richartz2015rotating}.
The effect was first measured in a water tank \cite{torres2017rotational} and subsequently in a (non-draining) photon superfluid \cite{braidotti2022measurement}.

Most experimental studies of the Hawking effect and superradiance to date have employed one of two types of simulator.
The first involves excitations analogous to sound waves (density perturbations) in a Bose-Einstein condensate (BEC), e.g. \cite{steinhauer2016observation,braidotti2022measurement}.
The second employs a classical liquid, e.g. room temperature water, where the excitations are surface waves \cite{weinfurtner2011measurement,euve2016observation,torres2017rotational}.
Since gravity and surface tension act as a restoring force, these are also known as capillary-gravity waves (where capillary is synonymous with surface tension).
In both types of simulator, the dispersive nature of these waves complicates the analogy with black holes.
In relativity, all massless particles propagate at the speed of light $c$. Similarly, in fluids, it is often the case that long wavelength modes propagate at a constant (i.e. frequency independent) speed $c$ -- the ``analogue speed of light''.
However, at short wavelengths, the waves probe the microscopic details of the fluid and the propagation speed becomes frequency dependent.
If short wavelengths propagate slower than $c$ (e.g. deep water gravity waves) the theory is called \textit{sub-luminally} dispersive, in analogy with relativity, whereas if they propagate faster than $c$ (e.g. capillary waves and Bogoliubov modes in a BEC) the theory is called \textit{super-luminally} dispersive.
In the case of capillary-gravity waves, the theory is a mixed type, containing a sub-luminal part at intermediate wavelengths (the deep water regime) and a super-luminal part as the wavelength goes to zero (the capillary regime).
The effect of each type of dispersion relation on superradiance around a draining vortex was investigated in \cite{patrick2020superradiance,patrick2021rotational}.
The effect of mixed dispersion in rotating systems that superradiate has yet to be understood.

Through the analogy, many dynamical processes in supersonic fluid flows can be interpreted in terms of their black hole scattering counterparts.
For instance, the Hawking effect in a trans-sonic BEC gives rise to the \textit{black hole laser} instability \cite{corley1999black,coutant2010black,steinhauer2014observation}.
The Hawking effect is also responsible for the phenomenon of the zero frequency undulation that appears when a fluid accelerates over a submerged obstacle \cite{coutant2014undulations}.
In rotating systems, the relativistic phenomenon of rotational superradiance is analogous to the fluid mechanical notion of over-reflection in shear flows \cite{acheson1976over}.
In quantum fluids, superradiance causes large vortices to decay into clusters of single vortex quanta \cite{giacomelli2020ergoregion,patrick2022quantum,patrick2022origin} whilst in classical fluids, it is related to the rotating polygon instability of hollow core, free surface vortices \cite{mougel2017instabilities}.

In this work, we investigate the occurrence of another well-known black hole instability in a draining water tank: the black hole bomb (BHB) \cite{cardoso2004black}. This was first studied by Press and Teukolsky \cite{press1972floating} and involved a black hole surrounded by an idealised spherical mirror. The essence of the BHB is that waves in a certain frequency range get (superradiantly) scattered outward near the boundary of the ergosphere then get partially or completely reflected back in at some location far away from the black hole. These two reflecting locations form a cavity which can trap the waves, and specific frequencies which fit inside the cavity (by satisfying the appropriate boundary conditions) will be resonances/bound states of the system. If the resonance lies within the superradiant bandwidth, it will be unstable and grow exponentially in time. Hence, the BHB is simply a superradiant bound state which occupies a region outside the black hole's ergosphere.

In gravity, a natural reflection mechanism arises for light massive fields, which are prevented from propagating at infinity by the effect of their own mass. This has led to intense interest on ultralight axions, which are speculated to arise in various standard model extensions \cite{marsh2016axion}. These fields can form clouds around black holes through the superradiant instability, sapping rotational energy from the system and dissipating into gravitational waves. This process leads to distinct observational signatures, raising the intriguing possibility of using black holes as observatories for axions \cite{arvanitaki2015discovering,brito2015black,brito2017gravitational,baumann2019probing}. However, it has been argued that nonlinear interactions could saturate growth of the cloud before it manages to spin down the black hole \cite{fukuda2020aspects}.
It is therefore pertinent to develop a more thorough understanding of nonlinear effects on BHB--type instabilities. In this regard, analogue models of the BHB in fluids (which possess their own unique nonlinearities) could help shed light on the influence of nonlinearities in a general sense, and can be implemented experimentally to test predictions.

BHBs in analogue systems have already been discussed, e.g. in \cite{berti2004quasinormal,ciszak2021acoustic}.
However, these works do not take the inherent dispersion or dissipation of the fluid into account, which will be present in any real experiment.
For instance, the detection of superradiance in \cite{torres2017rotational} was performed using deep water gravity waves, which are strongly dispersive. Subsequent theoretical analysis \cite{patrick2020superradiance} demonstrated that the character of superradiance (i.e. the amount of amplification and the bandwidth of amplified frequencies) can differ significantly from the simple long wavelength systems studied by many authors. Furthermore, damping (e.g. due to viscosity) could significantly alter the instability timescales or, in the most extreme cases, fully quench the instability and stabilise the system.
In view of this, our goal in this work will be to assess the impact of dispersion and dissipation on the BHB instability in analogue fluid systems.
Using a simplified model of a draining vortex, we will develop a calculational tool for computing unstable mode frequencies in strongly dispersive settings, laying the theoretical ground work for future experiments.

\subsection{Method}

To achieve this goal, the most suitable existing theory is the deep water theory of superradiance outlined in \cite{patrick2020superradiance}.
The theory is based on a Wentzel-Kramers-Brillouin (WKB) approximation, which considers wavelengths shorter than the scale over which the background varies.
The method reduces a complicated wave scattering problem governed by partial differential equations (PDEs) to a far simpler algebraic equation which governs the ray trajectories through phase space. The rays are the paths traced out by a collection of points along a wavefront of short wavelength (high frequency) waves and are analogous to spacetime geodesics in relativity.
Since conventional relativistic theories are second order in space and time derivatives, the solution of the linearised theory at a given frequency will be a superposition of two waves, representing the two allowed directions of propagation (in-going and out-going).
Hence, in the WKB theory, there will be two rays and intersections of the rays (caustics) are the locations where the wave scatters off the inhomogeneous background medium.
Similarly, the long wavelength (or shallow water) limit of linear fluid dynamics is second order in space and time derivatives and the two solutions correspond to the in/out-going waves in the regions where the flow velocity is weak. 
For shorter wavelengths, however (which is the natural regime for WKB techniques), higher order spatial derivatives in the linear fluid equations become important and the general solution becomes a superposition of more than two waves.
Correspondingly, the WKB theory contains more than two rays.
One then has to consider all possible intersections between neighbouring rays to fully understand the scattering of waves off the inhomogeneous background.
This was the approach taken in \cite{patrick2020superradiance} (and subsequently \cite{patrick2020quasinormal,patrick2022quantum}) which introduced a method of schematically sketching the interactions between different rays to facilitate the computation of scattering coefficients.
These scattering diagrams will be the principle tool used in this work as we attempt to understand the dispersive BHB around a draining vortex.

However, the deep water theory in \cite{patrick2020superradiance} contains unphysical negative energy modes (with divergent wavenumber $k$) asymptotically far from the vortex core and is therefore not suitable for an analysis of the BHB, which is dependent on the nature of reflections at the outer boundary.
These unphysical modes arise when the (sub-luminal) deep water theory is applied beyond its regime of validity. 
Surface tension regulates the theory by making it super-luminal at large $k$, so that in the weak field (i.e. slow flow) limit there are only two propagating modes outside the vortex corresponding to the usual in/out-going ones.
It is these modes which get reflected at the (mirror) boundary and give rise to the BHB.
Therefore, the primary aim of this work will be to establish the WKB theory of capillary-gravity waves around a draining vortex, and derive all the allowed scattering diagrams.
Using the diagrams, we will identify the necessary conditions for superradiant amplification then apply these to the calculation of oscillation frequencies and growth rates of the BHB modes.

\subsection{Outline and summary of results}

In Section~\ref{sec:system}, we start by reviewing the equations of fluid mechanics in the context of the draining vortex, before deriving the equations of motion for capillary-gravity waves.
Then in Section~\ref{sec:WKB}, we apply the WKB method to these equations and introduce the necessary concepts to study the scattering of rays in the inhomogeneous draining vortex background.
Next, we illustrate how the framework can be applied to the simpler case of shallow water waves in Section~\ref{sec:shallow}.
In this regime, the equations of motion can be cast in terms of an effective rotating black hole metric, allowing us to identify the superradiant instability as a BHB instability.
The main result is that the oscillation frequency $\omega_n$ is determined by a resonance condition of the form,
\begin{equation} \label{result1}
    \mathrm{Re}[S_\mathrm{path}(\omega_n)] = \pi\left(2n+\frac{1}{2}\right),
\end{equation}
where $S_\mathrm{path}$ is the phase integral for the radial part of the mode (which has to be evaluated numerically) along a trapped path consisting of a pair of rays, and $n=0,1,...$ indexes the different possible bound states.
We find an expression for the growth rate $\Gamma_n$ which has the intuitive interpretation that the exponential growth of the mode is equal to the amount of superradiant amplification i.e.
\begin{equation} \label{result2}
    e^{\Gamma_n T_\mathrm{path}(\omega_n)} = |\mathcal{R}(\omega_n)|,
\end{equation}
where $T_\mathrm{path}$ is the time elapsed along the rays in the trapping region, and $|\mathcal{R}|$ is the magnitude of the reflection coefficient characterising the scattering of an in-going ray with the inhomogeneous draining vortex flow.
When rotational superradiance occurs, we have $|\mathcal{R}|>1$ (i.e. amplification) and can immediately see that $\Gamma_n>0$ so that the eigenmode is unstable.
Since the destabilising mechanism is superradiant scattering, we identify these modes as the BHB modes.
We calculate the complex eigenfrequencies $\omega^\mathbb{C}_n=\omega_n+i\Gamma_n$ of the BHB modes for a range of parameters, comparing our WKB predictions with exact numerical calculations and finding excellent agreement.
We then identify the key features of the instability, allowing us to generalise the BHB to the dispersive regime.

In Section~\ref{sec:deep}, we analyse the scattering of capillary-gravity waves with the draining vortex, identifying all possible scattering diagrams in the same vein as \cite{patrick2020superradiance}. 
From the diagrams, we infer a condition for superradiant amplification of the form,
\begin{equation} \label{bandwidth}
    0<\omega<\omega_\star,
\end{equation}
where $\omega_\star$ is a characteristic location on the dispersion relation (to be discussed in detail).
We find in Fig.~\ref{fig:crit_C} that this band opens at a critical rotation, hence, there is a minimum required speed of rotation for superradiance to occur.
With this information, we obtain a condition for the existence of BHB modes in the system, accounting for both the dispersion and dissipation of the fluid.
The oscillation frequency is determined by a resonance condition which has the same form as the shallow water formula \eqref{result1}.
Intuitively, one would expect the formula \eqref{result2} for the growth rate to be modified by the amount of amplification on the right-hand side becoming suppressed.
This can happen in two ways: firstly, amplification can be weakened by dissipation and secondly, scattering with the extra dispersive modes can lead to energy escaping the trapping region.
We find that this expectation is indeed correct and that $\Gamma_n$ is governed by the relation,
\begin{equation} \label{result3}
    e^{\Gamma_nT_\mathrm{path}(\omega_n)} = e^{-\nu\mathrm{Re}[\Sigma_\mathrm{path}(\omega_n)]}\prod_a|X_a(\omega_n)|,
\end{equation}
where the $X_a$ are the different scattering coefficients along the trapped path consisting of a pair of two or more rays, and $\nu\Sigma_\mathrm{path}$ is the integral of the damping factor over the trapping region.
The different terms in this expression will be explained in detail in the main text and the origin of the formula in Appendix~\ref{app:res}.
We then calculate $\omega^\mathbb{C}_n=\omega_n+i\Gamma_n$ for some exemplary system parameters, finding in Fig.~\ref{fig:BS_maxinstab} that generally the instability is strongest in small systems (i.e. when the encircling mirror has a small radius) when the rotation speed is large and the drain is near the maximum allowed value that permits a trapping region outside the vortex.
We conclude in Section~\ref{sec:conc} by discussing future directions for our work.

\section{The system} \label{sec:system}

In this section, we present the  stationary background flow for our draining vortex before deriving a set of linearised equations governing small amplitude capillary-gravity waves. The key equations are the draining vortex flow profile \eqref{basic}, the linear equations of motion \eqref{lineqs} and statement of norm conservation in \eqref{cons_norm}.

\subsection{Basic equations}

\begin{figure}
\centering
\includegraphics[width=\linewidth]{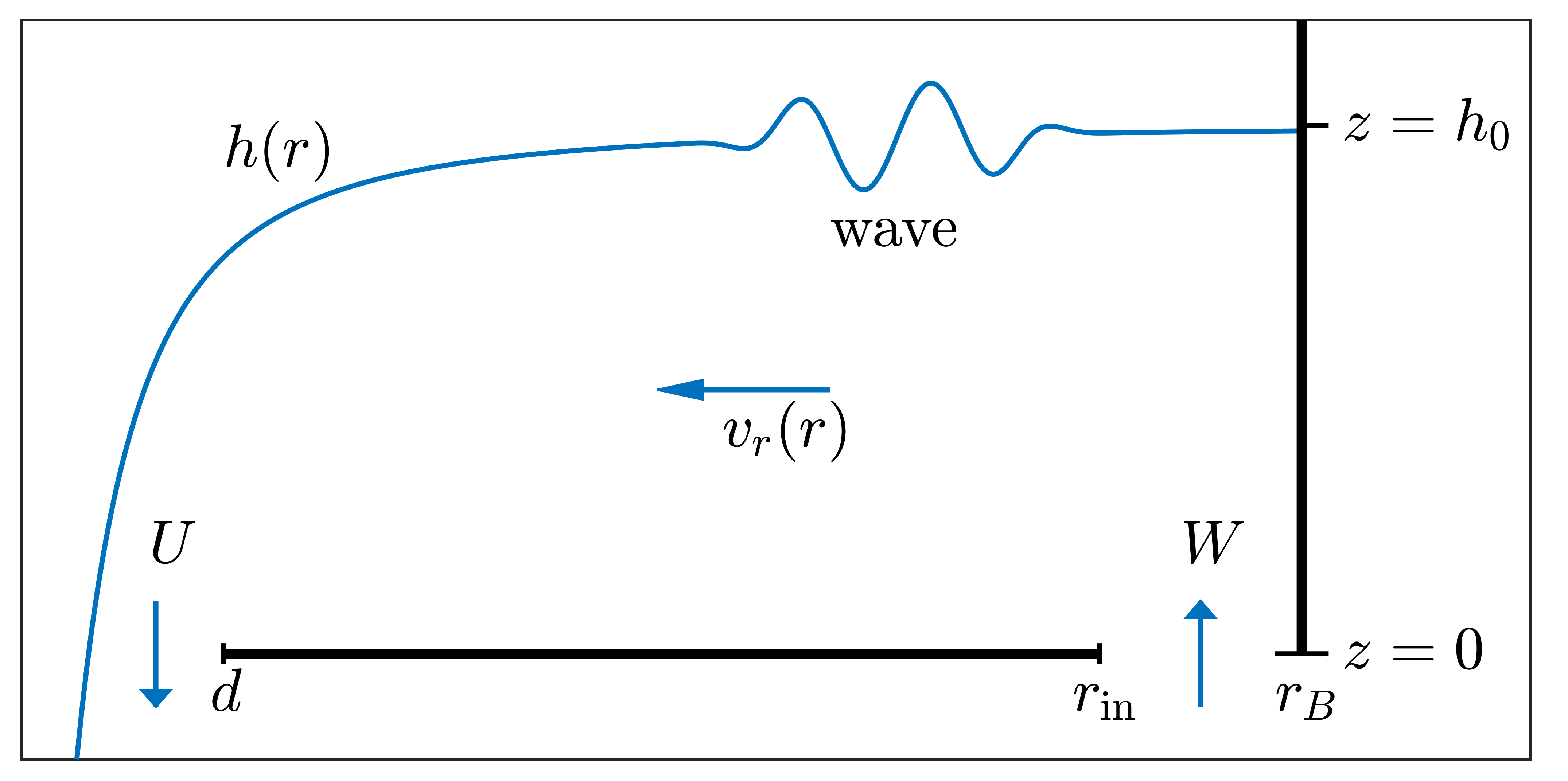}
\caption{A schematic illustration of our set-up. We display only a cross section in the $(r,z)$--plane. There is a hard wall boundary situated at $r=r_B$. Water enters through a cylindrical inlet at the bottom of the container near the outer wall, then moves inward with velocity $v_r(r)$ and exits through a hole of radius $d$ in the centre. The velocity field also has a component in $\theta$--direction, not shown. The surface is given by the function $h(r)$. In this work, we consider the dynamics of surface waves, also schematically depicted.} \label{fig:schem}
\end{figure}

We consider a cylindrical tank of water with a hole of radius $r=d$ in the centre of the plane at $z=0$.
Water drains toward the centre and is resupplied at the edge $r=r_B$ of the cylindrical tank.
The fluid has a free surface located at $z=h(x,y)$.
A schematic of the set-up is depicted on Fig.~\ref{fig:schem}.

The incompressible Navier-Stokes equations in three spatial dimensions are,
\begin{equation} \label{NavSto}
    (\partial_t+ \mathbf{v}\cdot\grad)\mathbf{v} - \nu\nabla^2\mathbf{v} = -\frac{\grad P}{\rho} + \mathbf{g},
\end{equation}
where $\mathbf{v}$ is the fluid velocity, $P$ is the pressure and (for water) the density $\rho\simeq 997\mathrm{kg m^{-3}}$ (which is taken to be constant at room temperature) and the kinematic viscosity $\nu\simeq 8.93\times10^{-7}\mathrm{m^2s^{-1}}$.
The acceleration due to gravity is $\mathbf{g} = -g\hat{\mathbf{e}}_z$ (with $g\simeq 9.81\mathrm{ms^{-2}}$ at the surface of the earth) and we assume no other external forces.
Since the fluid is incompressible, the velocity field satisfies,
\begin{equation} \label{DivFree}
    \grad\cdot\mathbf{v} = 0.
\end{equation}
We further assume an irrotational fluid satisfying,
\begin{equation} \label{CurlFree}
    \grad\times\mathbf{v} = 0 \quad \Rightarrow \quad \mathbf{v} = \grad\Phi,
\end{equation}
where $\Phi$ is the velocity potential.
Under these conditions, \eqref{NavSto} and \eqref{DivFree} can be reformulated as the Bernoulli and Laplace equations,
\begin{equation} 
    \partial_t\Phi + \frac{1}{2}\mathbf{v}^2 + \frac{P}{\rho} + gz = \mathrm{const}, \qquad \nabla^2\Phi = 0,
\end{equation}
where the relation \mbox{$\nabla^2\mathbf{v} = \grad(\grad\cdot\mathbf{v}) - \grad\times\grad\times\mathbf{v}$} has been used to drop the term proportional to $\nu$ and \mbox{$\mathbf{v}\cdot\grad\mathbf{v}=\frac{1}{2}\grad(\mathbf{v}^2)-\mathbf{v}\times\grad\times\mathbf{v}$} to re-express the advective term.

The boundary condition at the free surface is \cite{landau2013fluid,stepanyants2008stationary},
\begin{equation} \label{pressure_BC}
    P(z=h) = P_a - \sigma\grad_\parallel\cdot\left(\frac{\grad_\parallel h}{\sqrt{1+|\grad_\parallel h|^2}}\right) + 2\rho\nu\partial_z^2\Phi
\end{equation}
where $P_a$ is atmospheric pressure which we take to be zero, $\grad_\parallel$ is the gradient operator parallel to the $(x,y)$ plane and $\sigma\simeq 7.2\times10^{-2}\mathrm{kg s^{-2}}$ is the surface tension.
The second term on the right results from the curvature of the free surface and the final term encodes viscous friction within the fluid.
Bernoulli's equation at the free surface is then,
\begin{equation} \label{Bern}
\begin{split}
    & \Bigg[\partial_t\Phi + \frac{1}{2}\mathbf{v}_\parallel^2 + \frac{1}{2}v_z^2 + gh - 2\nu\nabla^2_\parallel\Phi \\
    & \qquad \qquad - \frac{\sigma}{\rho}\grad_\parallel\cdot\left(\frac{\grad_\parallel h}{\sqrt{1+|\grad_\parallel h|^2}}\right) \Bigg]_{z=h} = 0,
\end{split}
\end{equation}
whilst Bernoulli's equation in the bulk simply gives an expression for the pressure.
Note, the assumption that the fluid motion is irrotational means that the boundary condition on the tangential stress has been dropped. This leads to an underestimation of dissipation since we are not accounting for losses into the rotational degrees of freedom in the fluid.
A more thorough treatment for 1D deep water waves in a weakly viscous fluid leads to the replacement $\nu\to 2\nu$ in the results, see e.g. \cite{lucassen1968longitudinal}. Since we are only concerned with the generic effect of dissipation, we do not consider such losses here.

If the basic flow is assumed stationary and axisymmetric, it is natural to work in polar coordinates $(r,\theta)$. In the region where the fluid height is approximately uniform, the basic flow is given by,
\begin{equation} \label{basic}
    \mathbf{v} = -\frac{D}{r}\hat{\mathbf{e}}_r +\frac{C}{r}\hat{\mathbf{e}}_\theta, \qquad h \simeq h_0 = \mathrm{const},
\end{equation}
which is valid in the region $r\gg r_a\equiv \sqrt{(C^2+D^2)/2gh_0}$
We refer the reader to Appendix~\ref{app:flow} for a thorough discussion of the approximations that lead to \eqref{basic}.
Even working with these simplified profiles, solving the scattering problem over the full parameter space is a challenging task, especially with the inclusion of dispersive effects. Hence, order of magnitude estimates for the various parameters will be useful when it comes computing bound state frequencies.
We assume that $d$ and $r_B$ are respectively on the order of centimeters and tens of centimeters.
To get an estimate of the flow parameters achievable in such a set-up, we assume that (in practice) $r_a$ will be some significant fraction of the radius of the drain hole radius, say $r_a\sim\mathcal{O}(d)$.
Assuming $h_0$ in the range of millimeters to tens of centimeters, we find that $C$ and $D$ should roughly lie in the range $10$--$100\mathrm{cm^2s^{-1}}$, in accord with experimental realisations \cite{andersen2003anatomy,torres2017rotational}.
  
\subsection{Linear equations}

To study perturbations of the basic flow, we work in the approximation that gradients of the free surface can be neglected, i.e. $h'\simeq 0$. This approximation is valid suitably far from the vortex axis, i.e. $r\gg r_a$.
We now perturb the fluid variables as,
\begin{equation}
    \Phi\to\Phi+\phi, \qquad h\to h+\eta,
\end{equation}
where $\phi$ and $\eta$ are the velocity potential and height perturbations respectively.
\eqref{Bern} gives the first equation,
\begin{equation}
\left(\partial_t+\mathbf{v}_\parallel\cdot\grad_\parallel\right)\phi + g\eta - \frac{\sigma}{\rho}\nabla^2_\parallel\eta - 2\nu\nabla^2_\parallel\phi \Big|_{z=h} = 0.
\end{equation}
The second is obtained by solving Laplace's equation $\nabla^2\phi=0$ in the bulk.
In the region $r>d$, we can write a solution in terms of Fourier modes,
\begin{equation} \label{lin_sol}
    \phi = \int \frac{d^2\mathbf{k}}{(2\pi)^2}\phi_\mathbf{k}e^{i\mathbf{k}\cdot\mathbf{x}}, \qquad \phi_\mathbf{k} = A_\mathbf{k}\cosh(k z),
\end{equation}
where $k=|\mathbf{k}|$.
Using this, we can write,
\begin{equation} \label{vert_fluc}
    \partial_z\phi|_{z=h} = F(-i\grad_\parallel)\phi|_{z=h},
\end{equation}
where we have defined the operator,
\begin{equation} \label{Fk}
    F(\mathbf{k}) = k\tanh(hk).
\end{equation}
Now, perturbing the kinematic boundary condition (see \eqref{kinFS} in Appendix~\ref{app:flow}) we obtain,
\begin{equation}
    \partial_z\phi|_{z=h} + \eta\partial_zv_z  = (\partial_t+\mathbf{v}_\parallel\cdot\grad_\parallel)\eta,
\end{equation}
which can be reformulated when combining with \eqref{vert_fluc} as,
\begin{equation}
    (\partial_t + \grad_\parallel \cdot\mathbf{v}_\parallel)\eta - F(-i\grad)\phi|_{z=h} = 0,
\end{equation}
where the divergence operator in parentheses acts on both $\mathbf{v}_\parallel$ and $\eta$ to the right.
Finally, we drop the subscript $\parallel$ and $|_{z=h}$ and understand that all quantities in the linear theory live on a two dimensional plane at $z=h$. 
The linearised equations of motion are,
\begin{equation} \label{lineqs}
\begin{split}
\left(\partial_t+\mathbf{v}\cdot\grad\right)\phi + g\eta - \gamma\nabla^2\eta - 2\nu\nabla^2\phi = 0, \\
(\partial_t + \grad \cdot\mathbf{v})\eta - F(-i\grad)\phi = 0,
\end{split}
\end{equation}
where we have introduced $\gamma=\sigma/\rho$.
The term proportional to $\nu$ encodes the effect of dissipation.
The conservative part of the dynamics (i.e. excluding the term proportional to $\nu$) can be derived from the action,
\begin{equation} \label{action}
\begin{split}
    \mathcal{S} = & \ \int dt d^2\mathbf{x}\bigg[-\eta D_t\phi - \frac{1}{2}g\eta^2 \\
    & \qquad \qquad \qquad - \frac{1}{2}\gamma|\grad\eta|^2 - \frac{1}{2}\phi F(-i\grad)\phi \bigg],
\end{split}
\end{equation}
where $D_t=\partial_t+\mathbf{v}\cdot\grad$ is the material (convective) derivative on the free surface.

Although it is not the case for our draining vortex, if the basic flow were independent of the coordinates, \eqref{lineqs} admits an exact solution in terms of plane waves $e^{i\mathbf{k}\cdot\mathbf{x}-i\omega t}$ with dispersion relation,
\begin{equation}
    \omega = -i\nu k^2 +\mathbf{v}\cdot\mathbf{k} \pm \sqrt{(gk+\gamma k^3)\tanh(hk)-\nu^2 k^4},
\end{equation}
and $k=|\mathbf{k}|$. In Section~\ref{sec:WKB}, we will see how a local version of this dispersion relation can be obtained when $\mathbf{v}$ is not constant.
We will then be concerned with two limits.
The low $k$ limit of the theory is,
\begin{equation}
    \omega = \mathbf{v}\cdot\mathbf{k} \pm \sqrt{gh}k,
\end{equation}
that is, one can neglect the effects of surface tension, viscosity and deep water effects arising from the hyperbolic tangent.
We also consider the deep water limit $k\gg h^{-1}$, which leads to,
\begin{equation}
    \omega = -i\nu k^2 +\mathbf{v}\cdot\mathbf{k} \pm \sqrt{gk+\gamma k^3-\nu^2 k^4}.
\end{equation}
In this case, the depth of the fluid is effectively infinite (relative to the wavelength).
The price to pay is that, in the regime of short wavelengths, the effects of surface tension and viscosity are important and should be properly taken into consideration.

\subsection{Norm conservation}

Splitting the fluctuations into positive and negative frequency components,
\begin{equation}
    \phi\to\phi+\phi^*, \qquad \eta\to\eta+\eta^*,
\end{equation}
the action \eqref{action} becomes,
\begin{equation}
\begin{split}
    \mathcal{S} = & \ \int dt d^2\mathbf{x}\bigg[-\eta^* D_t\phi -\eta D_t\phi^* - g|\eta|^2 - \gamma|\grad\eta|^2 \\ & \quad  - \frac{1}{2}\phi F(-i\grad)\phi^* - \frac{1}{2}\phi^* F(-i\grad)\phi \bigg].
\end{split}
\end{equation}
To recover the equations of motion, we vary the action with respect to $\phi$ and $\eta$ independently.
This operation involves the commutation of derivative terms, hence, we define $F$ by its Taylor expansion about $k=0$,
\begin{equation} \label{taylor}
    F(\mathbf{k}) = \sum_{m=1}^\infty F_m k^{2m},
\end{equation}
where we have used the fact that $F$ is an even function of $k$.
This allows us to write,
\begin{equation}
\begin{split}
    \phi F(-i \grad) & \phi^* = \left[F(-i\grad)\phi\right]\phi^* - i\grad\cdot\Bigg(\sum_{m=1}^\infty F_m\times \\ & \sum_{n=0}^{2m-1}(-1)^n(-i\grad)^n\phi(-i\grad)^{2m-1-n}\phi^*\Bigg).
\end{split}
\end{equation}
Varying the action then leads to,
\begin{equation}
\begin{split}
    & \delta\mathcal{S} = \ \int dt d^2\mathbf{x}\Bigg\{\delta\eta^*\left[-D_t\phi-g\eta+\gamma\nabla^2\eta\right] + \\
    & \qquad \delta\phi^*\left[\partial_t\eta + \grad\cdot(\mathbf{v}\eta) - F(-i\grad)\phi\right] -\partial_t(\eta\delta\phi^*) \\ 
    & \qquad -\grad\cdot\bigg[\gamma\delta\eta^*\grad\eta-\mathbf{v}\delta\phi^*\eta + \frac{i}{2}\sum_{m=1}^\infty F_m \times \\
    & \sum_{n=0}^{2m-1}(-1)^n(-i\grad)^n\phi(-i\grad)^{2m-1-n}\delta\phi^* \bigg] + \mathrm{c.c.}\Bigg\}.
\end{split}
\end{equation}
The total derivatives can be rewritten as terms on the boundary where the variations are taken to vanish.
Setting $\delta\mathcal{S}=0$ then yields the equations of motion.
Alternatively, the total derivative terms can be used to deduce conserved currents of the system. 
Indeed, the action is invariant under continuous phase rotations of the form,
\begin{equation}
    \phi\to\phi e^{-i\alpha}\simeq \phi + \delta\phi, \qquad \delta\phi = -i\alpha\phi,
\end{equation}
where $\alpha$ is infinitesimal, and similarly for $\eta$.
Stripping out $\alpha$, we obtain a conservation equation for the norm current,
\begin{equation} \label{cons_norm}
    \partial_t\rho_n + \grad\cdot\mathbf{J}_n = 0,
\end{equation}
whose components are given by,
\begin{equation} \label{norm}
\begin{split}
    \rho_n = & \ 2\mathrm{Im}[\eta\phi^*], \\
    \mathbf{J}_n = & \ 2\mathrm{Im}\bigg[\gamma\eta^*\grad\eta+\mathbf{v}\phi^*\eta + \frac{i}{2}\sum_{m=1}^\infty F_m \times \\
    & \sum_{n=0}^{2m-1}(-1)^n(-i\grad)^n\phi(-i\grad)^{2m-1-n}\phi^*\bigg].
\end{split}
\end{equation}
This conserved current is related to the expression for the varying wave amplitude that we will shortly derive.

\section{WKB method} \label{sec:WKB}

In the previous section, we derived the set of equations obeyed by capillary gravity waves \eqref{lineqs} in the vortex flow field of \eqref{basic}. In this section, we present an approximate method of solving these equations. Key concepts will be the dispersion relation \eqref{disp} and transport equation \eqref{transport} which determine the local wavelength and amplitude of a wave. A formula is given in \eqref{refl_formula} that describes the local scattering of waves with the vortex and in \eqref{energy_balance}, we argue that norm conservation implies superradiant amplification for certain frequencies.

\subsection{Ray tracing} \label{sec:ray}

The equations of motion in \eqref{lineqs} are complicated to solve in practice.
However, simplifications occur if we consider the limit of short wavelengths.
In particular, if the local wavelength of the fluctuations is much smaller than the scale over which the basic flow varies, we can assume that the solution looks locally like a plane wave, with an amplitude and wavevector which vary gradually over the flow. 
Each point along a wavefront will then trace out a path as it moves through the system.
These paths are known as rays, and are the fluid mechanical/optical analogue of geodesics in curved spacetime.
This level of approximation, where only the ray trajectories are considered, is known as the \textit{geometric optics} limit. 
The geometric optics result then forms the leading order term in the WKB expansion, with the next-to-leading order term encapsulating variations of the wave amplitude.
More details about the method can be found in e.g. \cite{berry1972semiclassical,buhler2014waves,torres2018waves}.

To find the rays, we write,
\begin{equation} \label{wkb_ansatz}
    \begin{bmatrix}
    \phi \\ \eta
    \end{bmatrix} = \begin{bmatrix}
    \mathcal{A}(\mathbf{x},t) \\ \mathcal{B}(\mathbf{x},t)
    \end{bmatrix}
    e^{i\mathcal{S}(\mathbf{x},t)},
\end{equation}
where the amplitudes $\mathcal{A},\mathcal{B}$ vary smoothly with respect to the phase $\mathcal{S}$.
The wavevector $\mathbf{k}$ and frequency $\omega$ of the wave are defined by,
\begin{equation}
    \omega = -\partial_t\mathcal{S}, \qquad \mathbf{k} = \grad\mathcal{S}.
\end{equation}
Importantly, these can be functions of $\mathbf{x}$ and $t$ in inhomogeneous media.
Substituting \eqref{wkb_ansatz} into \eqref{lineqs} and neglecting derivatives of $\mathcal{A}$, $\mathcal{B}$, $\mathbf{k}$ and $\omega$ (to capture the smoothly varying assumption) we obtain a local dispersion relation (see Appendix~\ref{app:trans} for further details),
\begin{equation} \label{disp}
\begin{split}
    \Omega^2 = & \ \mathcal{D}(k), \qquad \Omega = \omega-\mathbf{v}\cdot\mathbf{k}, \\
    \mathcal{D}(k) = & \ f(k)F(k), \qquad f(k) = g+\gamma k^2,
\end{split}
\end{equation}
where $\mathcal{D}(k)$ is the dispersion function and $k=|\mathbf{k}|$ is the modulus of the wavevector. The inverse length scale $k_c=\sqrt{g/\gamma}$ determines the importance of surface tension relative to gravity. For room temperature water, we find $k_c\simeq 370\mathrm{m^{-1}}$. Surface tension therefore becomes important when the local wavelength is $\lesssim 1.7\mathrm{cm}$.

The rays are defined to be perpendicular to the wavefronts (i.e. surfaces of constant phase) hence, the ray equations are obtained by extremising $\mathcal{S}$.
To ensure that \eqref{disp} is satisfied, a constraint $\mathcal{H}=0$ is added to $\mathcal{S}$ to define an action for the rays,
\begin{equation}
    S = \int d\lambda \left(k_\mu\dot{x}^\mu-\xi\mathcal{H}\right)
\end{equation}
where \mbox{$x^\mu=(t,\mathbf{x})$}, \mbox{$k_\mu=(-\omega,\mathbf{k})$}, $\xi$ is a Lagrange multiplier, $\lambda$ is an arbitrary parameter along the rays and \mbox{$\dot{}=d/d\lambda$}.
Rescaling $\lambda\to\lambda/\xi$ absorbs the Lagrange multiplier.
The effective Hamiltonian is,
\begin{equation} \label{Ham_eff}
\begin{split}
    \mathcal{H} = & \ -\frac{1}{2}\left(\omega-\omega_D^+\right)\left(\omega-\omega_D^-\right), \\
    \omega_D^\pm = & \ \mathbf{v}\cdot\mathbf{k} \pm \sqrt{\mathcal{D}(k)},
\end{split}
\end{equation}
where $\omega_D^\pm$ are the upper ($+$) and lower ($-$) branches of the dispersion relation and the rays are the solutions of Hamilton's equations,
\begin{equation} \label{Hamilton}
    \dot{x}^\mu = \frac{\partial\mathcal{H}}{\partial k_\mu}, \qquad \dot{k}_\mu = -\frac{\partial\mathcal{H}}{\partial x^\mu}.
\end{equation}
When $\mathcal{D}(k)\sim k^2$, the rays coincide with geodesics of the effective spacetime geometry seen by shallow water waves (discussed in Section~\ref{sec:spacetime}).

\begin{figure}
\centering
\includegraphics[width=\linewidth]{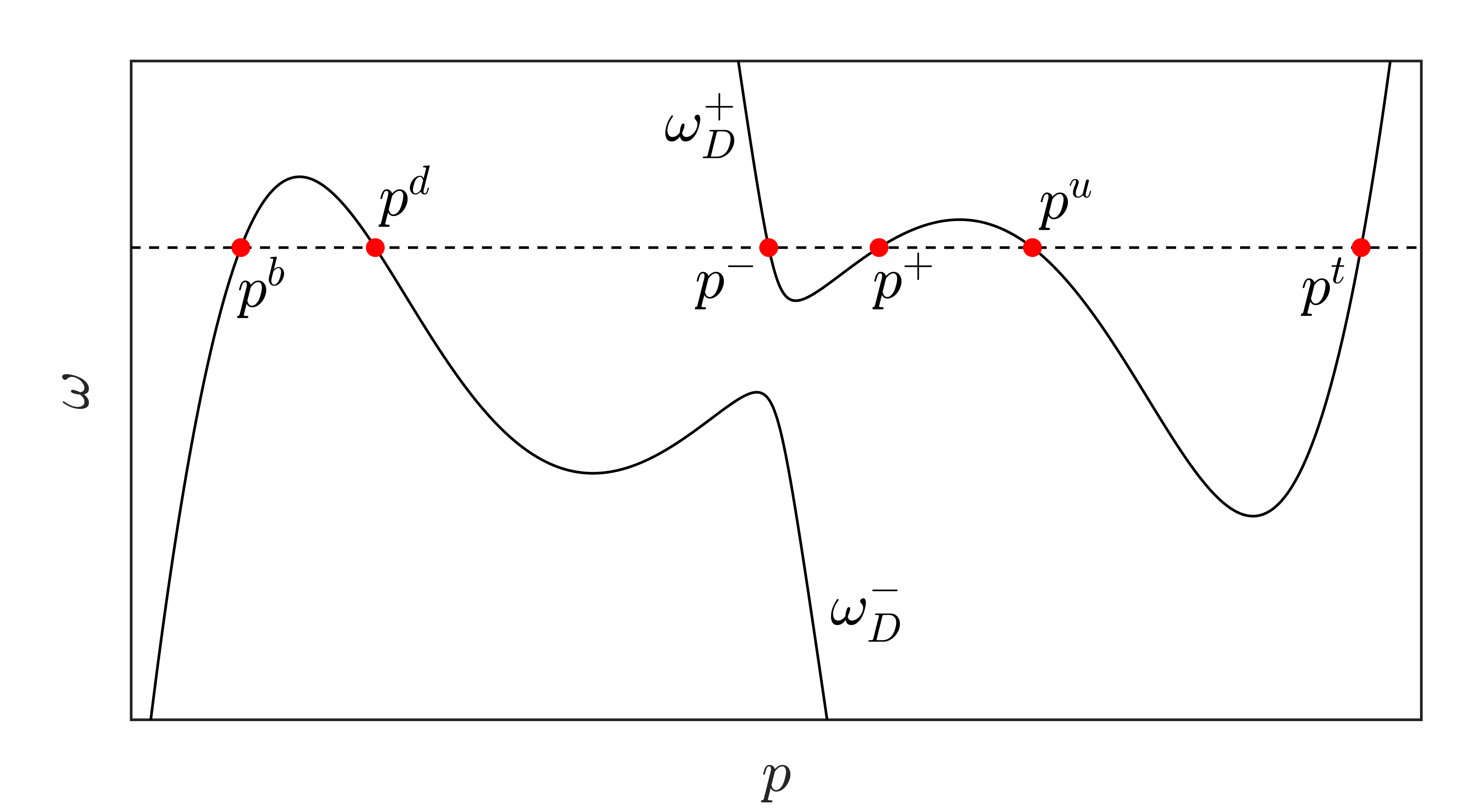}
\caption{A schematic of the dispersion relation for capillary-gravity waves.
The curves $\omega_D^\pm$ in \eqref{Ham_eff} at fixed $m$ and $r$ are shown as black curves.
The dispersion relation in \eqref{disp} is equivalent to $\omega=\omega_D^\pm$.
Hence, for a given frequency (see the dashed line for an example) the solutions of this equation define the different allowed roots $p^j$ (red dots).
For our system, there are a maximum of 6 real (propagating) roots at a fixed $r$. As $r$ is varied, the form of the $\omega^\pm_D(p)$ changes and the six roots will shift location. Note the $\omega_D^\pm$ are exaggerated here to clearly illustrate the six propagating roots. This schematic is representative of when surface tension is smaller than a critical value $\gamma_c$, discussed in detail in Section~\ref{sec:deep}.} \label{fig:disp}
\end{figure}

An important quantity in the following will be the group velocity, which we will soon see determines the direction along which the norm is transported.
The group velocity is defined as,
\begin{equation}
    \mathbf{v}_g \equiv \partial_\mathbf{k}\omega = \partial_\mathbf{k}\omega_D^\pm,
\end{equation}
where we have used the fact that the frequency satisfies $\omega=\omega_D^\pm$.
Hence, the group velocity is given by the gradient of the branches in the $\mathbf{k}$ direction at the points where $\omega=\omega_D^\pm$ has solutions.
Two useful properties of the group velocity are,
\begin{equation} \label{vg_rel}
    \partial_\mathbf{k}\mathcal{H} =  \mathbf{v}_g\Omega, \qquad \mathbf{v}_g = -\frac{\partial_\mathbf{k}\mathcal{H}}{\partial_\omega\mathcal{H}} = \frac{\dot{\mathbf{x}}}{\dot{t}}  =\frac{\partial \mathbf{x}}{\partial t}.
\end{equation}
where the latter follows from Hamilton's equations \eqref{Hamilton} and demonstrates that a point travelling through space along a ray parametrised by coordinate time moves in the direction of $\mathbf{v}_g$ as expected.

Solving Hamilton's equations yields the rays as function of the parameter $\lambda$. There exists, however, a more direct approach when the system has symmetries.
In particular, \eqref{basic} does not depend on $t$ or $\theta$.
Working in polar coordinates $k_\mu = (-\omega,p,m/r)$ and Hamilton's equations tell us that $\omega$ and $m$ are conserved along the rays.
Then, instead of solving $\dot{r}=\partial_p\mathcal{H}$ and $\dot{p}=-\partial_r\mathcal{H}$ for $r(\lambda)$ and $p(\lambda)$, one can search for the solutions $p(r)$ to $\mathcal{H}=0$, thereby solving the original dispersion relation in \eqref{disp}.
This amounts to searching for the intersections of $\omega_D^\pm(p)$ with $\omega=\mathrm{const}$ at fixed $m$ and $r$ (see Fig.~\ref{fig:disp}).

At each radius, $\mathcal{H}=0$ has solutions $p^j$ where $j=1...N$ and $N$ is an integer which depends on the functional form of $\mathcal{D}$.
The direction of travel along a ray is determined by the group velocity in the radial direction,
\begin{equation} \label{vgr_rad}
    v^j_{g,r}\equiv\hat{\mathbf{e}}_r \cdot\mathbf{v}^j_g=\partial_p\omega^\pm_D|_{p=p^j},
\end{equation}
and one selects $\pm$ depending on which branch $p^j$ is on.
For the dispersion relation \eqref{disp}, there are a maximum of 6 propagating modes at a given radius, $j\in\{b,d,-,+,u,t\}$.
The $p^\pm$ modes are the ones present in the low $k$ theory, whilst the others appear in pairs as a result of dispersion: $p^{d,u}$ arise from deep water effects \cite{patrick2020superradiance} whereas $p^{b,t}$ are due to surface tension \footnote{The labels are a nod to quark pairs in particle physics, i.e. up/down and top/bottom.}.
The modes are labelled in order of increasing $\mathrm{Re}[p^j]$ so that they appear from left to right on the dispersion relation when all 6 solutions are real, see Fig.~\ref{fig:disp}.
A subtlety of this construction is that, in many cases, not all 6 modes are simultaneously propagating at a given location. However, we will see in Section~\ref{sec:scattering} that there is a natural, unambiguous way to label the modes.

For the flow profile \eqref{basic} considered in this work, the solutions of the dispersion relation have the following asymptotic properties (these will be derived properly in Section~\ref{sec:scattering}).
Approaching the vortex axis, there are four propagating modes: an in-going/out-going pair on $\omega_D^+$ and another in-going/out-going pair on $\omega_D^-$.
Far away from the vortex core, there will be two propagating modes (one in-going and one out-going) which are on $\omega_D^+$ ($\omega_D^-$) for $\omega>0$ ($\omega<0$).
These are the parts of the solution capable of forming a BHB instability, provided a suitable condition for amplification is met.


\subsection{Transport equation}

In Appendix~\ref{app:trans}, we derive the following conservation equation describing the transport of the wave amplitude along the rays,
\begin{equation} \label{transport}
    \partial_t\left( \frac{\Omega\mathcal{A}^2}{f}\right) +  \grad\cdot\left(\frac{\mathbf{v}_g\Omega\mathcal{A}^2}{f}\right) + 2\nu k^2 \frac{\Omega\mathcal{A}^2}{f} = 0.
\end{equation}
(A similar equation for $\mathcal{B}$ can  be obtained using \mbox{$\mathcal{B} = i\Omega\mathcal{A}/f$}).
Note that the first two terms are equivalent to the conservation equation for the norm in \eqref{cons_norm} whilst the final term encodes dissipative losses.
Since the basic flow is independent of $(t,\theta)$, we can solve \eqref{transport} exactly by writing it as,
\begin{equation}
    \partial_r\log\left( \frac{rv_{g,r}\Omega\mathcal{A}^2}{f}\right) = -\frac{2\nu k^2}{v_{g,r}},
\end{equation}
and integrating,
\begin{equation} \label{amp}
    A = \frac{\alpha e^{i\varphi}}{|v_{g,r}\Omega f^{-1}|^{1/2}}\exp\left(-\nu\int\frac{dr}{v_{g,r}}k^2\right), \quad \mathcal{A} = \frac{A}{\sqrt{r}},
\end{equation}
where $\alpha$ is a complex valued integration constant. The slowly varying phase $\varphi(r)$ comes from the factor on the denominator whose modulus we have taken. In shallow water $\varphi$ is a constant and can be absorbed into $\alpha$.
Note that from Hamilton's equations \eqref{Hamilton}, we have the relation $dt=dr/v_{g,r}$.
Hence, the exponent is simply the damping coefficient $-\nu k^2$ integrated over the time elapsed along a given ray.
The mode amplitude therefore decays in the direction of $v_{g,r}$ as expected.
In the following, it will be easier to work with the amplitudes $A$, which are simply the full mode amplitudes $\mathcal{A}$ with $\sqrt{r}$ term factored out.
This term encodes the fact that as a radial wave propagates towards small $r$, its amplitude will increase by virtue of the fact that the norm density $\rho_n$ in \eqref{norm} has been focused onto a disk of smaller radius.
It is also useful to note that in regions where the $j^\mathrm{th}$--mode propagates, i.e. where $p^j\in\mathbb{R}$, the phase of $A$ can be entirely absorbed into the integration constant and we can write,
\begin{equation} \label{amp2}
    A^j  = \alpha^j\left|\frac{f^j}{\partial_p\mathcal{H}^j}\right|^\frac{1}{2}\exp\left(-\nu\int\frac{dr}{v^j_{g,r}}{k^j}^2\right),
\end{equation}
where we have used \eqref{vg_rel} to write $v_{g,r}\Omega=\partial_p\mathcal{H}$ and the superscript $j$ indicates that the amplitude is evaluated for a particular $p^j$.
In the evanescent region, the phase $\varphi(r)$ can acquire a non-zero value, although we will argue shortly that it does not contribute to any of the physics.

\subsection{Turning points} \label{sec:tps}

The amplitude \eqref{amp} clearly diverges at locations where $v_{g,r}$ vanishes.
By considering the branches of the dispersion relation in Fig.~\ref{fig:disp}, it is apparent at locations where $v_{g,r}\equiv\partial_p\omega_D^\pm=0$, two neighbouring $p^j$ can become degenerate and the corresponding rays will intersect.
The corresponding conditions on $\mathcal{H}$ are,
\begin{equation} \label{tp0}
    \mathcal{H}_\tp = 0, \quad \partial_p\mathcal{H}_\tp = 0,
\end{equation}
which, for fixed $(\omega,m)$ and a given set of background parameters, are solved by the pair $(r_\tp,p_\tp)$.
Hamilton's equations \eqref{Hamilton} imply that the points of intersection satisfy $\dot{r}=0$, i.e. they are turning points for the rays.
Physically, the ray is scattered (reflected) by the inhomogeneous nature of the medium; a ray which is initially travelling in one direction instantaneously comes to rest before being reflected as a different root of the dispersion relation.
During this reflection, the adiabatic assumption of the WKB approximation \eqref{wkb_ansatz} is violated since the local amplitude \eqref{amp} varies over a much shorter scale than the local wavenumber $p^j(r)$.

There exists a standard procedure to fix this pathological behaviour. At each location where $v_{g,r}=0$, one can expand the equations of motion \eqref{lineqs} in the vicinity of the turning point and find an exact solution for the locally degenerate roots.
This solution can then be mapped onto the WKB solution either side of the turning point (see Appendix~\ref{app:mat} for a detailed description of this procedure).
The result is a matrix equation relating the WKB amplitudes either side of the turning point, which we now state.

Let the intersecting rays be denoted $p^\Ri$ and $p^\Le$, such that $p^\Ri>p^\Le$ away from the turning point $r_\tp$. The labels $\Ri,\Le$ are placeholders for a neighbouring pair of $j\in\{b,d,-,+,u,t\}$ modes.
On the opposite side of $r_\tp$, the two modes will be a complex conjugate pair which we call $p^\downarrow$ (the mode which decays in the direction of increasing $r$) and $p^\uparrow$ (the mode which grows in the direction of increasing $r$).
These satisfy $\mathrm{Re}[p^\uparrow] = \mathrm{Re}[p^\downarrow]$ and $\mathrm{Im}[p^\uparrow] = -\mathrm{Im}[p^\downarrow]$.
When the pair are propagating for $r< r_\tp$ and evanescent for $r>r_\tp$, the local amplitudes in \eqref{amp} are related by,
\begin{equation} \label{tp1}
    \begin{pmatrix}
        A^\Ri_\tp \\ A^\Le_\tp
    \end{pmatrix} = e^{\frac{i\pi}{4}}\begin{pmatrix}
        1 & -\frac{i}{2} \\ -i & \frac{1}{2}
    \end{pmatrix} \begin{pmatrix}
        A^\downarrow_\tp \\ A^\uparrow_\tp
    \end{pmatrix},
\end{equation}
whilst in the reverse scenario (evanescent for $r< r_\tp$ and propagating for $r>r_\tp$) we have,
\begin{equation} \label{tp2}
    \begin{pmatrix}
        A^\uparrow_\tp \\ A^\downarrow_\tp
    \end{pmatrix} = e^{\frac{i\pi}{4}}\begin{pmatrix}
        \frac{1}{2} & -\frac{i}{2} \\ -i & 1
    \end{pmatrix} \begin{pmatrix}
        A^\Ri_\tp \\ A^\Le_\tp
    \end{pmatrix}.
\end{equation}
Each matrix equation is defined so that it takes mode amplitudes on the side of large $r$ and returns the same on the side of small $r$.

An important point is that whilst the prefactor in \eqref{amp} diverges at the turning points, the integral in the exponent containing a factor of $1/v_{g,r}$ does not.
This point is worth mentioning since the dissipation integral contributes to our expressions for the growth/decay rates, hence it is important to check that it is well-behaved.
This can be seen directly by expanding the effective Hamiltonian around the turning point, \mbox{$\mathcal{H}=\partial_r\mathcal{H}(r-r_\tp)+\frac{1}{2}\partial_p^2\mathcal{H}_\tp(p-p_\tp)^2$}, and using \eqref{vg_rel} to write $v_{g,r} = \Omega_\tp^{-1}\partial_p^2\mathcal{H}_\tp(p-p_\tp)=\Omega_\tp^{-1}[-2\partial_r\mathcal{H}_\tp\partial_p^2\mathcal{H}_\tp(r-r_\tp)]^{1/2}$, where the last equality follows from $\mathcal{H}=0$.
We can then evaluate the integral proportional to $\nu$ in \eqref{amp} in a small region around the turning point,
\begin{equation}
\begin{split}
    & \frac{\nu k_\tp^2\Omega_\tp}{\sqrt{-2\partial_r\mathcal{H}_\tp\partial_p^2\mathcal{H}_\tp}}\int^{r_\tp+\epsilon}_{r_\tp}\frac{dr}{\sqrt{r-r_\tp}} \\
    & \qquad \qquad \qquad \qquad = \frac{\nu k_\tp^2\Omega_\tp}{\sqrt{-\partial_r\mathcal{H}_\tp\partial_p^2\mathcal{H}_\tp}}\sqrt{2\epsilon},
\end{split}
\end{equation}
which is manifestly regular as $\epsilon\to 0$.

In general, there can be multiple turning points over the range in $r$, and to calculate the asymptotic scattering coefficients, one should take into account the reflections from each.
Our procedure to compute these amplitudes will follow that outlined in \cite{patrick2020superradiance,patrick2021rotational,patrick2022quantum} in that we only account for reflections at turning points, which give the dominant contributions to scattering.
Sub-dominant contributions can be obtained e.g. around saddle points \cite{torres2020estimate,patrick2020quasinormal} and complex turning points \cite{coutant2016imprint}.

There is a convenient method to identify the locations of the turning points which we will make use of in the coming sections. 
This method is equivalent to solving the conditions \eqref{tp0}, although it has a more natural interpretation in terms of the dispersion relation.
For fixed $m$ and $r$ (and a given background flow) one can identify the extrema $p_\ex$ of $\omega_D^\pm$ by solving $\partial_p\omega_D^\pm|_{p=p_\ex}$=0.
Since there are a maximum of three per branch (see Fig.~\ref{fig:disp}), we will label these $p^a_\ex$ with $a=1,2,3$, noting that the extrema on the lower branch are located at equal and opposite $p$ values to those on the upper branch. 
We label the extrema by their proximity to the $p=0$ axis, with $p_\ex^1$ ($-p_\ex^1$) being the closest on the upper (lower) branch.
By solving for all $r$, we can promote these solutions to functions of the radius, i.e. $p_\ex^a(r)$.
Substituting back into the appropriate branch of the dispersion relation, we obtain two sets of three curves,
\begin{equation} \label{tp_potentials}
    \omega^+_a = \omega^+_a(r,p_\ex^a(r)), \qquad \omega^-_a = \omega^-_a(r,-p_\ex^a(r)),
\end{equation}
which give the $\omega$ values at the extrema of the dispersion relation.
The utility of these curves is that once we select a particular frequency, we can solve,
\begin{equation} \label{tp_cond}
    \omega = \omega^\pm_a(r_\tp),
\end{equation}
to find all the turning points. We will see a simple example of this in the next section on shallow water waves.
However, it will be particularly useful in the analysis of deep water waves, where there is no simple analytic expression for the $r_\tp$.

At each turning point, the amplitudes are related by \eqref{tp1} or \eqref{tp2} depending on the orientation.
Between turning points, we assume that the different $p^j$ do not interact with each other and evolve adiabatically according to the WKB approximation.
Let us write the full solution for velocity potential perturbations $\phi$ at an initial location $r=r_b$ in the form,
\begin{equation} \label{expans1}
\begin{split}
    \phi_b = & \ \frac{e^{im\theta-i\omega t}}{\sqrt{r}}R_b, \\
    R_b = & \ \sum_j A_b^j(r) e^{i\int_\mathrm{ref} p^j_b dr} \equiv \mathbf{P}^\mathrm{ref}_b\cdot\mathbf{A}_b
\end{split}
\end{equation}
where $\phi_b=\phi(r_b)$ and similarly for the other quantities.
In the second equality for $R_b$, $\mathbf{P}^\mathrm{ref}_b$ is a row vector contain the reference phases of the $j$ modes at $r_b$ and $\mathbf{A}_b$ is a column vector containing the mode amplitudes.
The column vector is ordered so that modes with largest $\mathrm{Re}[p^j]$ appear toward the top.
We can construct the full solution at a new point $r=r_a$, provided the WKB approximation is valid everywhere for $r_a\leq r\leq r_b$, through the relation $R_a = \mathbf{P}^\mathrm{ref}_b\cdot \mathbf{A}_a$, where the amplitudes are related by,
\begin{equation} \label{shift}
    \mathbf{A}_a = \mathrm{diag(\mathcal{F}^j_{ab})}\mathbf{A}_b.
\end{equation}
The factors $\mathcal{F}^j_{ab}$ shift the phase and amplitude of the mode adiabatically according to \eqref{amp} and are given by,
\begin{equation} \label{shift2}
\begin{split}
    \mathcal{F}^j_{ab} = & \ \left|\frac{f^j_a\partial_p\mathcal{H}^j_b}{f^j_b\partial_p\mathcal{H}^j_a}\right|^\frac{1}{2}e^{i\left(\varphi^j_a-\varphi^j_b\right)} \exp\left(-i\int^{r_b}_{r_a}p^j dr\right)  \times \\
    & \qquad \qquad \qquad \qquad \exp\left(\nu\int^{r_b}_{r_a}\frac{dr}{v^j_{g,r}}{k^j}^2\right).
\end{split}
\end{equation}
The phases $\varphi^j$ are defined so that they go to zero on the turning points and thus, when $r_a$ and $r_b$ are turning points, the factor $\exp[i(\varphi^j_a-\varphi^j_b)]$ drops out of the expression above.
By \eqref{amp2}, the $\varphi^j$ are also zero in regions where $p^j\in\mathbb{R}$.
In this way, all information about the phase shifts at the turning points are contained in the integration constants $\alpha$ in \eqref{amp}. It is worth noting that one case where the $\varphi^j$ vanish identically is in the relativistic (non-dispersive) regime discussed in Section~\ref{sec:shallow}.

In general, we can describe the asymptotic scattering properties of the system through the transfer matrix $\mathcal{M}$,
\begin{equation} \label{scatterM}
    \mathbf{A}_0 = \mathcal{M}\mathbf{A}_\as,
\end{equation}
where $r_0$ ($r_\as$) is a point such that there are no turning points for $r<r_0$ ($r>r_\as$), ensuring that the scattering amplitudes computed at $r_0$ ($r_\as$) will be invariant moving toward smaller (larger) $r$.
The matrix $\mathcal{M}$ is built from combinations of the matrices in \eqref{tp1}, \eqref{tp2} and \eqref{shift}.
Its specific structure is determined by the precise number and location of turning points for a given $\omega,m$--mode.
We will see a specific example of how to construct $\mathcal{M}$ later on in Section~\ref{sec:scattering}.
The key point is that knowledge of the turning points for a given set of parameters uniquely determines $\mathcal{M}$, from which we can infer the scattering amplitudes.
Hence, to completely characterise the scattering of waves in our draining water tank, we should find all radial turning points $(r_\tp,p_\tp)$ by solving \eqref{tp0} over the whole parameter space of the equations in \eqref{lineqs}.
This procedure was previously carried out for shallow water ($\mathcal{D}=ghk^2$) \cite{patrick2021rotational}, weakly dispersive ($\mathcal{D}=ghk^2-gh^3k^2/3$) \cite{patrick2020quasinormal} and deep water ($\mathcal{D}=gk$) \cite{patrick2020superradiance} gravity waves, where the focus was on unbounded systems with $\nu=\gamma=0$.
The extension in this work to the case of bounded systems, with surface tension and dissipation properly accounted for, will enable a computation of unstable BHB growth rates.

\subsection{Local reflection coefficient}

An important case is when there are two turning points (say $r_a<r_b$) such that interacting modes are evanescent for $r_a<r<r_b$.
In this case, we can combine the matrices \eqref{tp1} and \eqref{tp2} with the shift factors \eqref{shift} to write,
\begin{equation} \label{refl_formula}
\begin{split}
    \begin{pmatrix}
        A^\Ri_a \\ A^\Le_a
    \end{pmatrix} = & \  \mathcal{N}_{ab}\begin{pmatrix}
        A^\Ri_b \\ A^\Le_b
    \end{pmatrix}, \\
    \mathcal{N}_{ab} = & \ \mathcal{F}^\downarrow_{ab}\begin{bmatrix}
        1+f_{ab}^2/4 & i(1-f_{ab}^2/4) \\ -i(1-f_{ab}^2/4) & 1+f_{ab}^2/4
    \end{bmatrix},
\end{split}
\end{equation}
where $\mathrm{Re}[p^\Ri]>\mathrm{Re}[p^\Le]$ and we have defined,
\begin{equation} \label{f12}
\begin{split}
    f_{ab} = & \ \exp\left(-\int^{r_b}_{r_a}\mathrm{Im}[p^\downarrow]dr\right) \times \\
    & \qquad \qquad \quad \exp\left(-i\nu\int^{r_b}_{r_a}\mathrm{Im}\left[\frac{{k^\downarrow}^2}{v_{g,r}^\downarrow}\right]dr\right).
\end{split}
\end{equation}
We will be particularly interested in the scattering of the long wavelength modes $p^\pm$.
For large $r$ and $\omega>0$, $p^+$ ($p^-$) is out-going (in-going) since $v_{g,r}^+>0$ ($v_{g,r}^-<0$) as $r\to r_\as$.
At small $r$, the same is true if $p^\pm$ are on the upper branch of the dispersion relation $\omega_D^+$ as $r\to r_0$.
However, for certain parameters ranges (to be deduced later) $p^\pm$ are solutions of $\omega=\omega_D^-$, in which case $p^+$ ($p^-$) is the in-going (out-going) mode as seen from $v_{g,r}^+<0$ ($v_{g,r}^->0$) as $r\to r_0$.

When $v^\Ri_{g,r}(r>r_b)>0$ and $v^\Le_{g,r}(r>r_b)<0$, the local reflection coefficient $\mathcal{R}_b = A^\Ri_b/A^\Le_b$ is given by,
\begin{equation} \label{refl_loc}
     \mathcal{R}_b = -i\left(\frac{1-f_{ab}^2/4}{1+f_{ab}^2/4}\right)^{\pm 1},
\end{equation}
where the $\pm$ corresponds to the sign of $v_{g,r}^\Ri(r<r_a)$.
Note that for $\nu=0$, the phase of $\mathcal{R}_b$ is entirely determined by the prefactor $-i$.
Treating $\nu$ perturbatively, we can expand $f_{ab}$ to show that,
\begin{equation} \label{R_nu0}
    \mathcal{R}_b = \mathcal{R}_b^{\nu=0}\left[1\pm\mathcal{O}\left(\nu |f_{ab}|^2\right)\right],
\end{equation}
where we have assumed that $f_{ab}$ is itself small, which is true in our case when $r_{a,b}$ are well separated.
Later, we present results to leading order in $\nu$ and $|f_{ab}|^2$, hence, reflection coefficients can be evaluated for $\nu=0$.

\subsection{Energy conservation}

The expression for the reflection coefficient in \eqref{refl_loc} can exceed unity in certain cases, implying amplification of scattered waves. This is precisely the phenomenon of rotational superradiance.
We can understand the origin of this amplification using energy considerations in the limit $\nu=0$.
First, we note that in stationary backgrounds (which is our case) the conserved energy current is related to the norm current in \eqref{cons_norm} by a factor of $\omega$.
For WKB modes, we show in Appendix~\ref{app:trans} that the components of the norm current for each WKB mode (labelled by $j$) is,
\begin{eqnarray} \label{WKBnorm}
    \rho^j_n = 2\frac{\Omega^j}{f^j}|A^j|^2, \qquad \mathbf{J}_n^j = \mathbf{v}^j_g \rho^j_n.
\end{eqnarray}
The second equality implies that the norm is carried in the direction of the group velocity.
The sign of the norm is determined by the comoving frequency $\Omega^j$, which is positive (negative) on $\omega_D^+$ and $\omega_D^-$.
Therefore, modes on $\omega_D^-$ have negative norm, and if they also have $\omega>0$ they have negative energy densities $\omega\rho_n$.
The region where this occurs is the dispersive analogue of the ergosphere around rotating black holes (discussed in Section~\ref{sec:shallow}).
In particular, its presence implies that there is an ingoing mode ($v_{g,r}^j<0$) with negative energy density, which contributes positively to the outgoing energy current.
This is ultimately the phenomenon responsible for superradiance.

To see this explicitly, we can rewrite \eqref{cons_norm} as a statement of energy conservation relating the various WKB modes evaluated near $r=0$ and $r\to r_\as$,
\begin{equation}
    \sum_{j=1}^4 \frac{\mathcal{H}'(p^j_0)}{f(p^j_0)}|A^j_0|^2 = \sum_{j=1}^2 \frac{\mathcal{H}'(p^j_\as)}{f(p^j_\as)}|A^j_\as|^2,
\end{equation}
where we have used the fact that there are 4 propagating modes approaching the vortex axis and 2 propagating modes far from the vortex, as discussed at the end of Section~\ref{sec:ray}.
Now, let us consider $\omega>0$ so that the modes far from the vortex are on $\omega_D^+$.
Using the expression for the amplitudes in \eqref{amp2} we can write the equation above as,
\begin{equation}
    |\mathcal{R}|^2 - 1 = \left|\frac{\alpha_0^{\mathrm{out}+}}{\alpha_\as^{\mathrm{in}+}}\right|^2 - \left|\frac{\alpha_0^{\mathrm{in}+}}{\alpha_\as^{\mathrm{in+}}}\right|^2 - \left|\frac{\alpha_0^{\mathrm{out}-}}{\alpha_\as^{\mathrm{in}+}}\right|^2 + \left|\frac{\alpha_0^{\mathrm{in}-}}{\alpha_\as^{\mathrm{in+}}}\right|^2,
\end{equation}
where we use the suffixes $\mathrm{in}\pm$ and $\mathrm{out}\pm$ to label the in-going and out-going modes on $\omega_D^\pm$ respectively (note that these labels are placeholders for the mode labels $j\in\{b,d,-,+,u,t\}$ which we will eventually solve for by drawing the scattering diagrams).
We have also introduced the asymptotic reflection coefficient \mbox{$\mathcal{R}=\alpha^{\mathrm{out}+}_\as/\alpha^{\mathrm{in}+}_\as$}, whose modulus coincides with the one introduced in \eqref{refl_loc}.
Our boundary conditions in the vortex core (discussed in detail in Appendix~\ref{app:BCs}) amount to assuming the vortex does not emit.
The condition above then becomes,
\begin{equation} \label{energy_balance}
    |\mathcal{R}|^2 - 1 = - \left|\frac{\alpha_0^{\mathrm{in}+}}{\alpha_\as^{\mathrm{in+}}}\right|^2 + \left|\frac{\alpha_0^{\mathrm{in}-}}{\alpha_\as^{\mathrm{in+}}}\right|^2.
\end{equation}
Depending on the background and wave parameters, the constraint that the vortex does not emit fixes one (but not both) of the terms on the right-hand-side to be zero. The condition on the parameters can then be expressed as a condition on $\omega$ such that when \eqref{bandwidth} is satisfied, $|\mathcal{R}|>1$ and we have superradiance.
When this occurs, \eqref{energy_balance} tells us that the vortex absorbs a negative energy, and energy conservation results in the amplification of the wave reflected to infinity. It will be the challenge of Section~\ref{sec:deep} to identify the characteristic frequency $\omega_\star$ in \eqref{bandwidth} that governs when this absorption of negative energy occurs.

\section{Shallow water waves} \label{sec:shallow}

So far, we have established a set of equations governing capillary-gravity waves around our draining vortex and presented a general method capable of handling effects of dispersion and dissipation.
In this section, we will apply our method in the shallow water regime where the analogy with fields around a black hole is exact.
In this way, we demonstrate the manner in which the instability of a draining vortex in a cylindrical container is equivalent to the BHB instability, before generalising this effect to situations where dispersion and dissipation are important.
The key results will be presented in Figs.~\ref{fig:complex1} and \ref{fig:vary1}, where we validate our procedure against standard numerical methods available to us in this regime.

In the low $k$ limit of the theory, $k^{-1}$ is assumed much larger than all other length scales in the problem (except $r_B$).
In this regime \eqref{Fk} becomes $F(k)\simeq hk^2$, and the effects of surface tension and viscosity can be neglected.
The effective Hamiltonian \eqref{Ham_eff} becomes,
\begin{equation} \label{Ham_eff1}
    \mathcal{H} = -\frac{1}{2}(\omega-\mathbf{v}\cdot\mathbf{k})^2 + \frac{1}{2}c^2k^2, \qquad c=\sqrt{gh},
\end{equation}
where $c$ is the shallow water wave speed.
The corresponding wave equation for velocity potential fluctuations is obtained from \eqref{lineqs},
\begin{equation} \label{waveqn_shal}
    (\partial_t+\grad\cdot\mathbf{v})(\partial_t+\mathbf{v}\cdot\grad)\phi - \grad\cdot(c^2\grad\phi) = 0.
\end{equation}
The magnitude of the group velocity in a frame comoving with the background fluid flow is $|\partial_\mathbf{k}\Omega|=c$, that is, all frequencies propagate at the same speed $c$; the system is \textit{non-dispersive}.
This greatly simplifies the analysis of \eqref{waveqn_shal} and provides a good starting point to illustrate the physics involved in the BHB instability.

\subsection{Analogy to black holes} \label{sec:spacetime}

In this subsection, we review the standard analysis of wave scattering in the analogue metric (e.g. \cite{schutzhold2002gravity,basak2003superresonance}). It is included for completeness so that we may (a) map the system at low $k$ to an analogue black hole and (b) identify the key features of the instability that should be present for the instability to be called a BHB. 
This is an important point since there is no strict notion of a horizon in the dispersive regime, however, there is still an instability which coincides with the BHB in the shallow water limit.

The analogy to the dynamics of fields in curved spacetime arises due to the equivalence of \eqref{waveqn_shal} with the Klein-Gordon equation,
\begin{equation} \label{KQeq}
    \frac{1}{\sqrt{-g}}\partial_\mu\left(\sqrt{-g}g^{\mu\nu}\partial_\nu\phi\right) = 0,
\end{equation}
\footnote{Note, the $g$ here is the determinant of the metric, not the gravitational acceleration used in \eqref{Ham_eff1} and elsewhere, which is conventionally denoted with the same letter.} where $\phi$ is a massless scalar field and the effective metric tensor $g_{\mu\nu}$ is given by,
\begin{equation} \label{PG}
\begin{split}
    & g_{\mu\nu}dx^\mu dx^\nu \\
    & \quad = c^2\left[-c^2 dt^2 + \left(dr-v_r dt\right)^2 + \left(r d\theta-v_\theta dt\right)^2\right].
\end{split}
\end{equation}
Note that when taking $h\simeq h_0$ (as we do in this work) the pre-factor $c^2$ is a constant and can be dropped.
Since \eqref{KQeq} is invariant under Lorentz transformations, we have an effective relativistic theory, with the shallow water wave speed $c$ playing the role of the speed of light.
Working with the basic flow field in \eqref{basic}, this metric has a horizon $r_h$ and an ergosphere $r_e$ at the locations,
\begin{equation} \label{rh_re}
    r_h = \frac{D}{c}, \qquad r_e = \frac{\sqrt{C^2+D^2}}{c},
\end{equation}
where horizon is the boundary of the region where the radial velocity exceeds the wave speed, $|v_r(r_h)| = c$, and the ergosphere is the boundary of the region where the magnitude of the total velocity field exceeds the wave speed, $|\mathbf{v}(r_e)|=c$.
Therefore, inside $r<r_h$ it is not possible for a wave satisfying \eqref{waveqn_shal} to move in the direction of increasing $r$.
Similarly, for $r<r_e$ waves satisfying \eqref{waveqn_shal} cannot move against the rotation of the vortex.
Inside the ergosphere, shallow water waves are permitted to have negative energy densities $\omega\Omega^j|\mathcal{A}^j|^2$. In particular if $\Omega<0$, the positive frequency mode crossing $r_h$ carries a negative energy in that direction, which results in amplification of the wave escaping to infinity.

It is possible to define the following dimensionless quantities,
\begin{equation} \label{rescale1}
    \frac{r}{r_h} \to r, \qquad \frac{ct}{r_h} \to t, \qquad \frac{C}{D} \to C,
\end{equation}
which amounts to setting $c=D=1$. In the text, we will work in these units to lighten the notation whilst in all figures, quantities will be dimensionful. This scaling will facilitate an efficient analysis of the parameter space accessible to shallow water waves around a draining vortex.

To study the scattering of waves described by \eqref{waveqn_shal}, the conventional method involves first transforming the field,
\begin{equation} \label{expans2}
    \phi = \sum_{\omega m}e^{im\theta-i\omega t}\exp\left(i\int\frac{\tilde{\omega}(r)dr}{r\Delta(r)}\right)\frac{R(r)}{\sqrt{r}},
\end{equation}
where the radial functions $R(r)$ are specific to each $\omega,m$ mode and we have defined,
\begin{equation}
    \tilde{\omega} = \omega-\frac{mC}{r^2}, \qquad \Delta(r) = 1-\frac{1}{r^2}.
\end{equation}
Next, the tortoise coordinate $r_*$ is defined as,
\begin{equation} \label{tortoise}
    r_* = r + \frac{1}{2}\log\left|\frac{r-1}{r+1}\right|,
\end{equation}
which maps $r\in(1,\infty)$ on to $r_*=(-\infty,\infty)$, i.e. the horizon is pushed to negative infinity.
The wave equation \eqref{waveqn_shal} then becomes a one-dimensional scattering problem in the radial direction
\begin{equation} \label{1d_eq}
    -\partial^2_{r_*}R + V(r)R = 0,
\end{equation}
where the effective potential is,
\begin{equation} \label{potential}
    V(r) = -\tilde{\omega}^2 + \Delta\left(\frac{m^2-1/4}{r^2}+\frac{5}{4r^4}\right).
\end{equation}
\eqref{1d_eq} has asymptotic solutions,
\begin{equation} \label{horizon_BC}
    R(r)\sim\begin{cases}
    A^-_h e^{-i\tilde{\omega}_hr_*}, \qquad \quad \quad \ \qquad r_*\to-\infty \\
    A^+_\infty e^{i\omega r_*} + A^-_\infty e^{-i\omega r_*}, \quad \ \ r_*\to\infty
    \end{cases}
\end{equation}
where we have imposed the boundary condition that waves are purely ingoing at the horizon and the subscript $h$ implies that $\tilde{\omega}$ is evaluated at $r=1$ in our dimensionless units.
Using the conservation of the Wronskian $W=R^*\partial_{r_*}R-R\partial_{r_*}R^*$, i.e. $\partial_{r_*}W=0$ which is equivalent to conservation of the norm current $\mathbf{J}_n$ in \eqref{norm}, the amplitudes can be shown to obey,
\begin{equation}
    |A_\infty^+|^2 + \frac{\tilde{\omega}_h}{\omega}|A_h^-|^2 = |A_\infty^-|^2.
\end{equation}
Hence positive frequency modes with $\omega<mC$ will be superradiantly amplified, i.e. $|A^+_\infty|>|A^-_\infty|$.
This relation is equivalent to the one in \eqref{energy_balance} as we will discuss below.

When the system is enclosed by an impenetrable boundary, superradiance can cause the vortex to become unstable: amplified waves are reflected from the boundary back into the vortex where they are further amplified, leading to a runaway process.
We will now show how the WKB method can be used to predict the frequencies of unstable modes formed in this way, before verifying the method's accuracy with by comparing with exact numerical solutions of \eqref{waveqn_shal}.

\subsection{WKB method} \label{sec:shal_WKB}

The calculation of the unstable mode frequencies proceeds in three steps.
Firstly, we find the radial momentum $p$ for the two shallow water modes and determine which of these represents out/in-going waves.
Next, we determine the reflection coefficient $\mathcal{R}$ for a wave incident on the vortex.
Finally, we use $p$ and $\mathcal{R}$ to deduce a resonance condition for waves which are scattered between the vortex and the boundary of the system. 
We assume here that $m\geq 0$ and allow $\omega$ to take on positive and negative values. The results for $m<0$ can then be inferred via the symmetry of the problem under $(\omega,m,p)\to(-\omega,-m,-p)$.

\subsubsection{Scattering}

In polar coordinates, the shallow water effective Hamiltonian \eqref{Ham_eff1} becomes,
\begin{equation}
    \mathcal{H} = -\frac{1}{2}\left(\omega-\frac{mC}{r^2}+\frac{p}{r}\right)^2 + \frac{1}{2}\left(p^2+\frac{m^2}{r^2}\right),
\end{equation}
where we have used the rescalings in \eqref{rescale1} to set $\mathcal{H}r_h^2/c^2\to\mathcal{H}$.
Since this is quadratic in $p$, $\mathcal{H}=0$ only has two solutions corresponding the modes labelled $p^\pm$ on Fig.~\ref{fig:disp}.
These can be expressed concisely as,
\begin{equation} \label{p_shal}
    p^\pm = \frac{\tilde{\omega}/r\pm\sqrt{-V}}{\Delta}, \qquad V = -\tilde{\omega}^2 + \Delta\frac{m^2}{r^2}.
\end{equation}
Note that the $V$ defined here is equivalent to that in \eqref{potential} in the limit $|m|\gg 1$, which is a formal requirement for the WKB approximation to hold (although good agreement with more precise methods is still obtained for the lowest $m$ modes as we shall soon see).
With this identification, \eqref{expans1} and \eqref{expans2} become equivalent for a single $\omega,m$ mode upon noting that the WKB approximation applied to \eqref{1d_eq} gives the phase factor as $\exp(\pm i\int\sqrt{-V}dr_*)$, where $dr_*=dr/\Delta$.

The expression for the amplitude in \eqref{amp2} is easily calculated for $\gamma=\nu=0$ using $\partial_p\mathcal{H}^\pm = \pm\sqrt{-V}$ which gives,
\begin{equation} \label{amp_shal}
    A^\pm = \alpha^\pm|V|^{-1/4}.
\end{equation}
This diverges for $V=0$, hence the turning points $r_\tp$ defined by \eqref{tp0} are given by the zeros of the potential.
Note that, in shallow water, the adiabatic amplitude change in \eqref{amp_shal} for the two modes $p^\pm$ is the same, something which won't be true for dispersive systems in general.
We also mention that the expression for $p^\pm$ in \eqref{p_shal} manifestly changes between real and complex when $V$ changes sign.

\begin{figure*} 
\centering
\includegraphics[width=\linewidth]{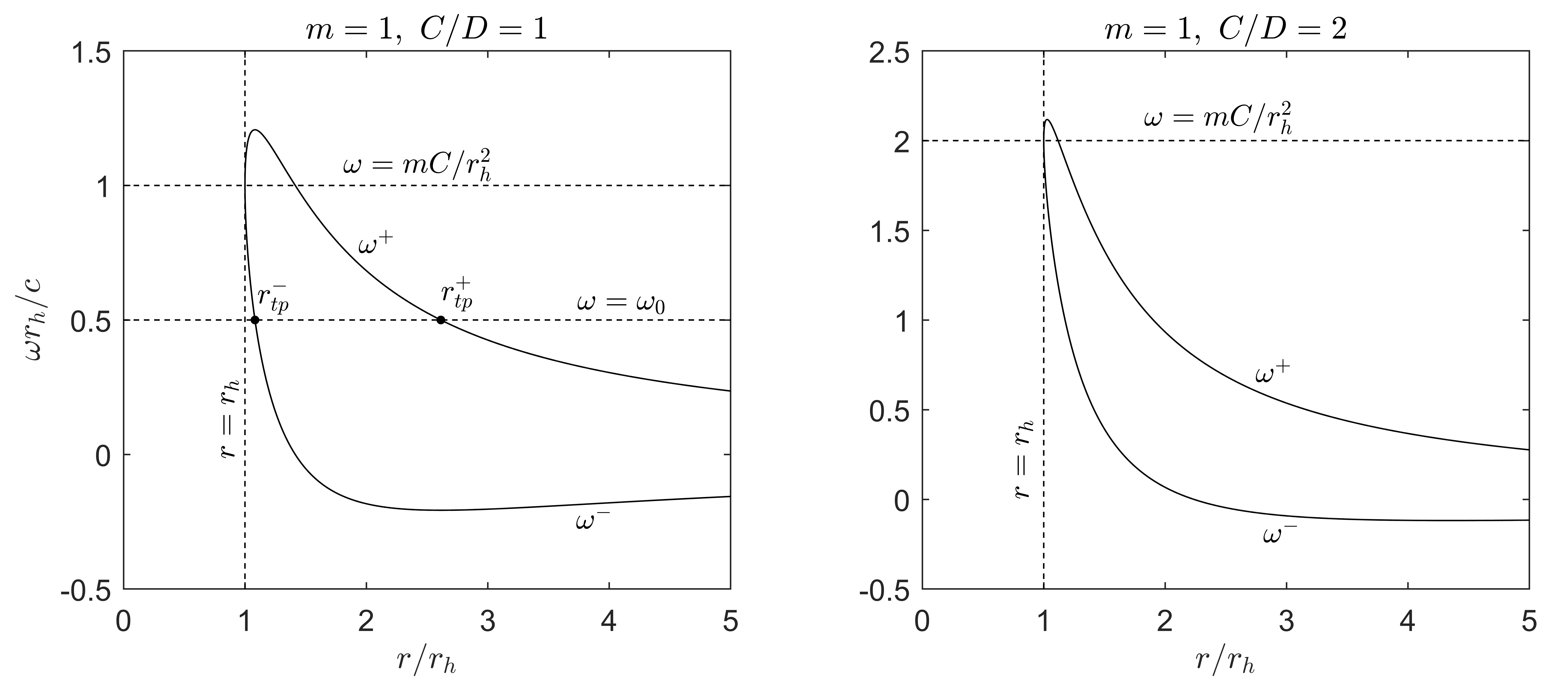}
\caption{The curves $\omega^\pm$ in \eqref{potential2} are illustrated for two different cases. In shallow water, the curves terminate at $r_h$ since $\omega_D^\pm$ have no extrema in the $p$ direction for $r<r_h$. Turning points are solutions of $\omega=\omega^\pm(r_\tp)$ and modes are evanescent for $\omega^-<\omega<\omega^+$. We display the $r_\tp$ for an example frequency $\omega_0$ in panel (a). The effect of increasing $C$ is to increase the asymmetric skew between $\omega^+$ and $\omega^-$ through the $mC/r^2$ term, whilst increasing $|m|$ results in an overall scaling by a factor of $|m|$.
The light ring frequencies $\omega_\lr^\pm$ are the extrema of $\omega^\pm(r)$.} \label{fig:omegapm}
\end{figure*}

To identify the turning points, we can write $V$ in the form,
\begin{equation} \label{potential2}
    V = -(\omega-\omega^+)(\omega-\omega^-), \qquad \omega^\pm = \frac{mC}{r^2}\pm\sqrt{\Delta\frac{m^2}{r^2}},
\end{equation}
where $\omega^\pm$ are defined in \eqref{tp_potentials} (the label $a$ is redundant here since in shallow water, $\omega_D^\pm$ each possess a maximum of one real extremum).
These curves are illustrated in Fig.~\ref{fig:omegapm}.
In shallow water, \eqref{tp_cond} admits closed form solutions for the two turning points,
\begin{equation} \label{tp_cf}
    r^\pm_\tp = \left[\frac{2m(C^2+1)}{2\omega C+m\mp2\sqrt{(\omega-\omega_\lr^+)(\omega_\lr^--\omega)}}\right]^{1/2},
\end{equation}
where the $\pm$ appearing on $r^\pm_\tp$ labels the location of the turning point such that $r^-_\tp<r^+_\tp$ and not the branch of the dispersion relation.
The extrema of $\omega^\pm$ in the $r$ direction are given by,
\begin{equation}
\begin{split}
    r_\lr^\pm = & \ \sqrt{\frac{2(C^2+1)}{C^2+1\pm C\sqrt{C^2+1}}}, \\
    \omega^\pm_\lr = & \ \frac{m(C\pm\sqrt{C^2+1})}{2},
\end{split}
\end{equation}
i.e. $\partial_r\omega^\pm|_{r=r^\pm_\lr}=0$ and $\omega_\lr^\pm = \omega^\pm(r_\lr^\pm)$, where the subscript stands for \textit{light-ring} to make the connection with the analogous concept around black holes \cite{cardoso2009geodesic}.
It is manifest from the expression in \eqref{tp_cf} that the $r_\tp^\pm$ are only real over the range $\omega_\lr^-<\omega<\omega_\lr^+$.
Using the relation $2m(C^2+1)={r_\lr^\pm}^2(2\omega_\lr^\pm C+m)$, the expression also makes it apparent that $r_\tp^\pm(\omega=\omega_\lr^+)=r_\lr^+$ and $r_\tp^\pm(\omega=\omega_\lr^-)=r_\lr^-$.

We clarify here the use of $\pm$ in the expressions above since there are three different meanings.
Firstly, in $\omega^\pm_\lr$ and $r_\lr^\pm$, the $\pm$ refers to the upper and lower branches of the dispersion relation $\omega_D^\pm$. 
Secondly, in $p^\pm$, the $\pm$ is just a label for the modes. For $\omega>\omega^+$, both $p^\pm$ are on the upper branch $\omega_D^+$ whilst for $\omega<\omega^-$ they are both on the lower branch $\omega_D^-$.
Thirdly, in $r_\tp^\pm$, the $\pm$ denotes ordering of turning points. Depending on the value of $\omega$, $r_\tp^\pm$ are either both on $\omega^+$, both on $\omega^-$, or $r_\tp^-$ is on $\omega^-$ and $r_\tp^+$ is on $\omega^+$. The latter case is when superradiance occurs.
These definitions are readily deducible by inspecting the left panel of Fig.~\ref{fig:omegapm}.

In terms of the potential, we can use \eqref{potential2} to show that $\partial_r V_{r=r_\lr}=0$ and $V(r_\lr,\omega_\lr)=0$.
In other words, $r_\lr$ is the location of the peak in the potential and $\omega_\lr$ is the frequency required for an $m$--mode to orbit the vortex on a circular trajectory at $r=r_\lr$.
This last fact can be deduced from the condition on the effective Hamiltonian \eqref{Ham_eff} at the light-ring,
\begin{equation} \label{lightring}
    \mathcal{H}_\lr=0, \quad \partial_p\mathcal{H}_\lr=0, \quad \partial_r\mathcal{H}_\lr=0,
\end{equation}
which, using Hamilton's equations \eqref{Hamilton}, implies $\dot{r}=0$ and $\dot{p}=0$.
That is, the ray which solves these equations has vanishing velocity and acceleration in the radial direction and therefore remains at $r=r_\lr$ indefinitely.
For our purpose, the light-ring is an important notion in determining the frequency ranges in which there are turning points (and therefore scattering within our approximation) and those without. Specifically, for $\omega_\lr^-<\omega<\omega_\lr^+$ there is a coupling between the two $p^\pm$ modes and in-going modes reflect will off the vortex.

The added utility of the curves $\omega^\pm(r)$ are that they allow us to infer the local energy density of the solutions.
A ray with frequency $\omega$ at a particular location $r$ lies on the upper (lower) branch of the dispersion relation $\omega_D^+$ ($\omega_D^-$) if the point $r,\omega$ is above $\omega^+$ (below $\omega^-$).
Using \eqref{WKBnorm}, one can infer that the sign of $\Omega$, and hence the sign of the WKB norm $\rho_n$ is positive (negative) on $\omega_D^+$ ($\omega_D^-$).
In stationary systems, the local energy density is simply $\omega\rho_n$ and thus, positive frequency modes on $\omega_D^+$ ($\omega_D^-$) have a positive (negative) energy flux in the direction of $v_{g,r}$ (see \eqref{WKBnorm}).
By inspecting Fig.~\ref{fig:omegapm} and noting that $\omega^\pm(r\to\infty)\to0$, we see that the condition for locally negative energy densities is fulfilled for modes at low $r$ in the frequency range $0<\omega<\omega^\pm(r_h)=mC/r_h^2$).
When the in-going wave at the horizon has a negative energy flux, conservation of energy means the wave reflected to infinity must have greater energy than the incident wave.
Hence, we recover the criterion for superradiant amplification $\omega<mC/r_h^2$ (for positive frequency waves).

\subsubsection{Reflection coefficient}

The WKB approximation allows us to find a semi-analytic expression for the reflection coefficient and therefore quantify the amount of amplification that occurs.
For this, we first need to know which of the modes $p^\pm$ is in-going on the horizon.
Evaluating \eqref{vgr_rad} for the two modes, we find,
\begin{equation}
    v_{g,r}^\pm = \frac{\pm\Delta\sqrt{-V}}{\tilde{\omega}\pm\sqrt{-V}/r},
\end{equation}
which approaching the horizon from outside (i.e. $\Delta\to 0^+$) becomes,
\begin{equation}
\begin{split}
    v^+_{g,r} & \ \overset{\Delta\to 0^+}{\sim} \begin{cases}
        \Delta/2, \qquad \ \ \, \tilde{\omega}_h>0, \\
        -\frac{2\tilde{\omega}_h^2}{\tilde{\omega}_h^2+m^2}, \quad \tilde{\omega}_h<0,
    \end{cases} \\
    v^-_{g,r} & \ \overset{\Delta\to 0^+}{\sim} \begin{cases}
        -\frac{2\tilde{\omega}_h^2}{\tilde{\omega}_h^2+m^2}, \quad \tilde{\omega}_h>0, \\
        \Delta/2, \qquad \ \ \, \tilde{\omega}_h<0. \\
    \end{cases}
\end{split}
\end{equation}
For $\tilde{\omega}_h>0$, we see that $p^+$ is out-going and $p^-$ is in-going just outside the horizon whilst for $\tilde{\omega}_h<0$, $p^-$ is out-going and $p^+$ is in-going.
Consistency with the horizon boundary condition in \eqref{horizon_BC} then tells us that the amplitude of the out-going mode should be set to zero.
Using the expression in \eqref{refl_loc} for the reflection coefficient, we obtain,
\begin{equation} \label{Refl_shal}
\begin{split}
    \mathcal{R} = & \ -i\left(\frac{1-e^{-2S_{-+}}/4}{1+e^{-2S_{-+}}/4}\right)^{\mathrm{sgn}(\tilde{\omega}_h)}, \\
    S_{-+} = & \ \int^{r_\tp^+}_{r_\tp^-}\frac{|\sqrt{-V}|dr}{\Delta}.
\end{split}
\end{equation}
Since $S_{-+}$ is real and positive, this approximate expression makes it manifest that $|\mathcal{R}|<1$ for $\tilde{\omega}_h>0$ whilst $|\mathcal{R}|>1$ for $\tilde{\omega}_h<0$.
This is consistent with \eqref{energy_balance}, where for $\tilde{\omega}_h<0$ ($\tilde{\omega}_h>0$), the in-going mode in the core is on $\omega_D^-$ ($\omega_D^+$) implying that the non-zero term on the right-hand zero of \eqref{energy_balance} is positive (negative) and the reflected wave is amplified (attenuated).

\begin{figure*} 
\centering
\includegraphics[width=\linewidth]{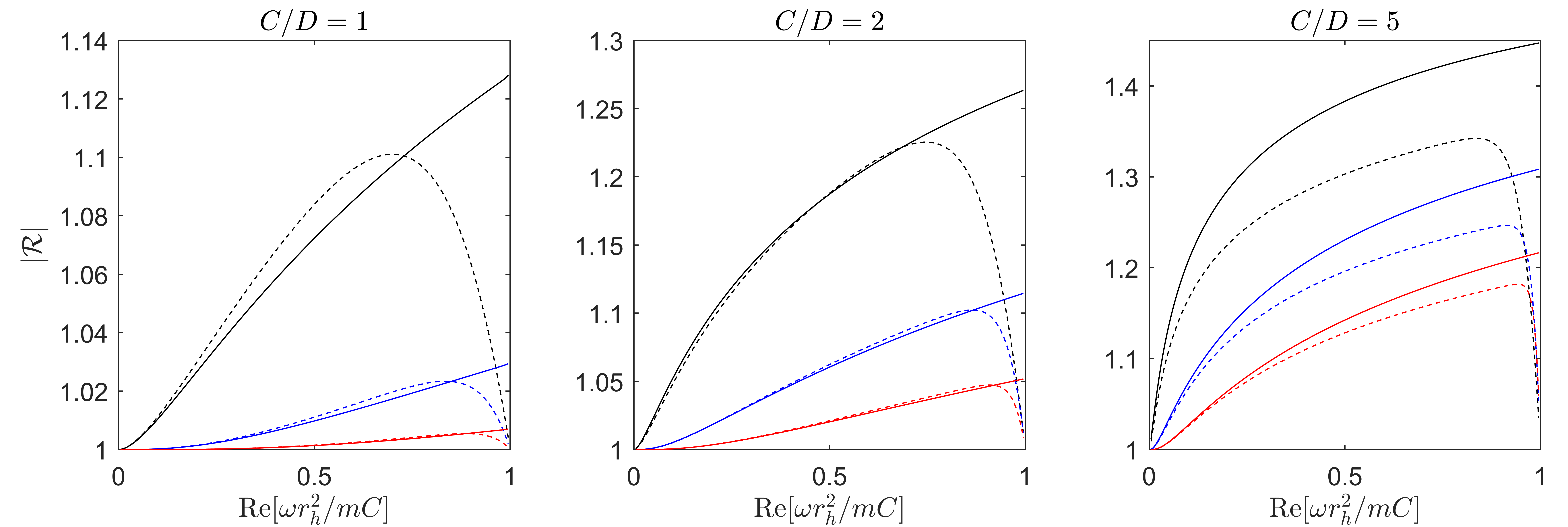}
\caption{Comparison between the WKB expression \eqref{Refl_shal} for the reflection coefficient (solid lines) and the exact value obtained numerically (dashed lines). Each panel represents a different dimensionless rotation parameter, and the horizontal axis is scaled so that the superradiant threshold coincides with 1. On each panel, we display the $m=1$ (black), $m=2$ (blue) and $m=3$ (red) results, which decrease in overall magnitude with $m$.
The agreement is best at low $C$.
Furthermore, the WKB result should not be trusted at the superradiant threshold since the geometric optics limit erroneously predicts a discontinuous change from amplification to absorption.} \label{fig:refl}
\end{figure*}

In Fig.~\ref{fig:refl}, we explicitly compare the expression in \eqref{Refl_shal} to the exact value obtained by numerically integrating the wave equation over the bandwidth of superradiant frequencies (see the Appendix A of \cite{patrick2020superradiance} for details of the integration).
We observe excellent agreement at low circulation whilst for higher values, our expression captures only the overall behaviour and order of magnitude.
The deviation occurs since for larger $C$, the tunnelling region between $\omega^+$ and $\omega^-$ becomes smaller for higher frequency (see Fig.~\ref{fig:omegapm} panel (b)) whilst WKB works best for well separated turning points.
Another feature that is not captured by the WKB approximation is that the correct $|\mathcal{R}|$ smoothly passes through unity crossing the superradiant threshold $\tilde{\omega}_h=0$.
The reason is that at the geometric optics level, the change from superradiant amplification to absorption corresponds to a discontinuous transition where the two modes $p^\pm$ transfer from the lower branch of the dispersion relation at small $r$ to the upper branch.
These differences between will be important in the next section when comparing the WKB growth rates of unstable modes with those obtained by a numerical computation.
One feature that is well captured by WKB is the decrease in the amount of amplification with $m$.
This can be seen from \eqref{Refl_shal} since, for a given $\omega/m$, $S_{-+}$ increases linearly with $|m|$ which means deviations of $|\mathcal{R}|$ from unity decrease.
Hence the most amplified modes in the system will occur for $m=1$.

\subsubsection{Bound states}

Having solved for the radial momenta $p^\pm$ and the reflection coefficient $\mathcal{R}$, we are poised to compute the resonance condition for trapped waves between the vortex and the boundary.
First we note that, in our formalism, this trapping can only occur when we have real turning points.
The frequency of any trapped modes must lie between the light-ring frequencies $\omega_\lr^-<\omega<\omega_\lr^+$.
A further constraint can be imposed by the requirement that the solution should be propagating approaching the boundary at $r_B$, leading to the constraint $\omega_\lr^-<\omega<\omega^-(r_B)$ for negative frequencies and $\omega^+(r_B)<\omega<\omega_\lr^+$ for positive frequencies.
The resonance condition can then be found by transporting the amplitude of the in-going mode at the boundary $A^-_B$ up to the turning point $r_\tp^+$ (according to \eqref{amp_shal} with the appropriate phase factor) where it gets reflected and transported back out to the boundary.
A detailed description of the computation can be found in Appendix~\ref{app:BCs}, where it is described as bound state A.
This leads to the condition for the real part of the frequency $\omega_n$,
\begin{equation} \label{res1_shal_WKB}
   S_{+B}(\omega_n) \equiv \int^{r_B}_{r_\tp^+(\omega_n)}\frac{\sqrt{-V(\omega_n)}dr}{\Delta} = \pi\left(n+\frac{1}{4}\right),
\end{equation}
where $n=0,1,...$ indexes the different modes which solve the condition.
A subtlety of the derivation is that we approximate the radial velocity at $r_B$ as zero so that the no penetration boundary condition implies the $\pm$ modes have equal amplitudes there. In reality, the resonance condition will be modified by either the small value of $v_r(r_B)$ or the inlet region where $v_r$ smoothly goes to zero on the boundary. However, we will soon see that the predictions of \eqref{res1_shal_WKB} agree exceptionally well with numerical calculations, suggesting that any differences due to the flow field over the inlet will be negligible.

Assuming the imaginary part of the complex eigenfrequency $\Gamma_n$ is a small correction to the real part, we find in Appendix~\ref{app:BCs} the expression,
\begin{equation} \label{res2_shal_WKB}
    \Gamma_n = \frac{\log|\mathcal{R}|}{2\partial_{\omega}S_{+B}}\Bigg|_{\omega=\omega_n},
\end{equation}
where $\mathcal{R}$ is given in \eqref{Refl_shal}.
Hence, we see explicitly that if the resonant frequency $\omega_n$ lies within the superradiant range where $|\mathcal{R}|>1$, $\Gamma_n>0$ and the bound state will be unstable.
Since we are only interested in the unstable modes, we can place constraints on values of the rotation parameter which result in an instability.
A minimum constraint for the existence of a bound state is that $r_\tp^+<r_B$ in \eqref{res1_shal_WKB} and, for superradiant frequencies, the minimum value of the outer turning point occurs at the superradiant threshold $\tilde{\omega}_h=0$. We can then invert the requirement $r_B>r_\tp^+(\omega=mC)\equiv\sqrt{C^2+1}/C$ to find the minimum $C$ value below which superradiant bound states cannot exist,
\begin{equation}
    C_\mathrm{min} = (r_B^2-1)^{-\frac{1}{2}}.
\end{equation}
For our analysis to apply, we must clearly have $r_B>1$ (i.e. the mirror is outside the horizon). Thus, as we bring the reflective boundary to its lowest allowed location, the rotation required for a superradiantly unstable mode tends to diverges.
In practice, this means that once we select a value of $r_B$, only $C>C_\mathrm{min}$ values are relevant to the study of the instability. Alternatively, if we fix $C$, the boundary should satisfy $r_B>\sqrt{C^2+1}/C$.

\begin{figure*} 
\centering
\includegraphics[width=\linewidth]{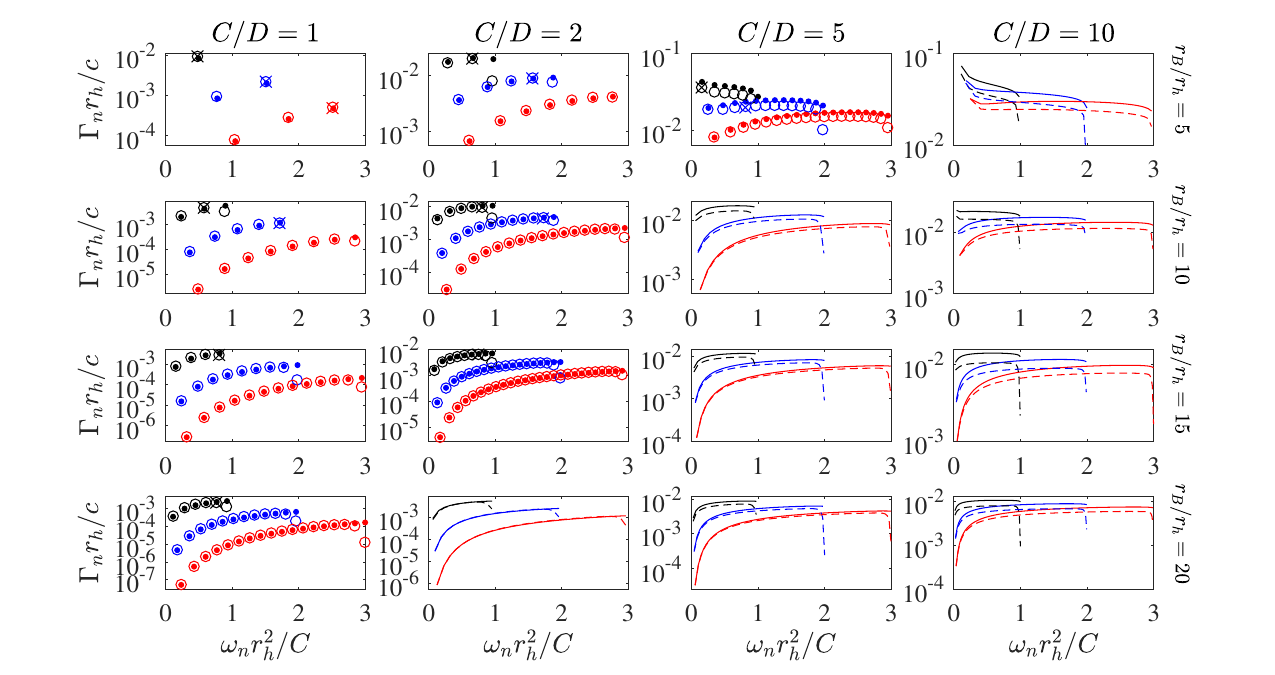}
\caption{We display the bound state frequencies $\omega^\mathbb{C}_n=\omega_n+i\Gamma_n$ obtained by solving the WKB conditions in \eqref{res1_shal_WKB} and \eqref{res2_shal_WKB} as solid points (or solid lines when there are sufficiently many states that the points become indiscernible). These are compared with numerical results of Section~\ref{sec:freqdom}, shown as open circles/broken lines. The real part $\omega_n$ is scaled by the dimensionless rotation so that the horizontal axis is the same for all panels.
The agreement of the real part is excellent in all cases whilst the imaginary part begins to deviate for large rotation.
Results of the time domain simulations from Section~\ref{sec:timedom} are displayed in a few cases as crosses and match the frequency domain simulations.
} \label{fig:complex1}
\end{figure*}

On Fig.~\ref{fig:complex1}, we display the complex frequencies \mbox{$\omega^\mathbb{C}_n = \omega_n+i\Gamma_n$} for the first three $m$--modes for a variety of $C$ and $r_B$ values. We find that the $m=1$ mode is the most unstable in all cases considered.
This is because reflection coefficient is largest for the lowest $m$--modes, as seen in Fig.~\ref{fig:refl}.
For a given $C$ and $r_B$, the spacing between modes is roughly constant, and unstable modes exist all the way up to the superradiant threshold. These two facts mean that, on a given panel, there are roughly $m$ times as many bound states in the $m>1$ channel compared with $m=1$ (at least in the limit of a large number of states).

\begin{figure*} 
\centering
\includegraphics[width=\linewidth]{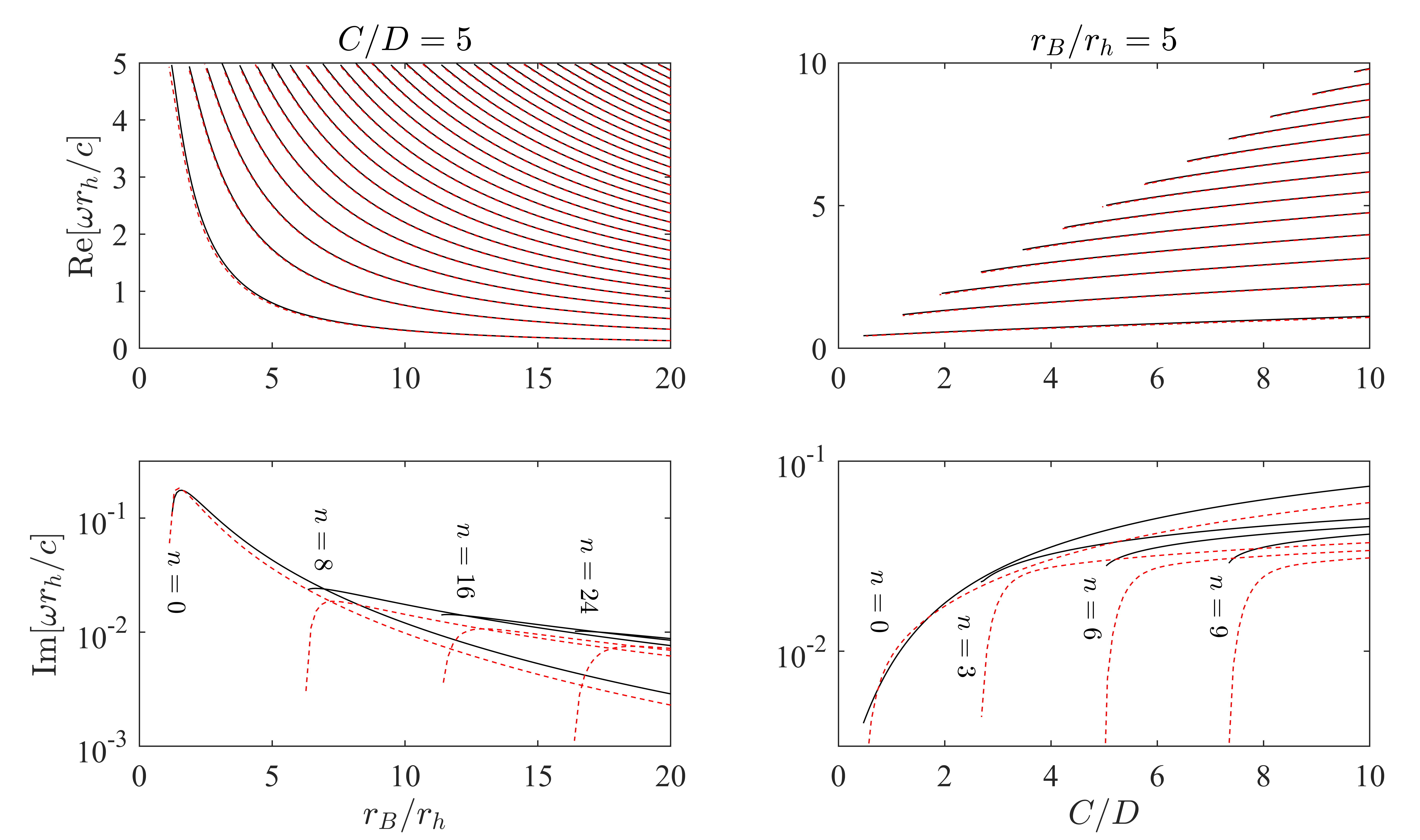}
\caption{Variation of the bound state frequencies with the cylinder size (left) and the dimensionless rotation (right). Only a few modes are displayed for the imaginary part to improve their visibility (on the plots of the real part, the modes are indexed $n=0,1,2,...$ as they appear from left to right). Increasing both $C/D$ and $r_Bc/D$ increases the number of bound states, with new modes first appearing at the superradiant threshold.
The WKB calculations are shown as solid black lines whilst numerical results are broken red. Agreement in the real part is excellent in all cases, whilst the imaginary part captures the order of magnitude and deviates significantly for modes near the superradiant threshold.} \label{fig:vary1}
\end{figure*}

We observe that increasing both $r_B$ and $C$ leads to the existence of more bound states. In the first case, increasing $r_B$ increases the density of states and the modes pack closer together in the allowed $\omega$--range.
In the second case, increasing $C$ raises the superradiant threshold leaving more room in the allowed $\omega$--range for the existence of trapped modes. 
This is further exemplified by Fig.~\ref{fig:vary1}. At fixed $C$, as we increase $r_B$ the oscillation frequency of each mode is $\sim 1/r_B$ asymptotically whilst at fixed $r_B$, they increase linearly with $C$ asymptotically.

There are two competing factors that determine the imaginary part. The first is that the factor $\log|\mathcal{R}|$ is larger for higher frequencies.
This leads to the increase of the growth rate with increasing $\omega_n$ at low $C$ and high $r_B$ on Fig.~\ref{fig:complex1}.
However, the time taken to orbit the cavity $2\partial_\omega S_{+B}$ goes to zero as the oscillation frequency goes to its lowest allowed value. 
If the frequency is small enough, this can outweigh the effect of the $\log|\mathcal{R}|$ and make small frequencies grow faster, e.g. in the top right panel of Fig.~\ref{fig:complex1}.
This can also be inferred from the lower left panel of Fig.~\ref{fig:vary1} where, as $r_B$ decreases, the $n=0$ mode eventually attains a larger growth rate than any of the higher $n$.

\subsection{Frequency-domain simulations} \label{sec:freqdom}

To validate the WKB used in the previous section, we compute numerically the spectrum of unstable modes in the shallow water regime by integrating \eqref{1d_eq} subject to in-going boundary conditions at the horizon up to \mbox{$r=r_B$}.
It is numerically more convenient to work with the radial coordinate $r$ rather than $r_*$, therefore the equation to be integrated is,
\begin{eqnarray}\label{eq:1d_in_r}
    \Delta^2 R''(r) + 2r^{-3}\Delta R'(r) + V(r) R(r) = 0.
\end{eqnarray}
The initial conditions are applied just outside the horizon, i.e. at $r=1+\epsilon$. To ensure that our solution satisfies the ingoing boundary condition, we write $R(r)$ as,
\begin{equation}\label{eq:ini_numerics}
    R(r) = e^{-i\tilde{\omega}_h r_*} \sum_{n=0}^N a_n (r-1)^n.
\end{equation}
The coefficients ${a_n}$ satisfy a 6--term recurrence relation given by,
\begin{eqnarray}
    \alpha_n a_{n+1} &+& 
    \beta_n a_n + 
    \gamma_n a_{n-1}  \nonumber\\
    &+&\delta_n a_{n-2} + 
    \epsilon_n a_{n-3} + 
    \lambda_n a_{n-4} = 0,
\end{eqnarray}
where the coefficients are explicitly,
\begin{eqnarray}
    \alpha_n &=& 8(1+n)(1 + n  - i\tilde\omega_h),\\
    \beta_n &=&-4 [1 + m^2 - 5 n^2 + n ( 3 + 8 i \tilde\omega_h)  \\
   & &- 2 C m \tilde\omega_h], \nonumber\\
    \gamma_n &=&  34 + 8 (C^2-1) m^2 - 48 n + 16 n^2\\ 
    & & + \tilde\omega_h [48i(1-n) + 32Cm], \nonumber\\
    \delta_n &=& 25 + 4 (5C^2-1) m^2 - 20 n + 4 n^2\\
    & & + \tilde\omega_h [32i(2-n) + 48Cm], \nonumber \\
    \epsilon_n &=& 16C^2 m^2 +8\tilde\omega_h[ 4Cm + i (3-n)],\\
    \lambda_n &=& 4 C m (C m + 2 \tilde\omega_h).
\end{eqnarray}
From the initial condition \eqref{eq:ini_numerics} with $N = 100$, we numerically integrate \eqref{eq:1d_in_r} from $r = 1 +\epsilon$, with $\epsilon = 10^{-5}$, up to $r=r_B$. We then extract the value of the derivative of the numerical solution at the wall $R'_{\mathrm{num}}(\omega,r=r_B)$. The numerical integration is performed using the Mathematica function NDSolve with the stiffness switching option.
Bound states frequencies are then defined by the condition,
\begin{equation}
    R'_{\mathrm{num}}(\omega,r=r_B) = 0.
\end{equation}
This condition is solved numerically using the FindRoot function with the WKB prediction as an initial guess.

The results are displayed on Figs.~\ref{fig:complex1} and \ref{fig:vary1}.
In all cases, we observe excellent agreement with the WKB prediction for the oscillation frequency.
This is because it is determined by the phase integral in a region of the flow where the velocity field of the central vortex is weakest. 
Consequently, the background varies smoothly and the WKB approximation works well. 
Conversely, agreement with the growth rate works well for reasonably low circulation but deviates notably for larger values.
The reason for this discrepancy can be traced back to the reflection coefficient in Fig.~\ref{fig:refl}.

\subsection{Time-domain simulations} \label{sec:timedom}

\begin{figure} 
\centering
\includegraphics[width=\linewidth]{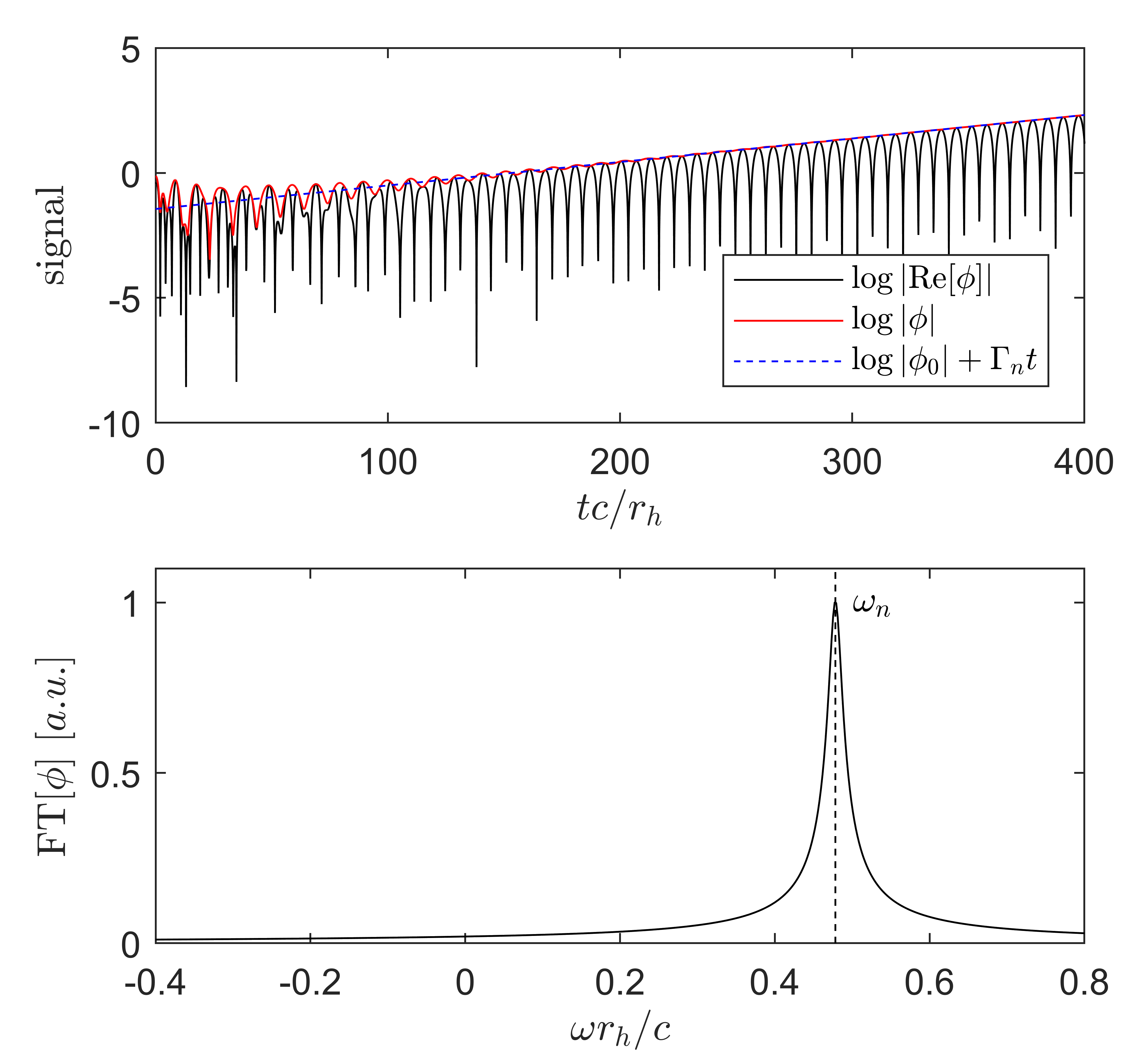}
\caption{The time development of the signal is extracted at the location $r/r_h=2$ and is plotted on a logarithmic scale in the upper panel. The growth rate $\Gamma_n$ is extracted by fitting the signal at late times to a straight line. The oscillation frequency $\omega_n$ is obtained by finding the peak in Fourier space (lower panel). The simulation parameters were $m=1$, $C=1$, $r_B=5$.} \label{fig:timedomain}
\end{figure}

The 1D equation in \eqref{1d_eq} can be simulated in the time-domain by restoring the time derivatives,
\begin{equation}
\begin{split}
    \partial_t\begin{pmatrix}
        R \\ \Pi
    \end{pmatrix} = & \ \begin{pmatrix}
        -imC/r^2 & 1 \\ \partial_{r_*}^2-U(r) & -imC/r^2
    \end{pmatrix}\begin{pmatrix}
        R \\ \Pi
    \end{pmatrix}, \\
    U(r) = & \ \Delta(r)\left(\frac{m^2-1/4}{r^2}+\frac{5}{4r^4}\right),
\end{split}
\end{equation}
where $\Pi=(\partial_t+imC/r^2)R$ and $r$ is understood as an implicit function of $r_*$ via \eqref{tortoise}.
This can be simulated using a $4^\mathrm{th}$--order Runge-Kutta algorithm.
At the inner and outer boundaries, we impose absorbing and reflective (Neumann) boundary conditions respectively,
\begin{equation}
    \left(\Pi-\partial_{r_*}R\right)|_{r_*=r_{*1}} = 0, \qquad \partial_{r_*}R|_{r_*=r_{*B}} = 0,
\end{equation}
where $r_{*1}$ is large and negative to approximate the location of the horizon (which is negative infinity) and $r_{*B}=r_*(r_B)$.
The initial condition is taken as a plane wave with wavenumber $\omega_0$ whose envelope is a Gaussian of width $\sigma_0$ centred at a location $r_{*0}$ near the outer boundary,
\begin{equation}
\begin{split}
    R(t=0) = & \ \exp\left[-\frac{(r_*-r_{*0})^2}{2\sigma_0^2}\right]e^{i\omega_0 r_*}, \\
    \Pi(t=0) = & \ 0.
\end{split}
\end{equation}
We use the parameters $\sigma_0=3$, $r_{*0}=3$ and $\omega_0$ is the frequency of the bound state with the largest growth rate in the WKB approximation.

In the time-domain, the fastest growing mode in the system will dominate the signal and be measurable at late times. We analyse the signal by plotting the time development at a single location, see Fig.~\ref{fig:timedomain}. 
The oscillation frequency of this mode is extracted by performing a Fourier transform, and the growth rate is found by fitting a straight line to the logarithm of the signal magnitude. 
On Fig.~\ref{fig:complex1}, we indicate the complex frequencies extracted in this manner with crosses.
When there are several nearby frequencies with comparable growth rates, the signal can exhibit a strong beating effect and it is not possible to reliably extract the growth rate with this method.
Hence, results are only presented for the cases where a single mode clearly dominates the response.
In each of these cases, we observe excellent agreement with results obtained in our frequency domain simulations.

\subsection{Validity}

In order for the shallow water analysis above to hold, the wavelength of the bound states must be much larger than the height of the fluid. Since the characteristic wavelength of the lowest $k$ modes in the system is set by $r_B$, the relevant condition becomes $h_0\ll r_B$.
A further approximation was that the curvature of the free surface was assumed negligible, i.e. $h'\ll 1$.
To see when this is satisfied, consider the basic velocity profile in \eqref{basic} and the height profile with the leading order curvature correction (see Appendix~\ref{app:flow}),
\begin{equation} \label{h_approx}
    h = h_0\left(1-\frac{r_a^2}{r^2}\right).
\end{equation}
The location of horizon will be shifted in this case but it will still be $\mathcal{O}(D/\sqrt{gh_0})$ provided the circulation is not too large.
The constraint on $h'$ then leads to the condition,
\begin{equation}
    r_h\gg r_s \equiv \left(\frac{D^2+C^2}{g}\right)^\frac{1}{3}.
\end{equation}
This condition improves for small $C$, and works best when $C=0$.
In that case, it can be reformulated as,
\begin{equation} \label{Dshallow}
    D \gg \sqrt{gh_0^3}.
\end{equation}
Altogether, we end up with a hierarchy of scales,
\begin{equation} \label{hierarchy}
    h_0 \ll r_s \ll r_h < r_B.
\end{equation}
For $r_B$ on the order of tens of centimeters, the requirement above is satisfied for $h_0$ on the order of millimeters.
For these low water heights, the viscous boundary layer is expected to spread throughout the whole fluid volume, at least near the drain where the flow velocities are large.
Furthermore, the drain is powered by gravity itself, hence, there is a complicated relation between $D$ and $h_0$.
A basic argument is the following. Starting from initial values $D_i$ and $h_{0i}$ and increasing the flow rate, the system would reach a new equilibrium with final values $D_f>D_i$ (to compensate the increased flow rate) and $h_{0f}>h_{0i}$ (due to the influx of more water). 
In other words, one cannot increase $D$ arbitrarily without also increasing $h_0$, thereby making the criterion above impractical to satisfy.

\subsection{Key features}

We conclude this section by highlighting the key features of the BHB which should be present when generalising to the dispersive regime. These are:
\begin{itemize}
    \item A region inside the vortex permitting negative energy waves.
    \item A mechanism for absorbing negative energies.
\end{itemize}
The first criterion is realised by the ergosphere in \eqref{rh_re} which leads to positive frequency modes with negative norm, i.e. $\Omega<0$ in the WKB framework. Hence, this is the feature to anticipate when we extend to more general scenarios.
The second criterion is provided in shallow water by the analogue horizon, however, the practical requirement is the in-going boundary condition into the centre of the vortex in \eqref{horizon_BC}.
This could be achieved for a more generic absorption mechanism without the strict notion of a horizon.
In our case, for example, the draining vortex has a hole in the middle which is a sink for the wave norm.
Furthermore, one could even relax the purely in-going boundary condition and require only that the out-going waves are sufficiently low amplitude that they can be treated as a higher order effect (see e.g. related work in \cite{torres2022imperfect,patrick2024quantum}).

\section{Deep water waves} \label{sec:deep}

To recap, so far we have derived a set of equations which govern the propagation of surface waves around a draining vortex (Section \ref{sec:system}), presented a general general method for studying the scattering of such waves (Section~\ref{sec:WKB}) and demonstrated that these methods accurately predict the unstable mode frequencies in the shallow water regime (Section~\ref{sec:shallow}).

However, the difficulty in realising shallow water conditions in practice motivates the study of the opposite regime $hk\gg 1$, known as the deep water regime.
When the wavelength becomes suitably short, the effect of surface tension and viscosity becomes non-negligible. 
The main difference compared with the low $k$ theory is that in deep water, one has to keep track of all 6 modes shown in Fig.~\ref{fig:disp}.
This significantly enhances the number of possible scattering processes and thereby greatly increases the complexity of the problem.

Our approach in this section will be to systematically analyse all scattering processes allowed by the deep water dispersion relation (which will be shown in Fig.~\ref{fig:diagrams}) before identifying those relevant to the BHB instability in a physically motivated parameter range.
The main result will be a compact expression governing the unstable mode frequencies and growth rates in \eqref{deep_oscil} and \eqref{deep_growth}, the properties of which are exemplified in the figures shown toward the end of the section.

\subsection{Simplifications}

We again work with the basic flow field in \eqref{basic} where the water height $h$ is treated as a constant. 
To completely characterise the scattering properties of the system, one must solve for the $p^j(r)$ at each point in the parameter space of \eqref{Ham_eff}.
Solving $\mathcal{H}=0$ with $\mathcal{D}(k)$ defined in \eqref{disp} is a daunting task, since one must find the roots of a transcendental equation over the 7--dimensional parameter space $(\omega,m,C,D,\gamma,g,h)$.
We can rescale $\mathcal{H}$ to effectively reduce the dimensionality of parameter space.
Following \cite{patrick2020superradiance}, we choose to remove $D$ and $g$ axes. We define,
\begin{equation}
    l = (D^2/g)^\frac{1}{3}, \qquad \tau = (D/g^2)^\frac{1}{3},
\end{equation}
and perform the following rescalings,
\begin{equation} \label{rescale2}
\begin{split}
    & \frac{r}{l}\to r, \quad \frac{h}{l}\to h \quad pl\to p, \quad \omega\tau\to\omega, \\ 
    & \frac{\gamma}{gl^2}\to\gamma \quad \frac{C}{D}\to C, \quad \mathcal{H}\tau^2\to\mathcal{H}, \quad \frac{\nu}{D}\to\nu,
\end{split}
\end{equation}
where, in anticipation of including the effects of viscosity at next to leading order, we have also rescaled $\nu$.
The result is equivalent to setting $D=g=1$:
\begin{equation}
\begin{split}
    \mathcal{H} = & \ -\frac{1}{2}\bigg[\left(\omega-\frac{mC}{r^2}+\frac{p}{r}\right)^2 -\left(k+\gamma k^3\right)\times \\ & \qquad \quad \tanh(hk)\bigg], \qquad k=\sqrt{p^2+\frac{m^2}{r^2}}.
\end{split}
\end{equation}
Unlike the shallow water regime, the magnitude of the group velocity $|\partial_\mathbf{k}\omega|$ is no longer independent of frequency.
In other words, different frequencies will travel at different speeds; the medium is \textit{dispersive}.
This is the physical reason for the appearance of extra spatial modes in the system compared with the shallow water case.

Considerable further simplification occurs if we work in the deep water limit $hk\gg 1$.
In this case, we can approximate $\tanh(hk)\to 1$ in the expression above, which reduces the parameter space to $(\omega,m,C,\gamma)$.
Furthermore, we needn't assume that the water height is constant in this regime; all deep water waves have wavelength much longer than the local water height, therefore, variations in $h$ become insignificant.
Although this is a simplification of the full problem, many of the scattering processes we will encounter result from the large $k$ properties of $\mathcal{H}$. Intuitively, this is because the large velocities achieved at small $r$ eventually blue-shift in-coming waves into the regime where $k\gg h^{-1}$.
Hence, qualitative features of the system in the $hk\to\infty$ limit will also carry over to the full dispersion relation.

\subsection{Scattering} \label{sec:scattering}

Since we are ultimately concerned with the BHB instability, which arises for $m>0$, we focus our attention on $m\neq 0$.
In this case, we can scale out $m$ from the effective Hamiltonian.
First we notice that $\mathcal{H}$ is symmetric under the transformation $(\omega,m,p)\to(-\omega,-m,-p)$.
Therefore, positive frequency waves with $m<0$ will scatter the same as the negative frequency waves with the opposite radial and angular momentum.
Without loss of generality, we can therefore fix $m>0$ and consider the frequency as ranging from negative to positive infinity.
Now we perform the following rescalings,
\begin{equation}
\begin{split}
    & r/m^\frac{1}{3}\to r, \quad \omega/m^\frac{1}{3}\to\omega \quad p/m^\frac{2}{3}\to p, \\
    & \gamma m^\frac{4}{3} \to \gamma \quad \mathcal{H}/m^\frac{2}{3}\to \mathcal{H},
    \end{split}
\end{equation}
and write the effective Hamiltonian as,
\begin{equation} \label{H3}
\begin{split}
    \mathcal{H} = & \ -\frac{1}{2}\left[\left(\omega-\frac{C}{r^2}+\frac{p}{r}\right)^2 - k-\gamma k^3\right], \\ 
    k = & \ \sqrt{p^2+\frac{1}{r^2}}.
\end{split}
\end{equation}
The parameter space is now $(\omega,C,\gamma)$.
As we proceed, our tactic will be to understand how the ray trajectories vary throughout the parameter space at fixed $\gamma$. We will then vary $\gamma$ from small to larger values to see how the structure of the parameter space changes.


\begin{figure} 
\centering
\includegraphics[width=\linewidth]{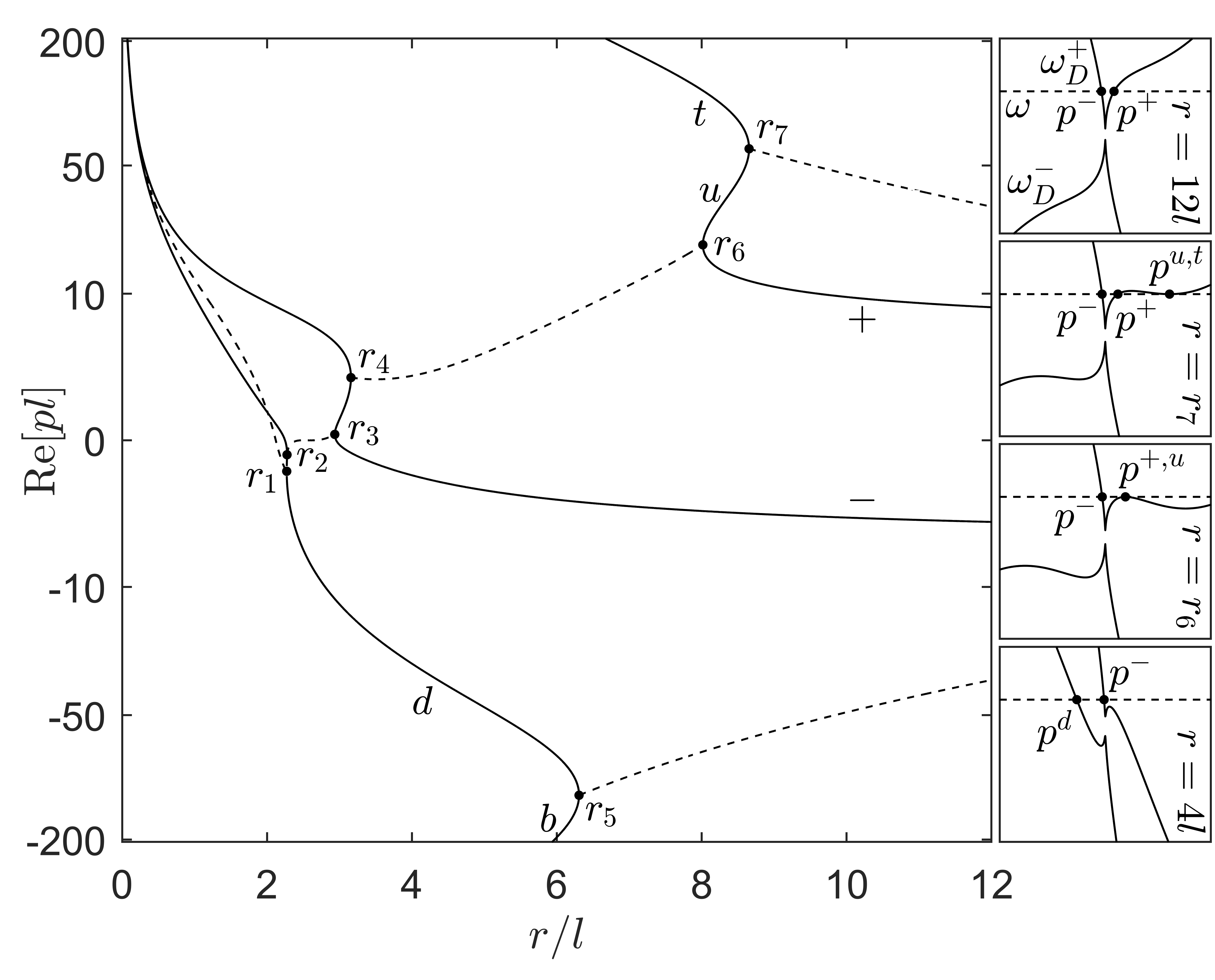}
\caption{An exemplary interaction between the 6 capillary-gravity wave modes for the parameters $C/D=15$, $\gamma m^\frac{4}{3}/gl^2=10^{-4}$ and $\omega\tau=2.3$. \textbf{Main panel:} On the vertical axis (which has non-uniform spacing), we show the real part of the radial wavevector of the 6 modes that solve $\mathcal{H}=0$. As $r$ is varied, the value of $p^j$ changes. Solid lines represent propagating modes with $p^j\in\mathbb{R}$. Neighbouring $p^j$ coincide at turning points (black dots) and are converted into evanescent modes with $p^j\in\mathbb{C}$ (dashed lines). Note that a single dashed line represents a complex conjugate mode pair. In the classification scheme of Fig.~\ref{fig:diagrams}, this diagram corresponds to Type VI scattering. \textbf{Small panels:} Propagating modes are shown as black points on the two branches of the dispersion relation $\omega_D^\pm$ for 4 illustrative radii (indicated on the right of the panels). The shape of the branches varies with $r$, changing the locations of the $p^j$ roots and generating the trajectories shown on the main panel.} \label{fig:scatter_example}
\end{figure}

In shallow water, there are two modes ($p^\pm$) which either superradiantly scatter, non-superradiantly scatter or do not scatter \cite{patrick2021rotational}.
These three possibilities cover the entire parameter space of the low $k$ theory.
In deep water, there are 6 propagating modes that can scatter with each other.
We show an example of this scattering for a particular choice of parameters in Fig.~\ref{fig:scatter_example}.
The diagram shows the trajectories of the WKB modes through the $(r,p)$--phase space.
There are 6 functions $p^j(r)$ from which one can reconstruct the WKB solution of the scattering problem using \eqref{expans1}.
Hence, drawing the scattering diagram is equivalent to knowing the general solution to the linear equations of motion for a given parameter set.
The mode amplitudes on the far left and right sides of the diagram are related by the full transfer matrix $\mathcal{M}$ defined in \eqref{scatterM}.
For example, the matrix corresponding to Fig.~\ref{fig:scatter_example} is,
\begin{equation} \label{M1}
\begin{split}
    \mathcal{M} = & \ \mathrm{diag}\big(\mathcal{F}^t_{01},\mathcal{F}^u_{01},\mathcal{F}^+_{01},\mathcal{F}^\uparrow_{01},\mathcal{F}^\downarrow_{01},\mathcal{F}^b_{01}\big)\tilde{t}_1 \\
    \times & \ \mathrm{diag}\big(\mathcal{F}^t_{12},\mathcal{F}^u_{12},\mathcal{F}^+_{12},\mathcal{F}^-_{12},\mathcal{F}^d_{12},\mathcal{F}^b_{12}\big)\tilde{t}_2 \\
    \times & \ \mathrm{diag}\big(\mathcal{F}^t_{23},\mathcal{F}^u_{23},\mathcal{F}^\uparrow_{23},\mathcal{F}^\downarrow_{23},\mathcal{F}^d_{23},\mathcal{F}^b_{23}\big)\tilde{t}_3 \\
    \times & \ \mathrm{diag}\big(\mathcal{F}^t_{34},\mathcal{F}^u_{34},\mathcal{F}^+_{34},\mathcal{F}^-_{34},\mathcal{F}^d_{34},\mathcal{F}^b_{34}\big)\tilde{t}_4 \\
    \times & \ \mathrm{diag}\big(\mathcal{F}^t_{45},\mathcal{F}^{\uparrow^+}_{45},\mathcal{F}^{\downarrow^+}_{45},\mathcal{F}^-_{45},\mathcal{F}^d_{45},\mathcal{F}^b_{45}\big)\tilde{t}_5 \\
    \times & \ \mathrm{diag}\big(\mathcal{F}^t_{56},\mathcal{F}^{\uparrow^+}_{56},\mathcal{F}^{\downarrow^+}_{56},\mathcal{F}^-_{56},\mathcal{F}^{\uparrow^-}_{56},\mathcal{F}^{\downarrow^-}_{56}\big)\tilde{t}_6 \\
    \times & \ \mathrm{diag}\big(\mathcal{F}^t_{67},\mathcal{F}^u_{67},\mathcal{F}^+_{67},\mathcal{F}^-_{67},\mathcal{F}^{\uparrow^-}_{67},\mathcal{F}^{\downarrow^-}_{67}\big)\tilde{t}_7 \\
    \times & \ \mathrm{diag}\big(\mathcal{F}^{\uparrow^+}_{7B},\mathcal{F}^{\uparrow^+}_{7B},\mathcal{F}^+_{7B},\mathcal{F}^-_{7B},\mathcal{F}^{\uparrow^-}_{7B},\mathcal{F}^{\downarrow^-}_{7B}\big)
\end{split}
\end{equation}
where the locations numbered 1 to 7 correspond to the turning points in Fig.~\ref{fig:scatter_example}.
The matrices $\tilde{t}_j$ are constructed from the transfer matrices in \eqref{tp1} and \eqref{tp2} and are given explicitly by,
\begin{equation*}
\begin{split}
    \tilde{t}_1 = & \ \begin{pmatrix}
        1 & 0 & 0 & 0 & 0 & 0 \\
        0 & 1 & 0 & 0 & 0 & 0 \\
        0 & 0 & 1 & 0 & 0 & 0 \\
        0 & 0 & 0 & \beta & \beta^* & 0 \\
        0 & 0 & 0 & \alpha^* & \alpha & 0 \\
        0 & 0 & 0 & 0 & 0 & 1 \\
    \end{pmatrix}, \ \  
    \tilde{t}_2 = \begin{pmatrix}
        1 & 0 & 0 & 0 & 0 & 0 \\
        0 & 1 & 0 & 0 & 0 & 0 \\
        0 & 0 & \beta^* & \alpha & 0 & 0 \\
        0 & 0 & \beta & \alpha^* & 0 & 0 \\
        0 & 0 & 0 & 0 & 1 & 0 \\
        0 & 0 & 0 & 0 & 0 & 1 \\
    \end{pmatrix}, \\
    \tilde{t}_3 = & \ \begin{pmatrix}
        1 & 0 & 0 & 0 & 0 & 0 \\
        0 & 1 & 0 & 0 & 0 & 0 \\
        0 & 0 & \beta & \beta^* & 0 & 0 \\
        0 & 0 & \alpha^* & \alpha & 0 & 0 \\
        0 & 0 & 0 & 0 & 1 & 0 \\
        0 & 0 & 0 & 0 & 0 & 1 \\
    \end{pmatrix}, \ \
    \tilde{t}_4 = \begin{pmatrix}
        1 & 0 & 0 & 0 & 0 & 0 \\
        0 & \beta^* & \alpha & 0 & 0 & 0 \\
        0 & \beta & \alpha^* & 0 & 0 & 0 \\
        0 & 0 & 0 & 1 & 0 & 0 \\
        0 & 0 & 0 & 0 & 1 & 0 \\
        0 & 0 & 0 & 0 & 0 & 1 \\
    \end{pmatrix} \\
    \tilde{t}_5 = & \ \begin{pmatrix}
        1 & 0 & 0 & 0 & 0 & 0 \\
        0 & 1 & 0 & 0 & 0 & 0 \\
        0 & 0 & 1 & 0 & 0 & 0 \\
        0 & 0 & 0 & 1 & 0 & 0 \\
        0 & 0 & 0 & 0 & \beta^* & \alpha \\
        0 & 0 & 0 & 0 & \beta & \alpha^* \\
    \end{pmatrix}, \ \
    \tilde{t}_6 = \begin{pmatrix}
        1 & 0 & 0 & 0 & 0 & 0 \\
        0 & \beta & \beta^* & 0 & 0 & 0 \\
        0 & \alpha^* & \alpha & 0 & 0 & 0 \\
        0 & 0 & 0 & 1 & 0 & 0 \\
        0 & 0 & 0 & 0 & 1 & 0 \\
        0 & 0 & 0 & 0 & 0 & 1 \\
    \end{pmatrix}, \\
    \tilde{t}_7 = & \ \begin{pmatrix}
        \beta^* & \alpha & 0 & 0 & 0 & 0 \\
        \beta & \alpha^* & 0 & 0 & 0 & 0 \\
        0 & 0 & 1 & 0 & 0 & 0 \\
        0 & 0 & 0 & 1 & 0 & 0 \\
        0 & 0 & 0 & 0 & 1 & 0 \\
        0 & 0 & 0 & 0 & 0 & 1 \\
    \end{pmatrix},
\end{split}
\end{equation*}
with $\alpha=e^{i\pi/4}$ and $\beta=\alpha/2$ defined for compactness. Mode mixing results from the off-diagonal elements in the $\tilde{t}_j$.
Imposing the boundary conditions at $r_0$ and $r_B$, $\mathcal{M}$ determines a resonance condition.
However, it is evidently extremely impractical to write down $\mathcal{M}$ like this and in practice, it is much simpler to identify the relevant rays that form the bound state loops in phase space and transport the modes around them (see Appendix~\ref{app:res}).

\begin{figure*} 
\centering
\includegraphics[width=.9\linewidth]{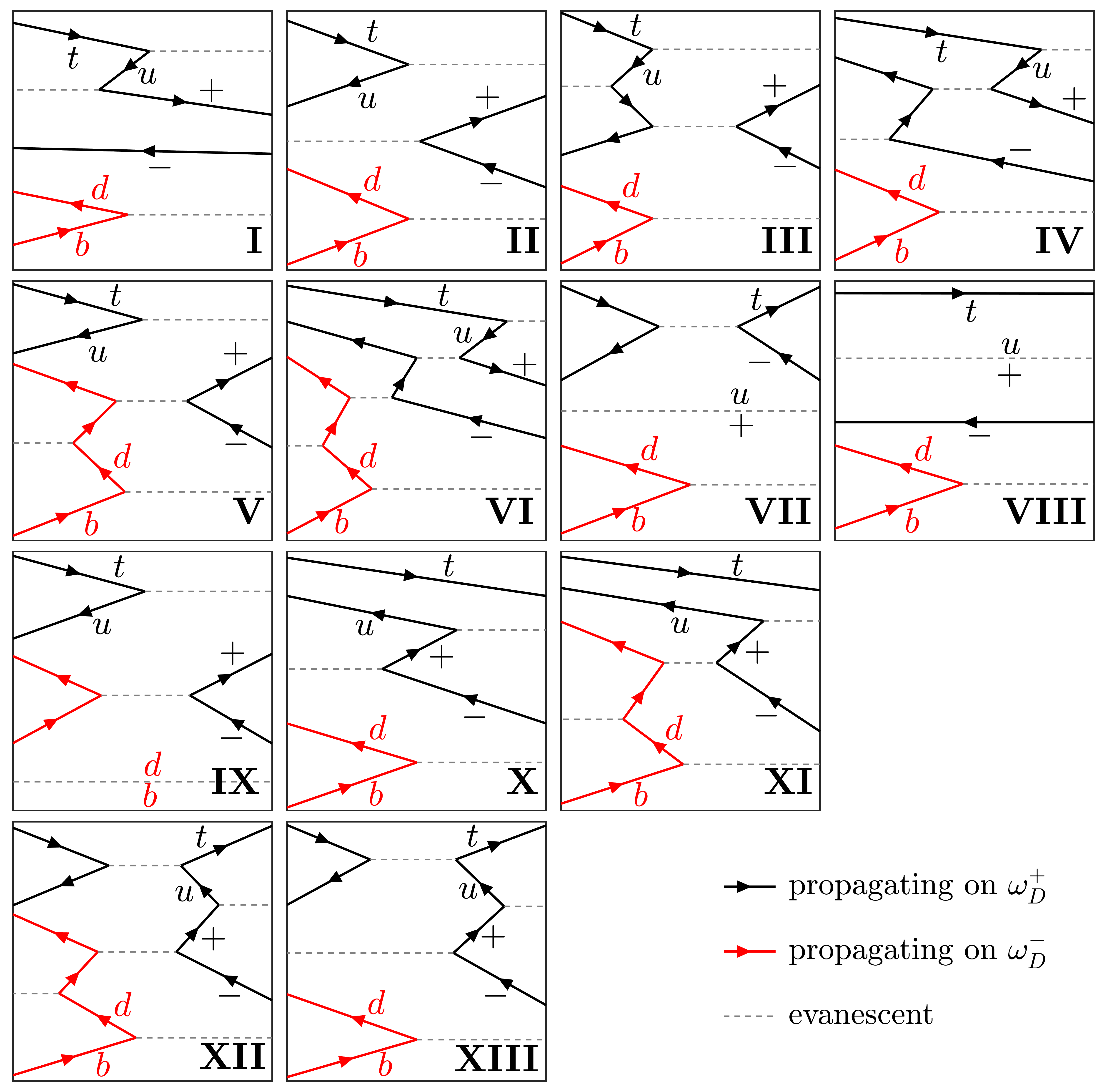}
\caption{Schematics illustrating the 13 different scattering possibilities allowed by the deep water system in \eqref{H3}. Solid black (red) lines represent modes propagating on the upper (lower) branch of the dispersion relation in the direction shown by the arrow. Dashed lines represent a pair of evanescent modes. The horizontal axis represents $r$ with the left and right sides representing the vortex core and spatial infinity respectively. The vertical axis represents $\mathrm{Re}[p]$, or more specifically, the relative values of $\mathrm{Re}[p]$ of propagating modes at a given $r$ (compare, e.g. the Type VI schematic with Fig.~\ref{fig:scatter_example}). The aim of the diagrams is to clearly indicate the locations of the turning points and which modes interact there.} \label{fig:diagrams}
\end{figure*}

Since there is a large range of mode interactions allowed by \eqref{H3}, it is helpful to classify them according to the turning point structure, i.e. the locations of the various $r_\tp$ and the modes that interact at each.
The different mode interactions can then be schematically illustrated as one of the 13 diagrams in Fig.~\ref{fig:diagrams}, which comprise an exhaustive list of the mode interactions allowed by \eqref{H3}.
The names of the diagrams (Types I--XIII) are chosen in the following way. The first 6 (I--VI) are modifications of the deep water diagrams in \cite{patrick2020superradiance} that regulate the theory by removing the propagating modes at infinity with divergent $k$. 
The next three (VI--IV) are characteristic of super-luminal dispersion relations, e.g. those studied in \cite{patrick2021rotational}. The last 4 (X--XIII) result from the mixed character of the capillary-gravity wave dispersion relation, which contains both sub- and super-luminal sections.

The diagrams are schematic illustrations of the $(r,p)$--phase space exemplified by Fig.~\ref{fig:scatter_example}.
These diagrams represent the real part of $p$ (vertical axis) as a function of $r$ (horizontal axis).
Solid lines indicate $p\in\mathbb{R}$ with the arrow indicating the direction of propagation as determined by the radial component of the group velocity.
Dashed lines indicate $p\in\mathbb{C}$.
Black (red) lines represent modes on $\omega_D^+$ ($\omega_D^-$) which have positive (negative) norm density $\rho_n$ \eqref{WKBnorm}.
The ordering of lines is such that $\mathrm{Re}[p]$ increases from bottom to top for propagating modes (note, this is not necessarily true for the dashed lines since evanescent modes can pass round propagating modes in the complex plane, although this behaviour is omitted since the aim of the diagrams is to clearly indicate the different mode interactions).
Bottom to top ordering here translates to the left to right ordering of the real roots of $\omega=\omega_D^\pm(p)$ (see Fig.~\ref{fig:disp}) and since the labels $j=\{b,d,-,+,u,t\}$ go from left to right on the dispersion relation, this translates to an unambiguous labelling convention for the modes in Fig.~\ref{fig:diagrams}.

\begin{figure*} 
\centering
\includegraphics[width=\linewidth]{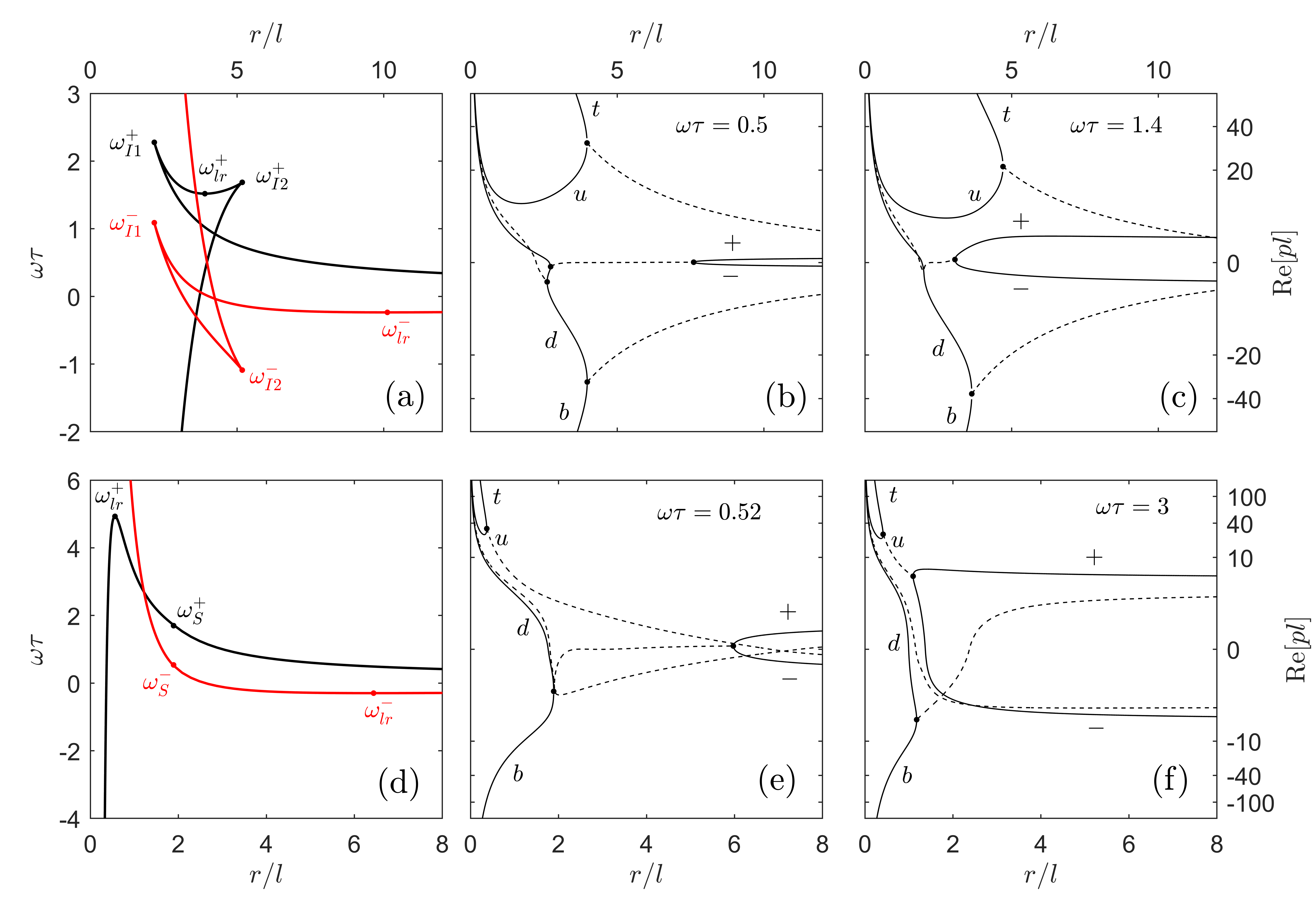}
\caption{In panel (a), we display examples of the curves defined in \eqref{tp_potentials} when $\gamma m^\frac{4}{3}/gl^2$ is small.
The specific parameters used are $C/D=8$ and $\gamma m^\frac{4}{3}/gl^2=10^{-3}$.
The characteristic frequencies defined in \eqref{charac1} are also indicated.
The black curves are the $\omega^+_a$, with $a=1$ extending to radial infinity, $a=2$ connecting the first curve at $\omega^+_{I1}$ and $a=3$ connecting the second curve at $\omega^+_{I2}$.
The red curves are the $\omega^-_a$ with the labelling the same.
In panels (b,c), we show the scattering of the 6 modes by plotting the $(r,p)$ phase space for two example frequencies. The turning points are shown as black points and are determined by the points of intersection between the curves on panel (a) and a line of constant $\omega$.
In panels (d-f), we display the same but for high $\gamma$, with the specific parameters $C/D=4$ and $\gamma m^\frac{4}{3}/gl^2=0.1$.
In this case, the characteristic frequencies are those in \eqref{charac2}.
The black curve is $\omega^+_1$ and the red curve is $\omega^-_1$.
Using the classification scheme of Fig.~\ref{fig:diagrams}, panels (b,c,e,f) correspond respectively to Type V, II, IX, VII scattering. 
} \label{fig:potential_deep}
\end{figure*}

Locations where solid lines connect to dashed lines, i.e. vertices, represent turning points $r_\tp$.
The number of modes is preserved either side of $r_\tp$ since the dashed lines actually represent two complex conjugate modes with the same $\mathrm{Re}[p]$.
Each diagram is unique in that it has a fixed structure of vertices \footnote{Note that for vertices connected by more than one line (or not connected) the relative locations of the $r_\tp$ are not necessarily captured by the diagrams. In other words, the diagrams can be deformed within the same classification provided: 1) vertices are neither created nor destroyed, 2) the direction (right or left) of the arrows is unaltered and 3) the ordering of propagating modes from top to bottom is respected. For example, the turning point between the $p^\pm$ modes on the Type II scattering schematic in Fig.~\ref{fig:diagrams} is not necessarily at larger $r$ compared to the turning point involving the $p^{u,t}$ modes (as can be seen by comparing with panel (c) of Fig.~\ref{fig:potential_deep}).}.
The scattering processes for $\omega<0$ are inverted versions (i.e. $p\to -p$) of those for $\omega>0$, which we denote with a superscript $\downarrow$.
Note, in the inversions Type VII$^\downarrow$ and VIII$^\downarrow$, it is more natural to label the modes which propagate as $r\to\infty$ as $p^\pm$. In this way, the mode labels are correct when transitioning between different diagrams as the frequency is varied.

Each diagram corresponds to a particular structure in the transfer matrix \eqref{scatterM} and therefore a particular computation of the asymptotic scattering coefficients or (in our case) bound state integrals.
However, to evaluate these integrals, we need quantitative values of the different $p^j$ between the various turning points. 
The first step in this procedure is to find accurate values for the $r_\tp$.
In deep water (unlike the low $k$ theory where we had a simple analytic formula \eqref{tp_cf}) there is no closed form expression for the $r_\tp$.
Instead, we consider the curves $\omega^\pm_a(r)$ defined in \eqref{tp_potentials} (see Appendix~\ref{app:tps} where we detail our procedure to calculate these).
Examples of the $\omega^\pm_a(r)$ are shown in Fig.~\ref{fig:potential_deep} for the case where the $\omega_D^\pm$ have three (low $\gamma$) and one (high $\gamma$) extrema respectively.
When there are three extrema (and therefore three curves $a=1,2,3$) the two which are real as $r\to\infty$ are $\omega_1^\pm$ and describe the scattering of the $p^\pm$ modes on the upper and lower branches of the dispersion relation respectively.
$\omega_1^+$ then joins onto $\omega_2^+$ (which determines how $p^{+,u}$ modes scatter) and $\omega_2^+$ in turn joins onto $\omega^+_3$ (which determines how $p^{u,t}$ modes scatter).
Similarly, $\omega_1^-$ joins onto $\omega_2^-$ (which determines how $p^{d,-}$ modes scatter) and $\omega_2^-$ in turn joins onto $\omega_3^-$(which determines how $p^{b,d}$ modes scatter).
When $\gamma$ is sufficiently large, there is only a single real extremum on each of the $\omega^\pm_D$ and the two corresponding curves are called $\omega^\pm_1(r)$.
With these functions in hand, we can find the $r_\tp$ by selecting a frequency and solving \eqref{tp_cond} graphically.

To determine which of the 13 scattering possibilities occurs for a particular frequency, we can calculate various characteristic frequencies $\omega(C,\gamma)$ which divide up the parameter space into regions where different scattering processes occur.
We defer a thorough discussion of how these frequencies arise to Appendix~\ref{app:freq} and simply quote the results here.
When $\gamma$ is smaller than the critical value $\gamma_c\simeq0.0522$, the relevant frequencies are,
\begin{equation} \label{charac1}
    \omega_\lr^+, \quad \omega_\lr^-, \quad \omega_{I1}^+, \quad \omega_{I1}^-, \quad \omega_{I2}^+, \quad \omega_{I2}^-,
\end{equation}
where the light-ring frequencies $\omega_\lr^\pm$ satisfy the same conditions \eqref{lightring} as in shallow water.
The other four frequencies are the primary ($\omega_{I1}^\pm$) and secondary ($\omega_{I2}^\pm$) inflection points, (where $\pm$ indicates whether these points are on $\omega_D^\pm$) which all satisfy the relations,
\begin{equation} \label{inflectionpoints}
    \mathcal{H}_I=0, \quad \partial_p\mathcal{H}_I=0, \quad \partial_p^2\mathcal{H}_I=0.
\end{equation}
These are the locations where the extrema of $\omega_D^\pm$ (in the $p$ direction) coalesce before moving off in the complex plane.
They therefore correspond to the locations where the $\omega^\pm_a$ meet and terminate.
They are important since the value of $\omega$ relative to the $\omega_I$ determines which mode is the propagating one moving in the direction of decreasing $r$.

Above $\gamma_c$, each branch of the dispersion relation has a single real extremum in the $p$ direction.
In this regime, $\mathcal{H}$ has no real inflection points and the transitions between different scattering processes are governed by a different set of frequencies.
In Appendix~\ref{app:freq}, we argue that the important frequencies are complex saddle points of $\mathcal{H}$ which satisfy,
\begin{equation} \label{saddlepoints}
\begin{split}
    \mathrm{Re}[\mathcal{H}_S] = & \ 0, \quad \mathrm{Im}[\mathcal{H}_S] = 0, \\ \mathrm{Re}[\partial_p\mathcal{H}_S] = & \ 0, \quad \mathrm{Im}[\partial_p\mathcal{H}_S] = 0.
\end{split}
\end{equation}
In this case there are four conditions as we need to solve for $(r_S,\mathrm{Re}[p_S],\mathrm{Im}[p_S],\omega_S)$.
There are two different saddle points for the upper and lower branches of the dispersion relation which we identify with $\pm$ below.
Hence, for $\gamma>\gamma_c$, the characteristic frequencies of the system are,
\begin{equation} \label{charac2}
    \omega_\lr^+, \quad \omega_\lr^-, \quad \omega_S^+, \quad \omega_S^-.
\end{equation}
On Fig.~\ref{fig:potential_deep}, we indicate the frequencies in \eqref{charac1} and \eqref{charac2} in panel (a) and (d) respectively.
Note that since $\omega_S^\pm$ (and the associated locations in the $(r,p)$--phase space) correspond to saddle points in the complex plane, they are not positioned directly on top of $\omega^\pm_1(r)$.

\begin{figure*}
\centering
\includegraphics[width=\linewidth]{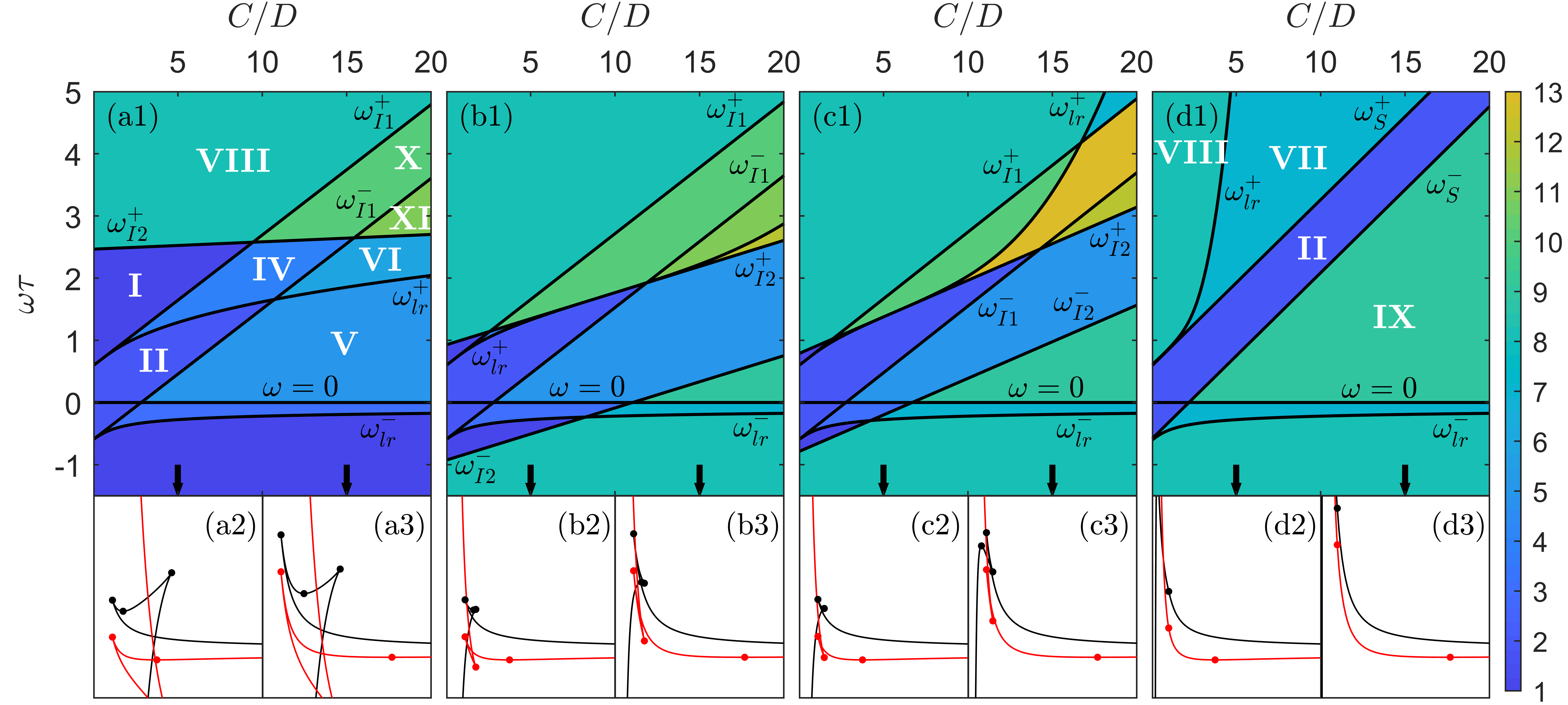}
\caption{$\gamma m^\frac{4}{3}/gl^2=\mathrm{const}$ slices of the full parameter space for $\gamma m^\frac{4}{3}/gl^2=10^{-4}$ (a1), $5\times10^{-3}$ (b1), $0.01$ (c1) and $0.08$ (d1). The plane is divided up by various characteristic frequencies into regions where different scattering processes occur. $\omega^\pm_\lr$ are the curves which are not linear in $C/D$ whereas $\omega^\pm_{I1,2}$ are linear in $C/D$, with $\omega^\pm_{I1}$ being the steeper of the two pairs.
Note that panel (d1) is the only case where $\gamma>\gamma_c$.
The colour axis indicates which of 13 processes occur in each region. We only include region labels for $\omega>0$ in panels (a1) and (d1) since the two central panels and the $\omega<0$ regions become obscured by including them. A detailed discussion of the different parameter space regions, with clearer diagrams, can be found in Appendix~\ref{app:param_plane}. In panels (a2,a3,b2,b3,...), we display the turning point curves $\omega^\pm_a$ at $C/D=5,15$ indicated by the black arrows, where the colour convention and solid points have the same meaning as in Fig.~\ref{fig:potential_deep}.} \label{fig:param_plane}
\end{figure*}

In Fig.~\ref{fig:param_plane}, we give examples of a $(\omega,C,\gamma=\mathrm{const})$ slice of the parameter space for various $\gamma$, and show how the characteristic frequencies divide the space into regions where the different scattering processes occur.
At our level of approximation, there is a discontinuous change in the mode scattering crossing the characteristic frequencies (this is the same phenomenon that lead to the errorenous jump in the reflection coefficient on Fig.~\ref{fig:refl} at the superradiant threshold).
In reality, these discontinuities will be smoothed over by mode scattering away from the turning points, which we do not account for.
Since Fig.~\ref{fig:param_plane} does not clearly indicate all the different regions, we provide a detailed discussion in Appendix~\ref{app:param_plane} of how varying $\gamma$ changes $(\omega,C)$ plane, giving the critical parameters where the plane undergoes a structural change.
Due to the symmetry $(\omega,m,p)\to(-\omega,-m,-p)$, the parameter space for $m<0$ can be obtained by inverting the $m>0$ parameter space through $\omega=0$, and also inverting all the scattering diagrams through $\mathrm{Re}[p]=0$.
Note that the arrows point in the same direction upon inversion since $v_{g,r}=\partial_p\omega$ does not change sign under this transformation.

To recap before moving on, in this section we characterized all possible scattering processes in the deep water system governed by \eqref{H3}. These are divided in 13 possible mode interactions which are summarized by the diagrams in Fig.~\ref{fig:diagrams}. In general, there are two propagating modes at spatial infinity (right side of the diagrams) and four approaching the vortex axis (left side). However, they differ in the modes which interact in the intermediate region and which modes propagate in the core and at infinity. The particular scattering configuration is then dictated by the dimensionless parameters $(\omega,C,\gamma)$, as shown in Fig.~\ref{fig:param_plane}.



\subsection{Superradiance}

Combining knowledge of the various scattering processes (Fig.~\ref{fig:diagrams}) with where they occur in the parameter space (Fig.~\ref{fig:param_plane}), we can now deduce a superradiance condition for deep water waves.
Superradiant amplification occurs for $\omega>0$ whenever the propagating modes on $\omega_D^+$ as $r\to\infty$ become propagating modes on $\omega_D^-$ at low $r$.
This can be seen by noting that when $p^\pm$ are on $\omega_D^-$, $p^+$ is the in-going mode, meaning that one takes the $-$ sign in the exponent of \eqref{refl_loc} and $|\mathcal{R}|>1$.
In Fig.~\ref{fig:diagrams}, the clearest signature is that the black part of the diagram connects to the red part of the diagram, which occurs for the following cases.

In Type V and IX scattering, all amplified radiation gets scattered to infinity because the out-going $p^+$ mode does not interact with dispersive $p^u$ mode.
In Type VI scattering, most of the amplified radiation is scattered back into the vortex and only a small amount tunnels out.
This can be seen because the out-going $p^+$ mode at infinity is connected to the superradiant part of the diagram by an evanescent (dashed) line.
For Type XI, all amplified radiation is scattered back into the vortex, which is a consequence of the out-going mode at infinity ($p^t$ in this case) not being connected to the superradiant part of the diagram.
For Type XII, most of the amplified radiation escapes (since the out-going mode is connected to the superradiant part by propagating lines) and only a small amount tunnels back into the vortex through the evanescent line on the $p^{u,t}$ modes.

\begin{figure} 
\centering
\includegraphics[width=\linewidth]{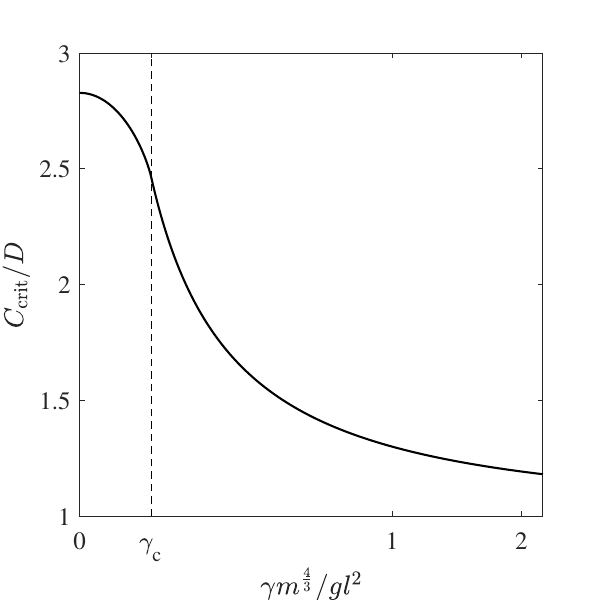}
\caption{The critical rotation is shown as a function of the dimensionless surface tension parameter. Superradiance cannot occur for rotations slower than this since modes propagating on $\omega_D^+$ at infinity do not reach $\omega_D^-$ in the vortex core.} \label{fig:crit_C}
\end{figure}

The criterion for amplification (of positive frequency modes) when $\gamma<\gamma_c$ is $\omega<\omega_{I1}^-$, i.e. the superradiant threshold is determined by an inflection point of $\omega_D^-(p)$.
However, the additional radiation is only guaranteed to reach infinity if $\omega<\omega_\lr^+$ is also satisfied (note that similar behaviour was found in the absence of surface tension in \cite{patrick2020superradiance}).
For $\gamma>\gamma_c$, superradiance occurs for $\omega<\omega_S^-$, i.e. a saddle point in the complex $p$--plane plays the role of the threshold.
Altogether, we can write,
\begin{equation} \label{superrad}
    0<\omega<\omega_\star = \begin{cases} 
    \mathrm{min}[\omega_{I1}^-,\omega_\lr^+], \qquad \gamma<\gamma_c \\
    \omega_S^-, \qquad \qquad \qquad \, \gamma>\gamma_c
    \end{cases},
\end{equation}
as a sufficient requirement for the amplified radiation to escape the vortex.
An interesting consequence of the bound above is that, below a critical rotation, the superradiant bandwidth closes up and amplification can no longer occur at any frequency.
This happens below (above) $\gamma_c$ when $\omega_{I1}^-$ ($\omega_S^-$) passes through zero.
In Fig.~\ref{fig:crit_C}, we show the critical rotation at which this happens as a function of the dimensionless surface tension parameter. As $\gamma\to 0$, the critical value goes to $2^{3/2}$, which agrees with the analytic prediction for deep water waves in the absence of capillary effects \cite{patrick2020superradiance}.
In the limit of large $\gamma$, we find numerically that the critical rotation tends to unity.

\subsection{Bound states}

Superradiance is the key ingredient in the BHB instability. We therefore only need to analyse the diagrams which satisfy \eqref{superrad} to see when we can have superradiantly unstable bound states.
We do this by looking in the parameter space of Fig.~\ref{fig:param_plane} to find any superradiant diagrams for the chosen parameters, then identifying the segment of the relevant diagram in Fig.~\ref{fig:diagrams} which forms a bound state between the vortex and the boundary.
This means that, starting from the in-going $p^-$ mode on the right edge of the diagram, one should be able to follow the arrows and end up back on the right edge as an out-going mode.
Once such a path has been identified, one then transports the amplitudes along the path and imposes the boundary condition.
In the present case, we also account for the $\nu$ dependent attenuation in \eqref{amp}.
This procedure is detailed in full in Appendix~\ref{app:res}.
The result is that the oscillation frequency $\omega_n$ satisfies a Bohr-Sommerfeld resonance condition,
\begin{equation} \label{deep_oscil}
\begin{split}
    \mathrm{Re}[S_\mathrm{path}(\omega_n)] = & \ \pi\left(2n+\frac{1}{2}\right), \\
    S_\mathrm{path} = & \ \int_\mathrm{path} p(r) dr,
\end{split}
\end{equation}
where $S_\mathrm{path}$ is the phase integral along the sequence of rays comprising the path of the bound state through $(r,p)$--phase space.
Assuming $\Gamma_n\ll\omega_n$, we find that the growth rate is given by,
\begin{equation} \label{deep_growth}
\begin{split}
    & \Gamma_n = \frac{\sum_a\log|X_a|-\nu\mathrm{Re}[\Sigma_\mathrm{path}]}{\partial_\omega \mathrm{Re}[S_\mathrm{path}]}\Bigg|_{\omega=\omega_n}, \\
    & \qquad \Sigma_\mathrm{path} = \int_\mathrm{path}\frac{k^2dr}{v_{g,r}},
\end{split}
\end{equation}
where $\Sigma_\mathrm{path}$ is the attenuation due to viscosity (which comes from the exponent containing $\nu$ in \eqref{amp2}) and the $X_a$ are different scattering coefficients along the path.
For Type V and IX scattering (which exhibit bound state A in Appendix~\ref{app:res}) there is a single superradiant reflection off the vortex so that $X_a=\mathcal{R}_1$.
For Type VI scattering (bound state B), there is a superradiant reflection followed by a regular transmission so that $X_a\in\{\mathcal{R}_1,\mathcal{T}_{23}\}$.
For Type XII scattering (bound state C), there is a superradiant reflection followed by two regular reflections so that $X_a\in\{\mathcal{R}_1,\mathcal{R}_2,\mathcal{R}_3\}$.
Upon noting that $T_\mathrm{path}=\partial_\omega\mathrm{Re}[S_\mathrm{path}]>0$ is the time taken for a point on the radial wavefront to complete a full orbit of the trapping region, the expressions in \eqref{deep_oscil} and \eqref{deep_growth} become equivalent to those presented in the introduction in \eqref{result1} and \eqref{result3} respectively.

The result in \eqref{deep_growth} demonstrates that the effect of the additional reflections/transmissions induced by dispersion is to reduce the magnitude of the instability.
Specifically, for bound states B and C we have $\log|\mathcal{R}_1\mathcal{T}_{23}|$ and $\log|\mathcal{R}_1\mathcal{R}_2\mathcal{R}_3|$ respectively, where $|\mathcal{R}_1|>1$ and all the rest are less than unity.
Hence, for an instability to occur, the amount of energy gained through superradiance must exceed the decrease in energy within the trapping at each additional reflection/transmission.

Furthermore, it is clear from \eqref{deep_growth} that the effect of viscosity is to further reduce the strength of the instability.
We can define a critical value of $\nu$ for which in the instability is quenched via,
\begin{equation}
    \nu_c(\omega_n) = \frac{\sum_a\log|X_a|}{\mathrm{Re}[\Sigma_\mathrm{path}]}\bigg|_{\omega=\omega_n},
\end{equation}
where this expression only makes sense when $\Gamma_n^{\nu=0}>0$.
The growth rate can then be expressed in terms of its value in the absence of viscosity,
\begin{equation} \label{nu_crit}
    \Gamma_n = \Gamma_n^{\nu=0}\left(1-\frac{\nu}{\nu_c}\right),
\end{equation}
where the factor in parentheses softens the instability for $\nu<\nu_c$ and completely quenches it for $\nu\geq\nu_c$.
This representation will be useful for comparing the undamped BHB modes to the damped ones.
Another useful representation of the damping term is,
\begin{equation}
    \frac{\mathrm{Re}[\Sigma_\mathrm{path}]}{T_\mathrm{path}} = \frac{\int_\mathrm{path} k^2 dt}{T_\mathrm{path}} = \langle \langle k^2 \rangle \rangle_\mathrm{path},
\end{equation}
where we have used $v_{g,r}=dr/dt$ which follows from \eqref{vg_rel} and the double angular brackets denote the average value over the path.
Thus, modes with larger $k$ will be more heavily attentuated by viscous damping.

\begin{figure*} 
\centering
\includegraphics[width=\linewidth]{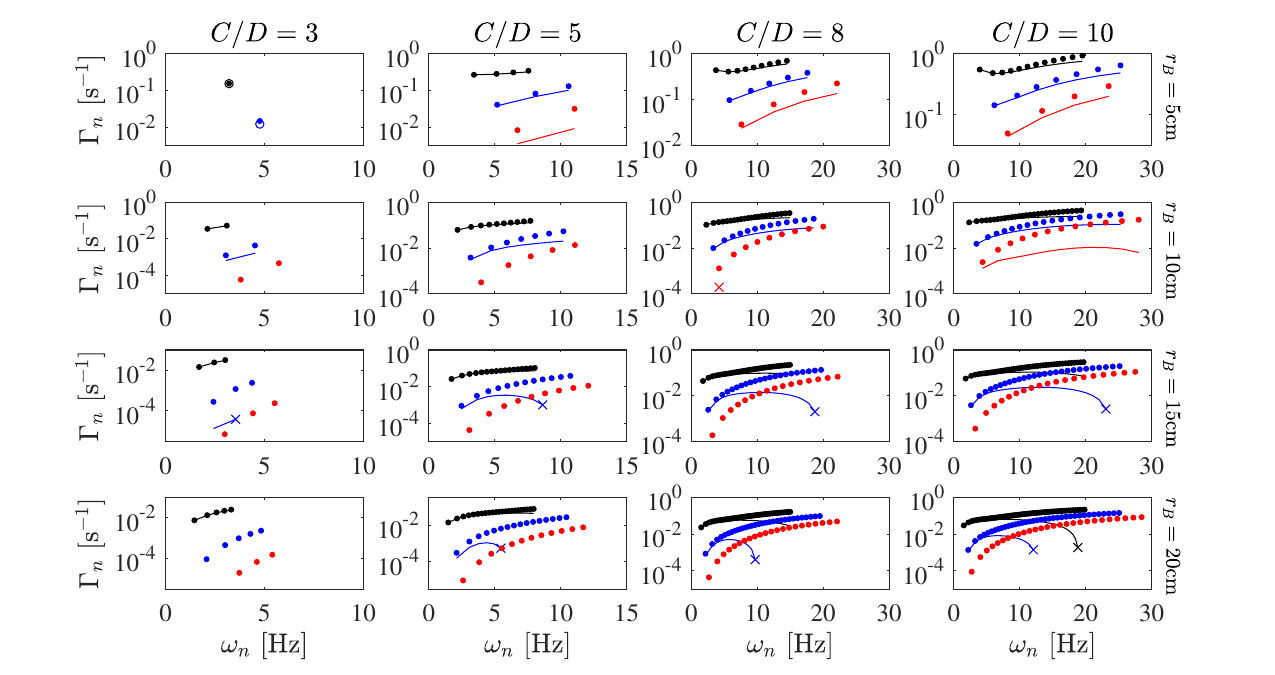}
\caption{BHB eigenfrequencies for $D=10\,\mathrm{cm^2s^{-1}}$ and a range of $r_B$ and $C/D$. We display results for $m=1,2,3$ in black, blue and red respectively.
The $\nu=0$ results are shown as solid points.
For $\nu\neq 0$, the eigenfrequencies have identical real parts to the undamped values but different imaginary parts through \eqref{nu_crit}. These are illustrated either as solid lines, or open circles where there is only a single point.
The lines terminate with crosses when the instability is quenched by viscosity (i.e. $\Gamma_n$ goes to zero).
No line is shown if every unstable mode in the undamped spectrum is stabilised by damping.
The trend is similar to the shallow water results in Fig.~\ref{fig:complex1} in the sense that larger $r_B$ and $C/D$ give rise to more modes and the growth rates are increased with rotation.
However, the functional dependence on the system's parameters is quite different from shallow water.} \label{fig:BS_complex}
\end{figure*}

\section{Examples}

\begin{figure*} 
\centering
\includegraphics[width=\linewidth]{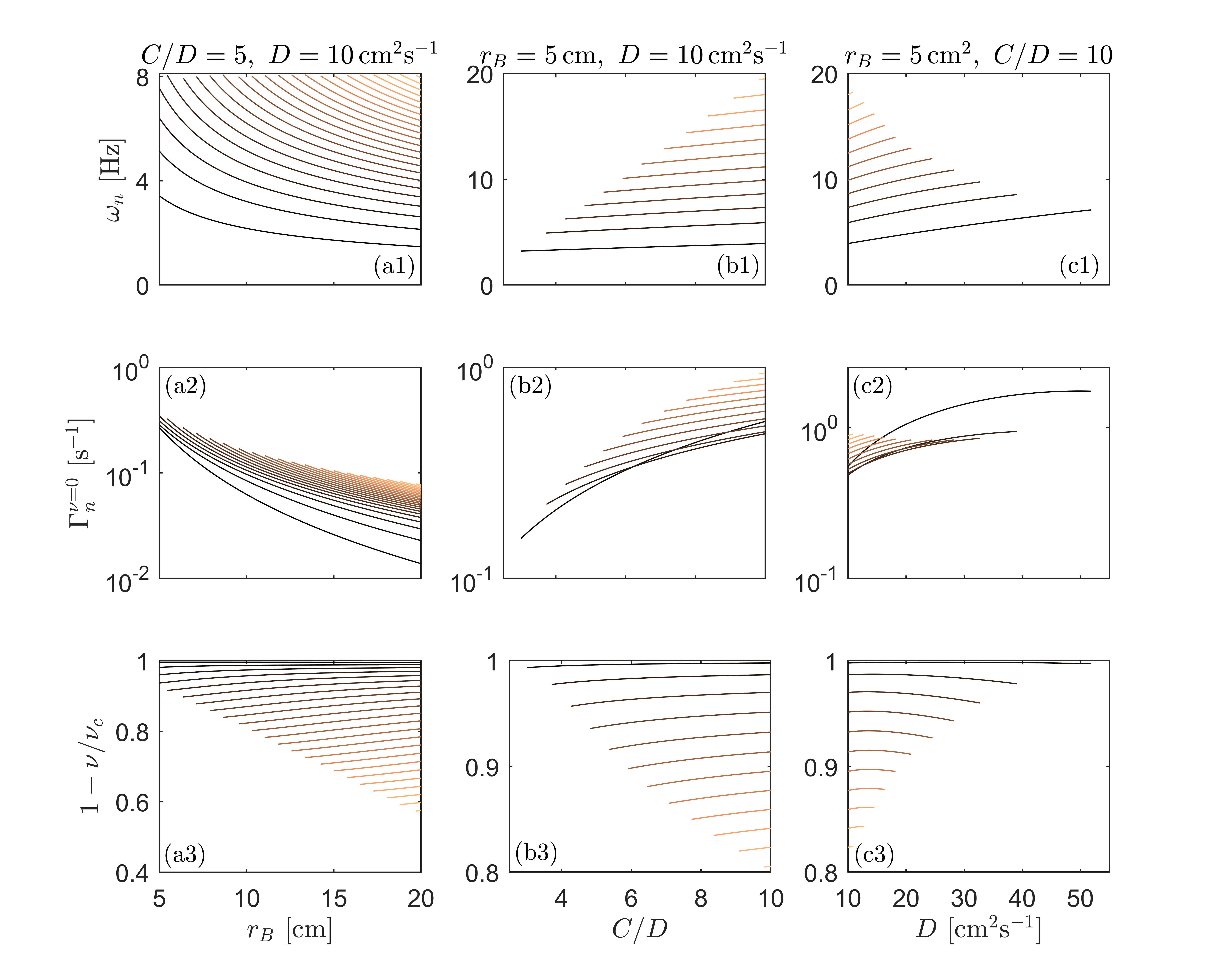}
\caption{Dependence of the BHB eigenfrequencies on $r_B$, $D$ and $C/D$ for $m=1$. Each column represents a different set of system parameters indicated at the top of the column.
The lines are colour coded so that the larger frequencies appear lighter orange.
The oscillation frequencies are shown in panels (a1,b1,c1), undamped growth rates in panels (a2,b2,c2) and the factor which multiplies the undamped growth rates when viscosity is non-zero in panels (a3,b3,c3).} \label{fig:BS_varyall}
\end{figure*}

We now compute $\omega^\mathbb{C}_n=\omega_n+i\Gamma_n$ for exemplary parameters to illustrate the effect of dispersion and dissipation on the BHB instability.
From hereon, we revert to using physical units so that our discussion is relevant to experiments.
For water at room temperature, the viscosity and surface tension parameters are $\nu=8.93\times10^{-7}\mathrm{m^2s^{-1}}$ and $\gamma = 7.22\times10^{-5}\mathrm{m^3s^{-2}}$ respectively.
We restrict our attention to trap sizes in the range $r_B\in[5,20]\,\mathrm{cm}$ and drain rates in the range $D\in[10,100]\,\mathrm{cm^2s^{-1}}$.
For our third variable, we take $C/D\in[1,10]$ since Fig.~\ref{fig:crit_C} shows that this ratio must be at least greater than unity.
Using these variables, we analyse the parameter space in Fig.~\ref{fig:param_plane}, finding that the relevant superradiant diagrams are Type V and Type IX.
This restriction comes from the chosen range for $C/D$, since the more exotic superradiant diagrams occur only at larger $C/D$ ratios.
Both diagrams for Type V and IX are concerned with the same class of bound state (bound state A in Appendix~\ref{app:BCs}) which scatters a single time with the vortex before reflecting from the boundary.
This greatly simplifies the analysis since we only need to locate the two turning points $r_{\tp1,2}$ where the $p^\pm$ modes scatter on $\omega_D^-$ and $\omega_D^+$ respectively.
Solving $\mathcal{H}=0$ for $p^\pm(r)$, we then have sufficient information to evaluate all the phase integrals in \eqref{deep_oscil} and \eqref{deep_growth}.
At a given selection of parameters $(r_B,D,C/D)$, these integrals are evaluated over the range of superradiant frequencies inside the trapping region.
The upper limit of this range is $\omega_\star$ defined in \eqref{superrad} whilst the lower bound is value of the turning point curve at the boundary, i.e. $\omega_1^+(r_B)$, in Fig.~\ref{fig:potential_deep}.

In the deep water regime, one has to solve \eqref{tp0} numerically for each frequency in the range (this step was trivial in shallow water where turning points are given analytically by \eqref{tp_cf}).
To do this, we first find an approximate solution to $\omega=\omega^-_1(r_{\tp1})$ and $\omega=\omega^+_1(r_{\tp2})$ graphically, where we can use the fact that $\omega_1^\pm(r\to\infty)\sim \pm 1/\sqrt{r}$ to identify which of the curves we need for $\gamma<\gamma_c$ (note that this implies the relation $(r_\tp,p_\tp)\sim(\omega^{-2},2\omega^5)$ in the $\omega\to 0$ limit).
We then take these solutions as initial guesses and solve \eqref{tp0} using Matlab's inbuilt \textit{fsolve} function. Once we have the turning points, $\mathcal{H}=0$ (with $\mathcal{H}$ given in \eqref{H3}) can be recast as a sixth order polynomial which is simple to solve for the six roots over the desired $r$--ranges.
Finally, we can solve \eqref{deep_oscil} for $\omega_n$ and then determine $\Gamma_n$ from \eqref{deep_growth}.

In Fig.~\ref{fig:BS_complex}, we display the locations of BHB eigenfrequencies in the complex plane for $D=10\mathrm{cm^2s^{-1}}$ and a range of $C/D$ and $r_B$.
We find that the strongest instabilities occur for the $m=1$ (as in the shallow water case) whereas the number of bound states per $m$--mode decreases with $m$ (unlike the shallow water case where it increased with $m$).
Viscous damping has a stronger effect on higher $\omega$ and $m$ modes, since these modes have larger $k$.
A consequence of this is some high $\omega,m$--BHB modes are completely stabilised by damping.

\begin{figure*} 
\centering
\includegraphics[width=\linewidth]{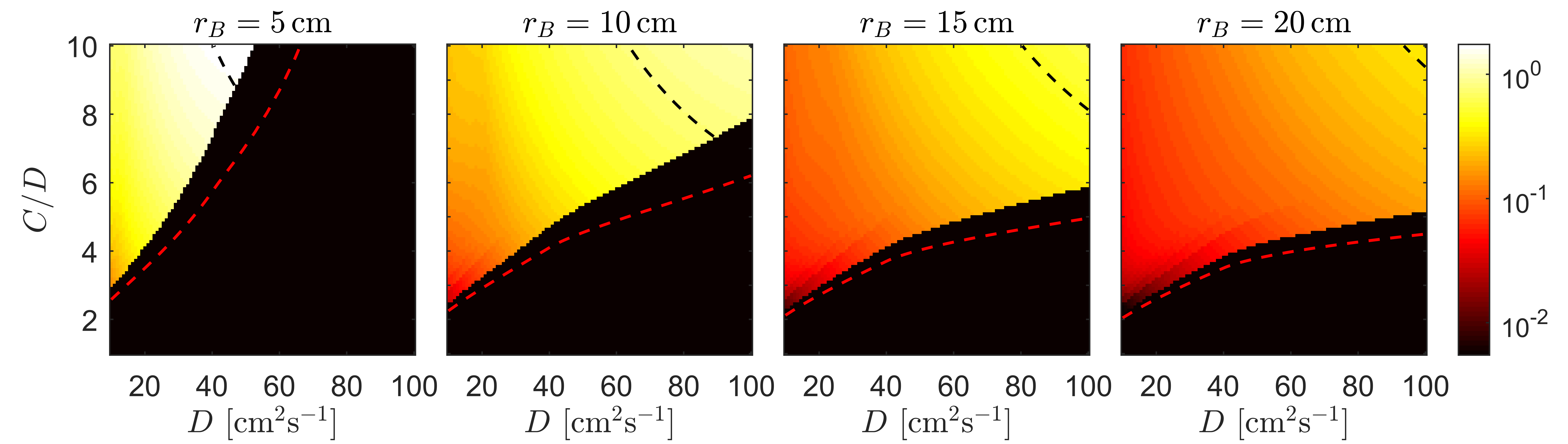}
\caption{Maximum growth rate in the $m=1$ mode as a function of $D$ and $C/D$ for various boundary locations.
The color axis gives $\mathrm{max}(\Gamma_n)$ in $\mathrm{s^{-1}}$ on a logarithmic scale, with the lighter shade indicating a stronger instability.
Hence, the dominant mode grows on a timescale which is between one and tens of seconds over the range of parameters probed.
The striped curves in the colour scale result from the dominance of different $n$, with $n=0$ being the strongest in the top right corner.
Black indicates that no BHB instability is found.
The red dashed line is the location at which the trapping region outside the vortex closes up.
The effect of viscosity is included in the growth rates.
When viscosity is neglected (not shown) the plots become slightly brighter at low $D$ since the strongest instabilities there occur at higher frequency which are attenuated by viscosity (see panels (c2,c3) of Fig.~\ref{fig:BS_varyall}).
The area of the plot where the ergosphere lies in a region of no water is delineated by a black dashed line, such that the plots only have meaning to its left.
The line is plotted for $h_0=12\mathrm{cm}$ and moves to the right as $h_0$ increases.} \label{fig:BS_maxinstab}
\end{figure*}

In Fig.~\ref{fig:BS_varyall}, we further elucidate the dependence on $r_B$, $D$ and $C/D$ by varying each variable continuously for fixed values of the others.
Here, we focus on the $m=1$ modes since these are the most unstable.
Like in the shallow water case, panel (a1) shows that increasing $r_B$ increases the density of states, thereby increasing the number of BHB modes that lie in the superradiant bandwidth.
Unlike shallow water, however, the oscillation frequencies approximately scale as $\omega_n\sim 1/\sqrt{r_B}$ in the limit of large $r_B$.
Panel (a2) reveals that stronger instabilities are achieved for smaller $r_B$, which is a consequence of the growth rate scaling inversely with the time taken to orbit the trap.
Increasing $C/D$ also has a similar effect as in shallow water.
Since $\omega_\star$ increases linearly with $C/D$ (as shown in Fig.~\ref{fig:param_plane}) at larger rotations there will be more modes inside the superradiant bandwidth, see panel (b1).
Panel (b2) shows that higher frequencies generally have larger growth rates since they are more amplified.
However, when the (amplifying) turning point gets close enough to $r_B$, the growth rate of low frequency modes can become quite large due to the orbit time decreasing.
Increasing $D$ in the dimensionless variables \eqref{rescale2} has the effect of decreasing the dimensionless mirror radius. 
The result is that the density of states decreases and modes leave the superradiant bandwidth, as shown in panel (c1).
In panel (c2), we see that the remaining low frequency modes have lower growth rates, however, once the trapping region becomes small enough the lowest frequency modes can attain much higher growth rates as the orbit time decreases. 
In panels (a3,b3,c3), we display the damping factor $1-\nu/\nu_c$. In all cases, we see that larger frequencies are damped more highly whereas the lowest frequencies are barely affected by damping at all.

An important consequence of these observations is the spatial structure of the dominant instability.
At low $r_B$ and high $D$, the dominant instability has low frequency and therefore low $n$ in \eqref{deep_oscil}.
Since $n$ is the number of nodes in the oscillatory region (i.e. outside the vortex) we can anticipate the appearance of the BHB.
At low $D$, the most unstable mode can have many radial nodes whilst at larger $D$, the mode with no nodes can come to dominate.
We comment that when the density of states is large, nonlinear interactions between modes may suppress rapid growth and prevent the instability from becoming large on short timescales.
By contrast, for a low density of states, there are less superadiant modes that can exchange energy through nonlinear interactions and the instability may be more dramatic.
Since this regime coincides with the dominance of the $n=0$ mode, we might expect that BHB manifests as a tilt of the entire free surface precessing in the same direction as the vortex rotation.

In Fig.~\ref{fig:BS_maxinstab}, we display the maximum growth rate (when viscosity is included) as a function of $D$ and $C/D$ for various system sizes.
Instabilities are present across the parameter space until the trapping region for superradiant modes closes up, which occurs for $\omega_\star=\omega^+_1(r_B)$.
The location where this is satisfied is shown as a dashed red line.
The reason the dashed red line does not overlap perfectly with the edge of the unstable patch is that the trapping region has to be wide enough for the first ($n=0$) mode to fit inside.
In general, we see that the draining vortex is most unstable when $r_B$ is smallest and when $D$ takes the maximum allowed value for a given $C/D$.
Both of these factors decrease the size of the trapping region, which decrease the orbit time, i.e. the term on the denominator of $\Gamma_n$ in \eqref{deep_growth}.
The instability can be made even stronger by increasing $C/D$, which increases the amount of superradiant amplification.
Only the low $D$ region of the parameter space (i.e. left edge of Fig.~\ref{fig:BS_maxinstab}) is affected by the inclusion of viscosity.
Without viscosity, the strongest instabilities at low $D$ will have high frequencies, as shown by Fig.~\ref{fig:BS_varyall} panel (c2).
These modes are strongly attentuated by viscosity, which renders the low $D$ region of parameter space more stable. 
We do not display this as the difference with Fig.~\ref{fig:BS_maxinstab} is hardly noticeable by eye.
For the BHB to manifest in a physical system, the region where the wave propagates into the core on $\omega_D^-$ (i.e. the ergosphere) should be in a region where there is water.
Approximating the free surface profile using \eqref{h_approx}, this gives a constraint $r_\tp>r_a$, where $r_\tp$ is the turning point inside the ergosphere.
This constraint is illustrated as a black dashed line on the figure for an example height of $h_0=12\mathrm{cm}$.

\begin{figure} 
\centering
\includegraphics[width=\linewidth]{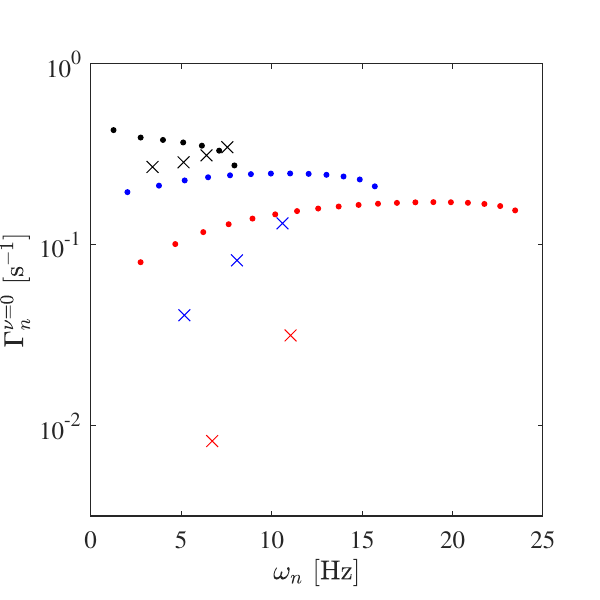}
\caption{Comparison of the BHB eigenfrequencies in shallow and deep water, represented as points and crosses respectively.
We use the parameters $r_B=5\,\mathrm{cm}$, $D=10\,\mathrm{cm^2s^{-1}}$, $C/D=5$, $\nu=0$ and the shallow water wave speed is $c=10\,\mathrm{cm\,s^{-1}}$. Results are displayed for $m=1,2,3$ in black, blue and red respectively.} \label{fig:compare}
\end{figure}

Finally, we illustrate the effect of the strong deep water dispersion on the BHB instability by explicit comparison with the shallow water case in Fig.~\ref{fig:compare}. 
This comparison is performed for the exemplary parameters $r_B=5\mathrm{cm}$, $D=10\mathrm{cm^2s^{-1}}$ and $C/D=5$. 
We compare the relevant panel in Fig.~\ref{fig:BS_complex} with the $r_B/r_h=5$ results in Fig.~\ref{fig:complex1} by choosing $r_h = 1\mathrm{cm}$, which corresponds to a height of $h\simeq1\mathrm{mm}$.
This is roughly in accordance with the hierarchy of scales in \eqref{hierarchy} given that $r_s\simeq 5\mathrm{mm}$, thereby justifying the use of the shallow water approximation (note that all the other panels with $C/D=5$ in Fig.~\ref{fig:complex1} would violate \eqref{hierarchy} for the chosen parameters).
Fig.~\ref{fig:compare} shows major differences between the two cases.
In deep water, the number of instabilities per $m$--mode decreases whilst in shallow water it increases.
Furthermore, the growth rates of the higher $m$--modes are much weaker than in shallow water.

\section{Summary and discussion} \label{sec:conc}

Using a WKB framework, we studied the scattering of capillary-gravity waves by a draining vortex, focusing in particular on the computation of the frequency of unstable modes which can be identified with the black hole bomb (BHB) instability.
The method was first applied in the shallow water regime where the analogy with black holes is exact (at the level of the scalar wave equation) and exact numerical methods were used to validate our approximate WKB approach.
We found that the WKB method was able to reliably predict the oscillatory component of the unstable mode frequency in all cases studied, whilst the prediction for growth rate deviated notably at large circulation.
We then studied the BHB instability in the deep water regime, accounting properly for the effects of surface tension and viscosity.
By considering all the different scattering possibilities for capillary-gravity waves in the system, we were able to identify a condition for superradiance.
Unlike shallow water, we found that superradiance only occurs when the circulation is above a critical multiple of the drain rate, where this multiple varied between $1$ and $2^{3/2}$ depending on the value of $m$ and the relative strength surface tension and the drain rate.
Provided the condition for amplification is met, and the space between the vortex and the boundary is large enough for the lowest excitation to fit, we found that the BHB instability is strongest when the radius of the confining mirror is small and the circulation and drain rates are large.
Under these conditions, we found that viscosity plays a negligible role.
However, unstable modes with higher frequencies (which arise at lower $D$ and larger $r_B$ and $C/D$) are affected more strongly and can grow at a fraction of the undamped rate.
Furthermore, higher $m$--modes can be completely stabilised by viscosity.

In both the shallow and deep water regimes, the $m=1$ mode was found to be the most unstable, which is a consequence of superradiance being strongest for the lowest $m$--modes in the system.
The number of unstable modes then depended on the flow parameters and the size of the system.
In the case where many unstable modes with comparable growth rates exist, we expect energy transfer between modes to lead to complicated free surface dynamics (e.g. beating) which may inhibit the growth of the dominant instability as it dissipates into the other slower-growing modes.
This scenario bears some similarity with nonlinear effects predicted for astrophysical BHBs, such as inference between different bound states capable of disguising the instability \cite{dolan2013superradiant}.
Nonlinearities also induce couplings between different $m$--modes \cite{dolan2013superradiant}
and can even lead to a collapse of the instability followed by a subsequent explosion \cite{yoshino2012bosenova}.
The study of nonlinear effects in our system therefore warrants further investigation.

By contrast, when there is only a single unstable mode, one would expect nonlinear effects to only become important at late times once the instability has reached a critical amplitude to excite decaying (stable) excitations.
This would then be the optimal regime to measure the BHB instability in an analogue experiment.
Since $n=0$ mode is expected to be the most unstable across a large parameter range, the appearance of the BHB would be a tilt of the entire free surface which rotates at the oscillation frequency of the unstable mode, i.e. a precession of the entire vortex and its free surface.

The main aim of this work has been to establish a framework capable of describing the influence of dispersion and dissipation on draining vortex instabilities.
Although we took a simplistic approach to modelling the fluid flow, our method can easily be extended to account for more realistic flow profiles. For example, one could study how surface waves are affected by the complicated 3D structure of the draining vortex base flow measured in experiments, e.g. \cite{mulligan2018understanding}. This could be achieved, for example, using empirically modelled velocity profiles that capture experimental observations.

Besides the analogy with rotating black holes, draining vortices are of inherent interest in fluid mechanics, e.g. due to the complex 3D nature of the flow near the drain \cite{andersen2003anatomy,andersen2006bathtub,yukimoto2010structure} and their role in various engineering applications \cite{odgaard1986free,hite1994velocity,stepanyants2008stationary,mulligan2016effects}.
There is a vast literature on instabilities around draining vortices, covering instabilities associated with vorticity induction starting from a non-rotating initial state \cite{mizushima2014bathtub}, rotating polygonal configurations in the open air-core \cite{mougel2017instabilities}, dip formation over the drain \cite{zhou1990axisymmetric}, excitation of Kelvin waves along the vortex core \cite{vladimirov1983experimental}, shear and centrifugal effects at the edge of viscous vortex cores \cite{gallaire2003instability} and formation of toroidal Taylor-like vortices in the bulk \cite{mulligan2018understanding}.
To the best of our knowledge, however, a precessing motion of the entire free surface (which is the signature of the $n=0$, $m=1$ BHB instability) has not been reported.
This could be because BHB instabilities are not present in the region of parameter space where experiments to date have been performed.
For example, scroll-type vortex chambers \cite{mulligan2016effects} have non-circular boundaries which are not suitable to excite the BHB unstable mode, whereas circularly symmetric set-ups are typically considered in the rotating frame \cite{andersen2003anatomy} where vorticity of the background flow plays an important role.
In order to test this hypothesis, a better understanding of superradiant instabilities in realistic flow models in required.
Our work therefore stands to push understanding of fluid mechanical vortex flows into unexplored regimes.

\section*{Acknowledgements}

SP would like to thank Ruth Gregory for helpful discussions.
This work was supported by the Science and Technology Facilities Council through the UKRI Quantum Technologies for Fundamental Physics Programme [Grant ST/T005858/1].

\bibliographystyle{apsrev4-2}
\bibliography{PRD_final.bbl}

\appendix
\numberwithin{equation}{section}

\section{Basic flow} \label{app:flow}

Here we provide a detailed derivation of the approximate flow profile in \eqref{basic}.
Under the assumption of axisymmetry, \eqref{DivFree} gives
\begin{equation} \label{cont}
    \frac{1}{r}\partial_r(rv_r) + \partial_z v_z = 0,
\end{equation}
whilst from \eqref{CurlFree} we have,
\begin{equation}
    v_\theta(r) = \frac{C}{r},
\end{equation}
where $C$ is the circulation constant.
The boundary condition on $v_r$ is the no penetration condition at the wall of the container $r=r_B$.
We also have a no-penetration condition at the base of the container except at the inlet (outlet) where we assume a constant inflow (outflow),
\begin{equation}
\begin{split}
    v_r(r=r_B,z) = & \ 0, \\
    v_z(r,z=0) = & \ -U\Theta(d-r) + W\Theta(r-r_\mathrm{in}),
\end{split}
\end{equation}
with $U$ and $W$ positive constants (see Fig.~\ref{fig:schem}).
Following \cite{stepanyants2008stationary}, we now make the simplifying assumption that the radial flow is independent of the $z$ coordinate.
Solving \eqref{cont} for $v_z$, we find,
\begin{equation} \label{vz}
    v_z(z) = -U\Theta(d-r)+W\Theta(r-r_\mathrm{in}) - \frac{z}{r}\partial_r(rv_r).
\end{equation}
The kinematic boundary condition at the free surface is,
\begin{equation} \label{kinFS}
    v_z(z=h) = (\partial_t+\mathbf{v}_\parallel\cdot\grad_\parallel)h = v_r\partial_r h,
\end{equation}
which is the statement that a fluid element at the free surface remains there, and the second equality follows our assumption that $h$ is a function of only $r$.
We can then combine \eqref{kinFS} and \eqref{vz} to write,
\begin{equation}
\frac{1}{r}\partial_r(hrv_r) = -U\Theta(d-r)+W\Theta(r-r_\mathrm{in}),
\end{equation}
which is solved for $v_r$,
\begin{equation} \label{vr_full}
    v_r(r) = -\frac{Dh_0}{rh}\times\begin{cases}
        r^2/d^2, \qquad \, \  r<d \\
        1, \qquad \qquad \ d\leq r<r_\mathrm{in} \\
        \frac{r_B^2-r^2}{r_B^2-r_\mathrm{in}^2}, \qquad \, r\geq r_\mathrm{in}
    \end{cases}
\end{equation}
where $D=Ud^2/2h_0$ is the drain constant which we take to be a positive number and $h_0$ is the asymptotic fluid height.
Applying the divergence theorem to \eqref{DivFree} over a volume $V$ gives $\int_{\partial V}\mathbf{v}\cdot d\mathbf{S}=0$ where $\partial V$ is the boundary of the volume.
Considering a volume element above the inlet, we find that the drain constant can be expressed as $D=Q/2\pi h_0$ where $Q=\pi(r_B^2-r_\mathrm{in}^2)W$ is the flow rate.
In practice, we will assume a large container such that $v_r(r_\mathrm{in})$ is very small and can be taken to be zero.
In this way, we can omit the region $r_\mathrm{in}\leq r<r_B$ from our analysis and work entirely in the region $r<r_\mathrm{in}$ where the functional form of $v_r$ is simpler.
This amounts to taking the limit $r_\mathrm{in}\to r_B$ such that $Q$ remains constant.

The final step is to solve the Bernoulli equation for the function $h(r)$.
As a first approximation, we can neglect the effects of surface tension and internal friction on the basic flow. \eqref{Bern} in the region $d\leq r<r_\mathrm{in}$ gives,
\begin{equation} \label{Bern2}
    \frac{C^2}{r^2}h^2 + \frac{D^2}{r^2}h_0^2(1+h'^2) + 2gh^2(h-h_0) = 0,
\end{equation}
where $h'=\partial_r h$.
We can solve \eqref{Bern2} by expanding in powers of $1/r$ and equating coefficients.
Writing,
\begin{equation} \label{height}
    h = h_0\left[1-\frac{r_a^2}{r^2} -h_4\frac{r_a^4}{r^4} - h_6\frac{r_a^6}{r^6} + \mathcal{O}\left(\frac{r_a^8}{r^8}\right)\right],
\end{equation}
we find,
\begin{equation}
\begin{split}
    r_a = & \ \sqrt{\frac{C^2+D^2}{2gh_0}}, \qquad r_h = \frac{D}{\sqrt{gh_0}}, \\
    h_4 = & \ \frac{r_h^2}{r_a^2}, \qquad h_6 = \frac{h_4}{2}\left(2h_4+3\right).
\end{split}
\end{equation}
Note the vertical velocity term proportional to $h'^2$ in \eqref{Bern2} only contributes at $\mathcal{O}(r_a^8/r^8)$, therefore its effect is absent from the above expression.
Denoting $r_\mathrm{min}$ the radius where $h(r_\mathrm{min})=0$, we find $Q=\pi(d^2-r_\mathrm{min}^2)U$.
For convenience in the modelling, we can take the limit  $d\to r_\mathrm{min}$ in such a way that $Q$ remains constant, so that \eqref{height} applies over the entire window of analysis.
The radial velocity is then given by,
\begin{equation}
\begin{split} \label{vr_expans}
    v_r = & \ -\frac{D}{r}\bigg[ 1+\frac{r_a^2}{r^2} + (1+h_4)\frac{r_a^4}{r^4} \\ 
    & \quad + (1+2h_4+h_6)\frac{r_a^6}{r^6} + \mathcal{O}\left(\frac{r_a^8}{r^8}\right) \bigg],
\end{split}
\end{equation}
Working in the region where $r\gg r_a$, the basic flow is given by the expressions \eqref{basic} in the main text, with the understanding that we should take $v_r(r_B)\simeq 0$ for consistency with the boundary condition.
The next to leading order corrections in $r_a/r$ increase the radial velocity in the small $r$ region whilst decreasing the height.
The analysis in this work is performed using the simple flow profiles in \eqref{basic} such that the wave analysis is tractable (semi-)analytically. However, we expect qualitative features of our conclusions to hold when corrections in powers of $r_a/r$ are taken into account.

\section{WKB expansion} \label{app:trans}

Here we provide a detailed derivation of the dispersion relation \eqref{disp} and the transport equation \eqref{transport}.
We will make use of the fact that a derivative of order $n$ (where we take $n$ to be even) acting on the WKB mode in \eqref{wkb_ansatz} gives,
\begin{equation}
    (-i\grad)^n\phi = \left[\mathbf{k}^n\mathcal{A} - \frac{i}{2\mathcal{A}}\grad\cdot\left(n\mathbf{k}^{n-1}\mathcal{A}^2\right)+...\right]e^{i\mathcal{S}},
\end{equation}
up to next-to-leading order (NLO).
In particular, this means that,
\begin{equation}
    F(-i\grad)\phi = \left[F(k)\mathcal{A} - \frac{i}{2\mathcal{A}}\grad\cdot\left(\partial_\mathbf{k}F\mathcal{A}^2\right)+...\right]e^{i\mathcal{S}}.
\end{equation}
Up to NLO, the linearised equations in \eqref{lineqs} applied to the ansatz \eqref{wkb_ansatz} give,
\begin{equation} \label{WKB_lineqs}
\begin{split}
    & \bigg[-i\Omega\mathcal{A} + f(k)\mathcal{B} + D_t\mathcal{A} \\
    & \qquad \qquad - \frac{i}{2\mathcal{B}}\grad\cdot\left(2\gamma\mathbf{k}\mathcal{B}^2\right) + 2\nu k^2\mathcal{A}\bigg]e^{i\mathcal{S}} = 0, \\
    & \bigg[-i\Omega\mathcal{B} - F(k)\mathcal{A} + \left(\partial_t+\grad\cdot\mathbf{v}\right)\mathcal{B} \\
    & \qquad \qquad \qquad \quad + \frac{i}{2\mathcal{A}}\grad\cdot\left(\partial_\mathbf{k}F\mathcal{A}^2\right)\bigg]e^{i\mathcal{S}} = 0,
\end{split}
\end{equation}
where the term proportional to $\nu$ is treated as a NLO term and $f(k)=g+\gamma k^2$ as in the main text.
Acting on the first equation with $D_t$ and inserting the second equation, we obtain,
\begin{equation}
\begin{split}
    & [-\Omega^2+f(k)F(k)]\mathcal{A} - f\left(\partial_t+\grad\cdot\mathbf{v}\right)\mathcal{B} \\
    & - \frac{if}{2\mathcal{A}}\grad\cdot\left(\partial_\mathbf{k}F\mathcal{A}^2\right) + D_t(f\mathcal{B}-i\Omega\mathcal{A}) -i\Omega D_t\mathcal{A} \\
    & - \frac{\Omega}{2\mathcal{B}}\grad\cdot\left(2\gamma\mathbf{k}\mathcal{B}^2\right) - 2i\nu\Omega k^2\mathcal{A} = 0.
\end{split}
\end{equation}
The first term is the only leading order term.
Setting this to zero gives the dispersion relation in \eqref{disp}.
The leading order relation between the amplitudes is found from \eqref{WKB_lineqs} to be $\mathcal{B}=i\Omega\mathcal{A}/f$.
Substituting into the remaining terms and multiplying by $i\mathcal{A}/f$ gives,
\begin{equation}
\begin{split}
& \partial_t\left(\frac{\Omega\mathcal{A}^2}{f}\right) +  \grad\cdot\left(\frac{\mathbf{v}\Omega\mathcal{A}^2}{f}\right) \\ 
& + \frac{1}{2}\grad\cdot\left[\left(\partial_\mathbf{k}F+\frac{2\gamma\mathbf{k}\Omega^2}{f^2}\right)\mathcal{A}^2\right] + 2\nu k^2\frac{\Omega\mathcal{A}^2}{f} = 0.
\end{split}
\end{equation}
To make further progress, we use $\Omega^2=f(k)F(k)\equiv\mathcal{D}(k)$ to rewrite,
\begin{equation}
\partial_\mathbf{k}F+\frac{2\gamma\mathbf{k}\Omega^2}{f^2} = \frac{f\partial_\mathbf{k} F+F\partial_\mathbf{k}f}{f} = \frac{\partial_\mathbf{k}\mathcal{D}}{f} = \frac{2\Omega\partial_\mathbf{k}\Omega}{f}.
\end{equation}
Finally, using the definition of the group velocity in the lab frame $\mathbf{v}_g=\mathbf{v}+\partial_\mathbf{k}\Omega$, we obtain the transport equation \eqref{transport} in the main text.

Another route to the transport equation is to substitute the WKB ansatz \eqref{wkb_ansatz} directly into the components of the norm current.
For the norm density we obtain,
\begin{equation} \label{WKB_norm}
    \rho_n = 2\frac{\Omega|\mathcal{A}|^2}{f},
\end{equation}
and for the current we get,
\begin{equation} \label{WKB_curr}
\begin{split}
    \mathbf{J}_n = & \ \gamma\mathbf{k}|\mathcal{B}|^2 -i\mathbf{v}\mathcal{A}^*\mathcal{B} \\
    & - \frac{1}{2}\sum_{m=1}^\infty F_m\sum_{n=0}^{2m-1}(-1)^n\mathbf{k}^n\mathbf{k}^{2m-1-n}|\mathcal{A}|^2 + \mathrm{c.c.} \\
    = & \ \frac{\gamma\mathbf{k}\Omega^2|\mathcal{A}|^2}{f^2} + \frac{\mathbf{v}\Omega|\mathcal{A}|^2}{f} \\
    & \qquad \quad + \frac{1}{2}\sum_{m=1}^\infty F_m 2m\mathbf{k}^{2m-1}|\mathcal{A}|^2 + \mathrm{c.c.} \\
    = & \ \frac{\mathbf{v}\Omega + \gamma\mathbf{k}F+\frac{1}{2}f\partial_\mathbf{k}F}{f}|\mathcal{A}|^2 + \mathrm{c.c.} \\
    = & \ 2\frac{\mathbf{v}_g\Omega|\mathcal{A}|^2}{f},
\end{split}
\end{equation}
where we have assumed $\mathbf{k}\in\mathbb{R}$.
Hence, the conservation equation for the norm \eqref{cons_norm} reduces to the transport equation \eqref{transport} in regions where the waves are propagating for the case $\nu=0$.

\section{Transfer matrices} \label{app:mat}

In this appendix, we provide a detailed derivation of the two transfer matrices in \eqref{tp1} and \eqref{tp2}.
The analysis can also be found in \cite{patrick2020superradiance}, although here we provide more details on the intermediate steps.
The spirit of the calculation is to solve the equations of motion exactly in the vicinity of the turning point, then map the asymptotic behaviour onto the WKB modes either side.
In this way, one obtains a relation between oscillatory and evanescent mode amplitudes at $r_\tp$.
In the first step, we expand the effective Hamiltonian around the point $(r_\tp,p_\tp)$ in phase space,
\begin{equation} \label{Ham_loc}
    \mathcal{H} = \partial_r\mathcal{H}(r-r_\tp) + \frac{\mathcal{H}''_\tp}{2}(p-p_\tp)^2 + ...
\end{equation}
where we have used \eqref{tp_cond} which defines the turning point and written $\mathcal{H}''_\tp=\partial^2_p\mathcal{H}_\tp$ for shorthand.
Next we promote $p\to-i\partial_r$ and let this act on the radial part $R(r)$ of the velocity potential perturbation defined in \eqref{expans1}. Let $R(r) = e^{ip_\tp (r-r_\tp)}\psi(r)$ and define a new variable $s=|Q|^{1/3}(r-r_\tp)$ where $Q=2\partial_r\mathcal{H}_\tp/\mathcal{H}''_\tp$.
With this definition, $s>0$ ($s<0$) corresponds to the right (left) of the turning point, and local form of the wave equation is,
\begin{equation} \label{airy}
    \partial_s^2\psi = \pm s\psi.
\end{equation}
The fact that this is second order in spatial derivatives is a consequence of only two distinct solutions of the dispersion relation mixing at a turning point.
The general solution of \eqref{airy} is,
\begin{equation} \label{airy_sol}
    \psi = C_A\mathrm{Ai}(\pm s) + C_B\mathrm{Bi}(\pm s).
\end{equation}
The asymptotic behaviour of these solutions is,
\begin{equation} \label{airy_asymp}
\begin{split}
    \mathrm{Ai}(s) & \overset{+\infty}{\sim}\frac{1}{2\sqrt{\pi}|s|^{1/4}}e^{-\frac{2}{3}s^{3/2}}, \\
    & \overset{-\infty}{\sim}\frac{1}{2\sqrt{\pi}|s|^{1/4}}\left(e^{-\frac{2i}{3}(-s)^{3/2}+\frac{i\pi}{4}}+e^{\frac{2i}{3}(-s)^{3/2}-\frac{i\pi}{4}}\right), \\
    \mathrm{Bi}(s) & \overset{+\infty}{\sim}\frac{1}{\sqrt{\pi}|s|^{1/4}}e^{\frac{2}{3}s^{3/2}}, \\
    & \overset{-\infty}{\sim}\frac{1}{2\sqrt{\pi}|s|^{1/4}}\left(e^{\frac{2i}{3}(-s)^{3/2}+\frac{i\pi}{4}}+e^{-\frac{2i}{3}(-s)^{3/2}-\frac{i\pi}{4}}\right).
\end{split}
\end{equation}
The aim now is to match this behaviour onto that of the WKB modes. In particular, we want to know the spatial profile of the modes as we move a small distance away from the turning point.
In this appendix, we will call the two interacting modes $p^\pm$ (such that $\mathrm{Re}[p^+]>\mathrm{Re}[p^-]$) although in the main text they are labelled $p^{\Ri/\Le}$ and can correspond to any neighbouring pair in the sequence $j\in\{b,d,-,+,u,t\}$.
Solving \eqref{Ham_loc}, we find,
\begin{equation}
    p^\pm = p_\tp \pm \sqrt{Q(r_\tp-r)}.
\end{equation}
We first analyse the case $Q>0$, which corresponds to the $+$ sign in \eqref{airy}.
Using the shift factors in \eqref{shift}, the two WKB modes a small distance $\delta r$ to the left of $r_\tp$ can be written,
\begin{equation}
\begin{split}
    A^\pm_\tp \mathcal{F}^\pm_{-\delta r} = & \ \left|\frac{q^\pm(r_\tp)}{q^\pm(r_\tp-\delta r)}\right|^\frac{1}{2}A^\pm_\tp e^{-i\int^{r_\tp}_{r_\tp-\delta r}p^\pm(r') dr'} \\
    = & \ \left|\frac{f_\tp}{\mathcal{H}''_\tp Q^{1/3}}\right|^\frac{1}{2}e^{ip_\tp(r-r_\tp)}\frac{\alpha_\tp^\pm}{|s|^{1/4}}e^{\mp\frac{2i}{3}(-s)^\frac{3}{2}},
\end{split}
\end{equation}
where we have used \eqref{amp} to introduce $\alpha^\pm_\tp$. For shorthand, we have defined $q=\partial_p\mathcal{H}/f$ and used the subscript $-\delta r$ to indicate that the shift factor is evaluated between the points $r_\tp-\delta r$ and $r_\tp$, where $\delta r=r_\tp-r$.
On the left, the general solution will be a superposition of the two modes above.
By comparing this with \eqref{airy_asymp}, we find,
\begin{equation}
    \begin{split}
        \alpha^+_\tp = & \ e^{i\pi/4} C'_A + e^{-i\pi/4} C'_B,  \\
        \alpha^-_\tp = & \ e^{-i\pi/4} C'_A + e^{i\pi/4} C'_B,
    \end{split}
\end{equation}
and we have factored out an overall constant by defining,
\begin{equation}
    C_j = \left|\frac{4\pi f_\tp}{Q^{1/3} \mathcal{H}''_\tp}\right|^{1/2}C'_j.
\end{equation}
Similarly, on the right of the turning point, we have,
\begin{equation}
\begin{split}
    A^{\upharpoonleft\!\downharpoonright}_\tp/\mathcal{F}^{\upharpoonleft\!\downharpoonright}_{+\delta r} = & \ \left|\frac{q^{\upharpoonleft\!\downharpoonright}(r_\tp)}{q^{\upharpoonleft\!\downharpoonright}(r_\tp+\delta r)}\right|^\frac{1}{2}A^{\upharpoonleft\!\downharpoonright}_\tp e^{i\int^{r_\tp+\delta r}_{r_\tp}p^{\upharpoonleft\!\downharpoonright}(r') dr'} \\
    = & \ \left|\frac{f_\tp}{\mathcal{H}''_\tp Q^{1/3}}\right|^\frac{1}{2}e^{ip_\tp(r-r_\tp)}\frac{\alpha_\tp^{\upharpoonleft\!\downharpoonright}}{|s|^{1/4}}e^{\pm\frac{2}{3}s^{3/2}},
\end{split}
\end{equation}
where $\delta r = r-r_\tp$ and we take the $+$ ($-$) sign for the mode which grows (decays) in the direction of increasing $r$ labelled by $\uparrow$ ($\downarrow$).
By comparing with \eqref{airy_asymp}, we find,
\begin{equation}
    \alpha^\uparrow_\tp = 2C'_B, \qquad \alpha^\downarrow_\tp = C'_A.
\end{equation}
Altogether, we find,
\begin{equation}
    \begin{pmatrix}
        \alpha^+_\tp \\ \alpha^-_\tp
    \end{pmatrix} = e^{\frac{i\pi}{4}}\begin{pmatrix}
        1 & -\frac{i}{2} \\ -i & \frac{1}{2}
    \end{pmatrix} \begin{pmatrix}
        \alpha^\downarrow_\tp \\ \alpha^\uparrow_\tp
    \end{pmatrix},
\end{equation}
which is the matrix relation in \eqref{tp1} expressed in terms of the $\alpha_\tp$ rather than the $A_\tp$.

We now consider the opposite situation where $Q<0$.
In this case, the propagating region is on the right of the turning point.
A small distance to the right, we have,
\begin{equation}
\begin{split}
    A^\pm_\tp/\mathcal{F}^\pm_{+\delta r} = & \ \left|\frac{q^\pm(r_\tp)}{q^\pm(r_\tp+\delta r)}\right|^\frac{1}{2}A^\pm_\tp e^{i\int^{r_\tp+\delta r}_{r_\tp}p^\pm(r') dr'} \\
    = & \ \left|\frac{f_\tp}{\mathcal{H}''_\tp Q^{1/3}}\right|^\frac{1}{2}e^{ip_\tp\delta r}\frac{\alpha_\tp^\pm}{|s|^{1/4}}e^{\pm\frac{2i}{3}s^{3/2}},
\end{split}
\end{equation}
where $\delta r=r-r_\tp$.
Taking $s\to-s$ in \eqref{airy_asymp} and comparing with the above, we get,
\begin{equation}
    \begin{split}
        \alpha^+_\tp = & \ e^{i\pi/4} C'_B + e^{-i\pi/4} C'_A,  \\
        \alpha^-_\tp = & \ e^{-i\pi/4} C'_B + e^{i\pi/4} C'_A.
    \end{split}
\end{equation}
On the left, the WKB modes are,
\begin{equation}
\begin{split}
    A^{\upharpoonleft\!\downharpoonright}_\tp\mathcal{F}^{\upharpoonleft\!\downharpoonright}_{-\delta r} = & \ \left|\frac{q^{\upharpoonleft\!\downharpoonright}(r_\tp)}{q^{\upharpoonleft\!\downharpoonright}(r_\tp-\delta r)}\right|^\frac{1}{2}A^{\upharpoonleft\!\downharpoonright}_\tp e^{-i\int^{r_\tp}_{r_\tp-\delta r}p^{\upharpoonleft\!\downharpoonright}(r') dr'} \\
    = & \ \left|\frac{f_\tp}{\mathcal{H}''_\tp Q^{1/3}}\right|^\frac{1}{2}e^{ip_\tp\delta r}\frac{\alpha_\tp^{\upharpoonleft\!\downharpoonright}}{|s|^{1/4}}e^{\pm\frac{2}{3}(-s)^{3/2}},
\end{split}
\end{equation}
where $\delta r = r_\tp-r$ and we take the $-$ ($+$) sign for the mode which grows (decays) in the direction of increasing $r$ labelled by $\uparrow$ ($\downarrow$).
By comparing with \eqref{airy_asymp}, again with $s\to-s$, we find,
\begin{equation}
    \alpha^\uparrow = C'_A, \qquad \alpha^\downarrow = 2C'_B,
\end{equation}
so that altogether, we have,
\begin{equation}
    \begin{pmatrix}
        \alpha^\uparrow_\tp \\ \alpha^\downarrow_\tp
    \end{pmatrix} = e^{\frac{i\pi}{4}}\begin{pmatrix}
        \frac{1}{2} & -\frac{i}{2} \\ -i & 1
    \end{pmatrix} \begin{pmatrix}
        \alpha^+_\tp \\ \alpha^-_\tp
    \end{pmatrix},
\end{equation}
which is the matrix relation in \eqref{tp2} expressed in terms of the $\alpha_\tp$ rather than the $A_\tp$.

\section{Boundary conditions} \label{app:BCs}

It is reasonable to expect that at the hard wall boundary, the out-going propagating mode with amplitude $A_B^{+,t}$  will transfer all of its energy into the in-going propagating mode with amplitude $A_B^-$. However, due to the presence of additional dispersive modes in the system, this expectation is non-trivial to verify.
In this appendix, we provide a justification when the container is assumed to be large relative to the wavelength of modes near the boundary.

The wall of the container $r=r_B$ is a hard impenetrable wall, hence the boundary condition of velocity potential perturbations is,
\begin{equation} \label{vr_bc}
    \partial_r\phi|_{r=r_B}=0,
\end{equation}
for all $z$.
An additional boundary condition is needed when surface tension is included, due to the extra derivatives appearing in the equations of motion.
This can be derived by considering the origin of the surface tension following \cite{landau2013fluid}.
The air water interface is a surface $z=h(\mathbf{x}_\parallel)$ with total area $A=\int d^2\mathbf{x}_\parallel\sqrt{1+|\grad_\parallel h|^2}$.
The total work done in deforming the surface is,
\begin{equation}
    \delta W = -\int d^2\mathbf{x}_\parallel [P(z=h)-P_a]\delta h + \sigma \delta A,
\end{equation}
where the first term is the work needed to bring about a change in volume whilst the second encodes the contribution of surface tension.
In thermodynamic equilibrium, $\delta W=0$ and application of the divergence theorem leads to the condition on the pressure \eqref{pressure_BC} for $\nu=0$.
The additional constraint that $\delta W=0$ when $h$ is varied on the physical boundary $r=r_B$ of the system leads to the condition,
\begin{equation}
    \frac{\partial_r h}{\sqrt{1+|\grad_\parallel h|^2}}\Bigg|_{r=r_B}=0.
\end{equation}
For surface waves $\eta$ moving on an approximately flat interface, we therefore have,
\begin{equation} \label{eta_BC}
    \partial_r\eta|_{r=r_B} = 0.
\end{equation}
When $r_B$ is large, $\nabla_\parallel^2\sim\partial_r^2$ and the dynamics described by \eqref{lineqs} becomes effectively 1D.
Since $\mathbf{v}$ is small in this region and the free surface is approximately flat, we can take the solution to the radial equation to be a superposition of plane waves,
\begin{equation} \label{expand_phi}
    \phi(r\sim r_B,z=h) = e^{im\theta-i\omega t}\sum_j \mathcal{A}^j_B e^{ip^j (r-r_B)}
\end{equation}
where the $p^j\simeq k^j$ satisfy the dispersion relation \eqref{disp}, i.e. $\omega^2=(gk+\gamma k^3)\tanh(h_0k)$.
This has an infinite number of roots due to the hyperbolic tangent, hence, to fix the amplitudes $\mathcal{A}^j$, one needs to specify an infinite number of boundary conditions.
To keep track of these, we write the dispersion relation using \eqref{taylor} as,
\begin{equation} \label{disp_taylor}
    \omega^2 = (g+\gamma k^2)\sum_{n=1}^{N+1}F_n k^{2n},
\end{equation}
then take the limit $N\to\infty$.
The dispersion relation \eqref{disp_taylor} has $2N+4$ roots.
Two of these are real with equal and opposite $p$ and correspond to the two asymptotically propagating roots of \eqref{disp} (either $p^\pm$ or $p^{t,-}$ depending on the scattering diagram, see Fig.~\ref{fig:diagrams}).
The other $2N+2$ are purely imaginary and come in complex conjugate pairs (note that as $r$ is decreased, two of these pairs migrate down toward the real line where they may or may not become propagating, which is why there is a total of 6 modes on the scattering diagrams of Fig.~\ref{fig:diagrams}).
Hence, the problem is solved by supplying a total of $2N+4$ boundary conditions.

The condition in \eqref{vr_bc} is actually equivalent to $N+1$ boundary conditions owing to the fact that it holds for all $z$.
This can be seen by using \eqref{lin_sol} to write a Taylor expansion for velocity potential perturbations in the bulk,
\begin{equation} \label{expand_phi2}
\begin{split}
    \phi(r\sim r_B,z) = & \ \cosh(-iz\partial_r)\phi_0, \\
    = & \ \sum_{n=0}^N \frac{z^{2n}}{(2n)!}(-i\partial_r)^{2n}\phi_0,
\end{split}
\end{equation}
where $\phi_0=\phi(z=0)$ and the $N$ here is the same one appearing in \eqref{disp_taylor}.
Since $\partial_r\phi$ at the wall vanishes for all $z$, we must have,
\begin{equation} \label{N1_bcs}
    (-i\partial_r)^{2n+1}\phi_0|_{r=r_B}=0 \quad \mathrm{for}~n=0,1...N.
\end{equation}
To derive a condition on the amplitudes $\mathcal{A}^j$, we expand $\phi_0$ analogously to \eqref{expand_phi} with new amplitudes $\tilde{\mathcal{A}}^j$, noting that \eqref{expand_phi2} implies the relation $\mathcal{A}^j=\cosh(hk^j)\tilde{\mathcal{A}}^j$.
The $N+1$ conditions in \eqref{N1_bcs} can then be written,
\begin{equation}
    \sum_{j=\mathrm{pairs}} {k^j}^{2n+1}\mathrm{sech}(h k^j)\left(\mathcal{A}^{j+}_B-\mathcal{A}^{j-}_B \right) = 0,
\end{equation}
where we have used the fact that the wavevectors come in pairs of equal and opposite sign, i.e. $k^{j\pm}=\pm k^j$.
The boundary condition \eqref{eta_BC} gives the additional constraint,
\begin{equation}
    \sum_{j=\mathrm{pairs}} \frac{\omega k^j}{g+\gamma{k^j}^2}\left(\mathcal{A}^{j+}_B-\mathcal{A}^{j-}_B \right) = 0,
\end{equation}
so that we have a total of $N+2$ constraints on $N+2$ pairs of amplitudes.
These are simultaneously satisfied provided $\mathcal{A}_B^{j+}=\mathcal{A}_B^{j-}$.
This condition translates to the amplitudes $A^j$ in \eqref{amp} so that $A_B^{j+}=A_B^{j-}$, since the factor $\sqrt{r}$ is only the slowly varying part of the waveform.

The resonance condition for the natural frequencies of the vortex is determined upon specification of the other $N+2$ boundary conditions inside the vortex core.
Since we do not have a physical boundary here, we make the physically reasonable assumption that all evanescent modes which grow as $r\to 0$ (of which there are $N$) have zero amplitude, i.e. $A^{j\downarrow}_0=0$.
The final two conditions are provided by the assumption that the vortex does not emit any noise in equilibrium.
Hence, the amplitudes of the two out-going modes as $r\to 0$ can be set to zero.
With all $2N+4$ boundary conditions specified, the resonance condition is uniquely determined.

Since $2N-2$ of the modes are evanescent over the whole system (and do not mix with other modes) application of the boundary conditions above sets their amplitudes everywhere to zero.
The remaining boundary conditions can be summarised as,
\begin{equation} \label{boundaryconditions}
    r=r_B~
        \begin{cases}
            A_B^\mathrm{out} = A_B^\mathrm{in} \\
            A_B^{1\uparrow} = A_B^{1\downarrow} \\
            A_B^{2\uparrow} = A_B^{2\downarrow}
        \end{cases}, \qquad r\to0~ 
        \begin{cases}
            A_0^{1\mathrm{out}} = 0 \\
            A_0^{2\mathrm{out}} = 0 \\
            A_0^\downarrow = 0
        \end{cases}.
\end{equation}
Note that superscripts refer to the specific modes which are out/in-going or growing/decaying at a given boundary and a number is included since there are multiple evanescent mode pairs/out-going modes.
For example, in Type VI scattering (which is one of the more complicated diagrams) we have
$A_B^\mathrm{in}=A_B^-$, 
$A_B^\mathrm{out}=A_B^+$, 
$A_0^{1\mathrm{out}}=A_0^b$, 
$A_0^{1\mathrm{in}}=A_0^+$,
$A_0^{2\mathrm{in}}=A_0^u$ and 
$A_0^{2\mathrm{out}}=A_0^t$.
One of the evanescent pairs at $r=r_B$ is formed by the $b,d$--modes and the second pair is comprised of the $u,t$--modes.
As $r\to0$, there is a single evanescent pair formed by the $-,d$--modes.

We remark that a subtlety of this derivation is our assumption that the radial velocity vanishes on the boundary, which meant the solutions of the dispersion came in pairs of equal and opposite $p$. The no penetration condition at the wall of the container certainly enforces this in any real experiment. However, in our evaluation of the WKB phase integrals in the main text, we assume that $v_r=-D/r$ everywhere in the system to simplify the analysis. In this case, the wavevector pairs aren't exactly equal in magnitude on the boundary, which will affect the boundary condition.
For example, in shallow water we should instead have $p_B^+A^+_B+p_B^-A^-_B=0$ which reduces to the boundary condition above when  $r_B\to\infty$ and $p^-_B\to-p^+_B$.
However, the excellent agreement with numerical calculations of the real part of the eigenfrequencies in Figs.~\ref{fig:complex1} and \ref{fig:vary1} suggest that differences arising from the precise nature of the flow field over the inlet are negligible.
In this way, we assume that using $v_r=-D/r$ everywhere with the above boundary conditions gives a good estimate of the BHB eigenfrequencies in both the shallow and deep water regimes.
The calculation could be improved by including the inlet region in \eqref{vr_full} in the radial velocity, although this would come at the expense of adding an extra parameter (the width of the inlet) to the parameter space of the system.

\section{Parameter space}

\subsection{Turning points} \label{app:tps}

To identify the turning points of the deep water system \eqref{H3}, we first identify the functions $\omega^\pm_a(r)$ defined in \eqref{tp_potentials}.
Then the intersection of a line of $\omega=\mathrm{const}$ with these curves tells us where the turning points are and which modes mix there.
The first step is to find the extrema of $\omega_D^\pm(p)$ at a fixed radius.
Using $\partial_p\omega_D^\pm|_{p=p_\mathrm{ex}}=0$, we find that the $p_\mathrm{ex}$ satisfy $\pm 2k\sqrt{k+\gamma k^3}=pr(1+3\gamma k^2)$.
Squaring this, we find a polynomial equation for $k_\mathrm{ex}$,
\begin{equation}
\begin{split}
    & 9\gamma^2r^2k_\mathrm{ex}^6 -4\gamma k_\mathrm{ex}^5 + (6\gamma r^2-9\gamma^2)k_\mathrm{ex}^4 \\
    & \qquad -4k_\mathrm{ex}^3 + (r^2-6\gamma)k_\mathrm{ex}^2-1 = 0,
\end{split}
\end{equation}
from which we can calculate $p_\mathrm{ex}=\pm\sqrt{k_\mathrm{ex}^2-1/r^2}$.
Note that since $\omega_D^\pm$ have a maximum of three real extrema (see Fig.~\ref{fig:disp}) which come in pairs (i.e. if there is an extremum at $p_\mathrm{ex}$ on $\omega_D^+$ then there is also one at $-p_\mathrm{ex}$ on $\omega_D^-$) at least three of these solutions are complex and can be discarded.
The remaining $a$ real $p_\mathrm{ex}$ (where $a=1,2,3$ for $\gamma<\gamma_c$ and $a=1$ for $\gamma>\gamma_c$) are then substituted into $\omega_D^\pm(p)$ to find $\omega^\pm_a$ at that radius.
Repeating this procedure for all $r$ generates the curves $\omega^\pm_a(r)$.

\subsection{Characteristic frequencies} \label{app:freq}

Here, we identify the characteristic frequencies of the deep water system \eqref{H3} which divide up the $(\omega,C,\gamma)$ parameter space into regions where different scattering processes occur.
We begin by analysing \eqref{H3} when $\gamma$ is small.
In particular, $\gamma$ should be small enough that the low $p$ region of dispersion relation (where the term proportional to $k$ dominates) is reasonably large.
A possible form for the $\omega_D^\pm$ at an intermediate $r$ is sketched in Fig.~\ref{fig:disp}.
There are a maximum of 6 propagating solutions.
In this sketch, the two central modes lie on $\omega_D^+$, however, it is also possible for these modes to be on $\omega_D^-$.

\begin{figure}
\centering
\includegraphics[width=\linewidth]{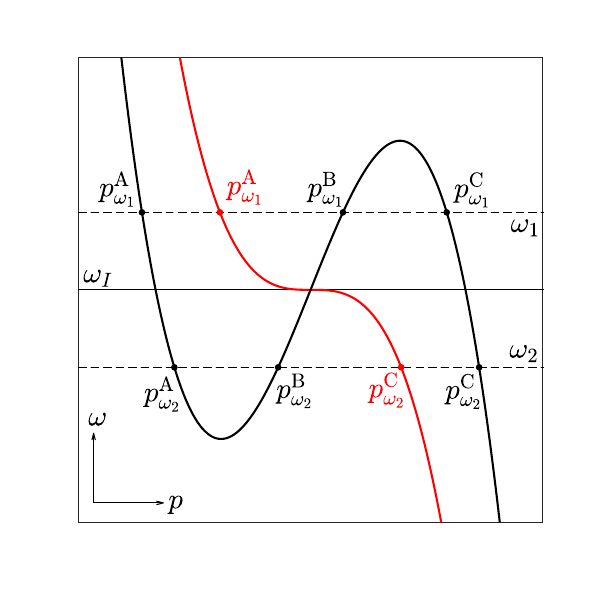}
\caption{The value of $\omega_D^\pm$ at inflection points determines which modes scatter.} \label{fig:inflection}
\end{figure}

Inflection points of the dispersion relation play an important role in scattering between the different $p^j$. Consider the diagram illustrated in Fig.~\ref{fig:inflection}, which shows one of the $\omega_D^\pm(p)$ in the vicinity of two extrema (i.e. points where $\partial_p\omega_D^\pm=0)$.
The branch is shown at say $r_I$ in red and at $r_I+\delta r$ in black.
In moving from the black to the red curve, the two extrema have coalesced into an inflection point $(r_I,p_I,\omega_I)$.
At $r_I+\delta r$, there are three real solutions, say $p^{A,B,C}$, whereas at $r_I$ there is only one.
If $\omega>\omega_I$, then $p^B$ and $p^C$ are degenerate (and therefore scatter) at their local extremum before moving off in the complex plane, so that $p^A$ is the propagating mode at $r_I$.
By contrast, for $\omega<\omega_I$, $p^A$ and $p^B$ scatter leaving $p^C$ as the propagating mode at $r_I$.
Hence, we see that value of $\omega$ relative to $\omega_I$ informs us which modes scatter at particular locations.

The $\omega_a^\pm$ meet (and terminate) at inflection points.
There are a total of 4 such points: $\omega^+_{I1}$ and $\omega^+_{I2}$ on the upper branch and $\omega^-_{I1}$ and $\omega^-_{I2}$ on the upper branch.
The primary inflection points $\omega_{I1}^\pm$ result from the deep water dispersion (in \cite{patrick2020superradiance} they are called the upper and lower critical frequencies) whilst the secondary inflection points $\omega_{I2}^\pm$ are a result of surface tension.

Fig.~\ref{fig:potential_deep} further illustrates another pair of important frequencies.
These are related to the light-ring we met in Section~\ref{sec:shal_WKB} and correspond to the extrema of the $\omega^\pm_j(r)$ in the radial direction, i.e. $\partial_r\omega^\pm_a|_{r=r_\lr}=0$ and $\omega^\pm_\lr=\omega^\pm_a(r_\lr)$, where the corresponding ray forms a closed circular orbit around the vortex.
The value of $\omega$ relative to $\omega_\lr$ informs us about scattering.
For example, in Fig.~\ref{fig:potential_deep}(a), the light-ring is on $\omega_2^+(r)$. In this case, the $p^{u,+}$ modes do not interact for $\omega$ just below $\omega_\lr^+$, whilst for $\omega$ just above $\omega_\lr^+$, they scatter and propagate in both directions moving away from $r_\lr$.
Hence, to characterise scattering at low $\gamma$, we need to calculate a total of 6 characteristic frequencies which are given in \eqref{charac1} of the main text.

At large $\gamma$, the parameter space undergoes a structural change.
As $\gamma$ is increased, there comes a critical value where $\omega_{I1}^+$ coalesces with $\omega_{I2}^+$ and $\omega_{I1}^-$ with $\omega_{I2}^-$.
Physically, this means that the $\omega_D^\pm(p)$ possess a single extremum each for all $r$.
The coalescence of inflection points is determined by the condition that $\mathcal{H}$ and its first three $p$-derivatives vanish.
Solving these conditions gives the critical value $\gamma_c\simeq 0.0522$.
The remaining three conditions give the location at which the coalescence occurs:
\begin{equation} \label{crit}
    (p_c,r_c,\gamma_c) = (1.105,2.033,0.0522).
\end{equation}
For $\gamma>\gamma_c$, there are a maximum of 4 propagating modes at any given $r$.
In particular, there are necessarily two as $r\to\infty$ and four as $r\to 0$.
Depending on the value of $\omega$, the modes which propagate at infinity can either end up on $\omega_D^+$ or $\omega_D^-$ as $r\to0$.
To find a condition for the threshold value $\omega_S$ marking the transition between these two cases, it is instructive to inspect the trajectories of the $p^j$ through the complex plane generated by varying $r$, see e.g. the diagram in Fig.~7 of \cite{patrick2021rotational} where these trajectories are shown explicitly.
One finds that, when the frequency crosses the threshold, there is a reconnection between the large and small $r$ segments of the diagram in the vicinity of a complex saddle point $p_S$.
We can then find the value of $\omega_S$ by solving for the saddle points of $\mathcal{H}$.
We can identify two such values $\omega_S^\pm$:
above $\omega_S^+$ (below $\omega_S^-$) propagating modes at infinity connect onto the upper (lower) branch of the dispersion relation at small $r$ as propagating modes.
Between $\omega_S^+$ and $\omega_S^-$, the propagating modes at infinity are evanescent as $r\to0$.
Hence, to characterise scattering at low $\gamma$, we need to calculate a total of 4 characteristic frequencies given in \eqref{charac2} of the main text.

At the threshold value of $\gamma_c$, there should be a smooth transition from the inflection points in \eqref{charac1} to the saddle points in \eqref{charac2}.
Indeed, as $\gamma\to\gamma_c$ from above, $p_S$ and $p_S^*$ move toward the real axis, such that at $\gamma=\gamma_c$ we have $p_S\in\mathbb{R}$.
Since two of the $p^j$ have the same value $p_S$ on the saddle point, four of the $p^j$ will have the real value $p_S$ when $\gamma=\gamma_c$.
Therefore, the Hamiltonian can be expanded to leading order as $\mathcal{H}\sim\partial_p^4\mathcal{H}_S(p-p_S)^4$ near this point. This is consistent with the vanishing of $\mathcal{H}$ and its first three derivatives at the merger of inflection points.
In other words, when inflection points of the dispersion relation coalesce, they produce saddle points which move off in the complex plane.
We therefore have,
\begin{equation}
\begin{split}
    & \omega_{I1}^+(\gamma_c)=\omega_{I2}^+(\gamma_c)=\omega_S^+(\gamma_c), \\
    & \omega_{I1}^-(\gamma_c)=\omega_{I2}^-(\gamma_c)=\omega_S^-(\gamma_c),
\end{split}
\end{equation}
which is true for all $C$.

\subsection{Numerical method}

Most of the work is in calculating the characteristic frequencies in \eqref{charac1} and \eqref{charac2}.
This is done numerically starting from a set of initial guesses.
The inflection frequencies are found by solving the conditions $\partial_p\omega_D^\pm=\partial_p^2\omega_D^\pm=0$ for $(p_I,r_I)$.
Both of these equations are independent of the rotation parameter hence the values $(p_I,r_I)$ will not depend on $C$.
We start by treating $\gamma$ perturbatively to get the following analytic expressions,
\begin{equation} \label{infl_guess}
\begin{split}
    (\omega^\pm_{I1},p^\pm_{I1},r_{I1}) \simeq & \ \left(\frac{C\pm 2^{3/2}}{2^{2/3}3},~\pm\frac{2^{1/6}}{3^{1/2}},~\frac{6^{1/2}}{2^{1/6}}\right), \\
    (\omega^\pm_{I2},p^\pm_{I2},r_{I2}) \simeq & \ \left[ \pm\frac{\alpha^{3/4}}{\gamma^{1/4}},~\pm\frac{\alpha^{1/2}}{\gamma^{1/2}} ,~\left(\frac{1}{9\alpha\gamma}\right)^{1/4}\right],
\end{split}
\end{equation}
with $\alpha=2/\sqrt{3}-1$, where at this order the primary inflection points coincide with the deep water results of \cite{patrick2020superradiance}.
$(p_I,r_I)$ are then found numerically by solving the simultaneous equations for some small value of $\gamma$ using \eqref{infl_guess} as an initial guess.
$\gamma$ is then increased in steps up to the value $\gamma_c$, each time using the previous numerical result as the initial guess.
This procedure generates $(p_I,r_I)$ as a function of $\gamma$.
The frequencies are then found by choosing a $C$ value and computing $\omega^\pm_I=\omega_D^\pm(p_I^\pm,r_I)$.

The procedure to find $\omega_\lr^\pm$ is simpler since the defining equations can be easily formulated as a single polynomial equation, which can be solved without needing an initial guess.
First, the conditions $\partial_r\mathcal{H}=\partial_p\mathcal{H}=0$ can be combined to write $p^\pm_\lr r^\pm_\lr = B_\pm \equiv C\pm\sqrt{C^2+1}$.
We can then formulate a polynomial equation for $k$,
\begin{equation}
    4\gamma k^5 -9\gamma^2B_\pm^2k^4+4k^3-6\gamma B_\pm^2k^2-B_\pm^2 = 0,
\end{equation}
where the solution we want has $k\in\mathbb{R}$ and $k>0$.
From here we can compute $r^\pm_\lr = \sqrt{B_\pm^2+1}/k^\pm_\lr$, $p^\pm_\lr = B_\pm/r_\lr^\pm$, then $p^\pm_\lr = B_\pm/r^\pm_\lr$ and lastly $\omega_\lr^\pm = \omega_D^\pm(p_\lr^\pm,r_\lr^\pm)$.
This procedure then is repeated over the $(\gamma,C)$ parameter space.

Finally, the saddle frequencies are obtained from $\mathcal{H}(x,y,r,\omega)$, where the arguments are treated as real with $p=x+iy$.
The defining simultaneous equations are $\mathrm{Re}[\mathcal{H}]=\mathrm{Im}[\mathcal{H}]=\mathrm{Re}[\partial_p\mathcal{H}]=\mathrm{Im}[\partial_p\mathcal{H}]=0$.
The starting value of $\gamma$ is chosen to be slightly above $\gamma_c$ and the values in \eqref{crit} are used as initial guesses.
The four equations are solved at an arbitrary value of $C$ to find the four arguments of $\mathcal{H}$.
From this we extract $(p^+_S,r_S)$, which are $C$ independent, as a function of $\gamma>\gamma_c$.
The frequency is then obtained by selecting a $C$ value and computing $\omega^+_S=\omega_D^+(p^+_S,r_S)$. Note that the complex conjugate $p_S^{+*}$ gives the same $\omega$ value.
The value on the lower branch is obtained from $p_S^-=-p_S^{+*}$ and $\omega^-_S=\omega_D^-(p^-_S,r_S)$.

\begin{figure*} 
\centering
\includegraphics[width=\linewidth]{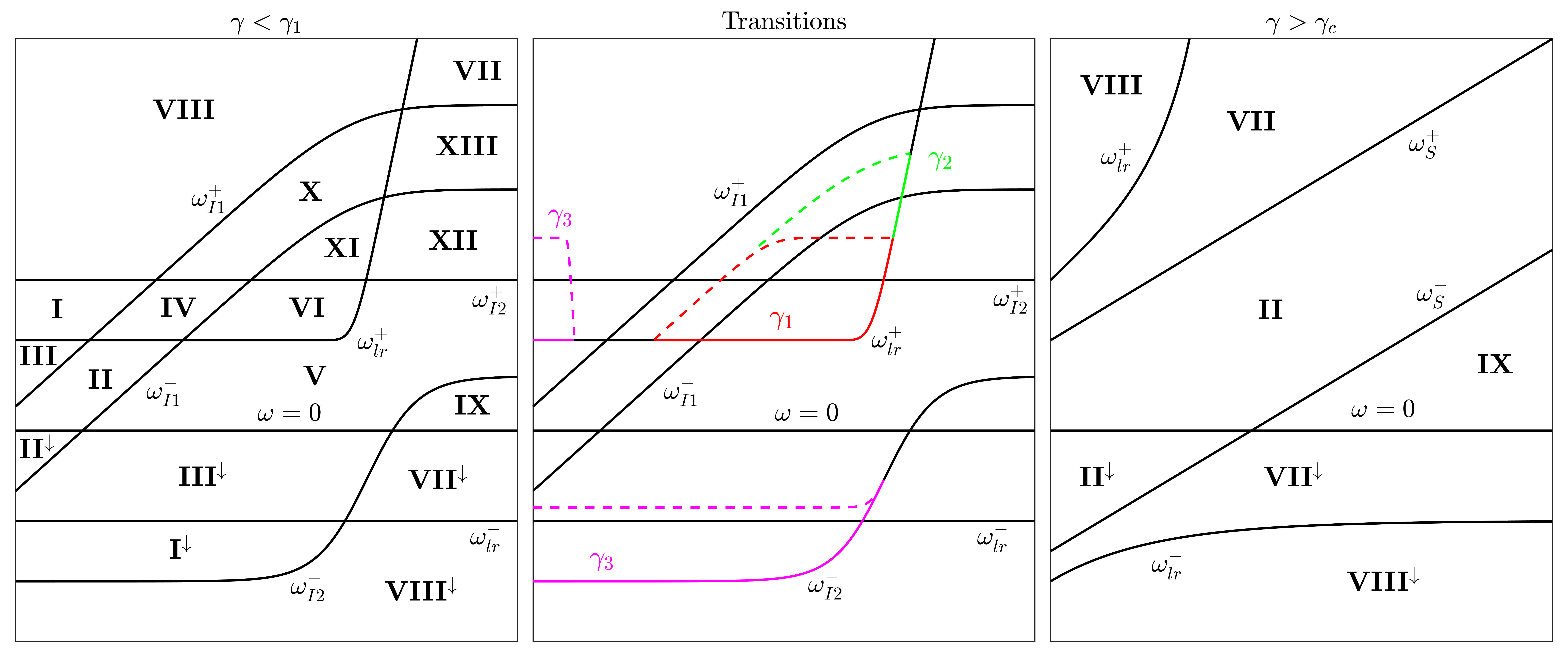}
\caption{We schematically depict in which regions of the $(\omega,C,\gamma)$ space different scattering processes occur. A $\gamma=\mathrm{const}$ slice is shown in the left panel in the limit that $\gamma$ goes to zero.
Increasing $\gamma$, a series of transitions occur at the values $(\gamma_1,\gamma_2,\gamma_3)=(3.96,7.45,45.1)\times 10^{-3}$, depicted in the central panel as the solid coloured lines moving to the coloured dashed lines with the corresponding labels $\gamma_{1,2,3}$. The structure of the parameter space is invariant for $\gamma>\gamma_c\simeq0.0522$ and is shown in the right panel.} \label{fig:trans}
\end{figure*}

\subsection{Parameter space regions} \label{app:param_plane}

To illustrate where in the parameter space the various scattering outcomes occur, the approach in Fig.~\ref{fig:param_plane} is to plot a slice $(\omega,C,\gamma=\mathrm{const})$ of the full parameter space, then vary $\gamma$ to see how the different regions change.
The problem with Fig.~\ref{fig:param_plane} is that there are small regions of the parameter space that are not visible within the resolution of the figure. 
Here we rectify this shortcoming.

On the left panel of Fig.~\ref{fig:trans}, we schematically illustrate the different regions on the plane $(\omega,C,\gamma=\mathrm{const})$, with $\gamma$ small.
Here small means that further decreasing $\gamma$ does not alter the qualitative structure of the parameter space, i.e. regions do not disappear from the plane although they may change shape.
At small $\gamma$, each of the 13 outcomes appears at least once in the plane.
Note that Fig.~\ref{fig:trans} is only a schematic; the different regions and their relative positioning are captured clearly, but not necessarily in their correct shape.
Consequently, neither the form of the characteristic $\omega$ curves, nor the precise locations of intersections between neighbouring curves, need be representative of reality. In general, these can change with $\gamma$.

As $\gamma$ increases, the parameter space undergoes a series of transitions, illustrated in the central panel of Fig.~\ref{fig:trans}.
During the first transition, which occurs at $\gamma_1\simeq3.96\times10^{-3}$ and $C_1\simeq11.5$, $\omega_\lr^+$ crosses the intersection between $\omega_{I1}^-$ and $\omega_{I2}^+$. This is depicted on Fig.~\ref{fig:trans} as the solid red line moving to the dashed red line.
When this occurs, Type VI scattering disappears from the plane and a second Type XIII region appears in the region bounded by the red dashed line and the lines $\omega^+_{I2}$ and $\omega^-_{I1}$.
The second transition occurs at $\gamma_2\simeq7.45\times10^{-3}$ and $C_2\simeq15.4$ when the two intersections between $\omega_\lr^+$ and $\omega_{I1}^-$ coalesce and vanish. This is shown on Fig.~\ref{fig:trans} as the solid green line moving to the dashed green line. 
Type XI scattering disappears from the plane during this transition.
The third transition arises when the intersection between $\omega_\lr^-$ and $\omega_{I2}^-$ crosses through $C=0$ and reappears as an intersection between $\omega_\lr^+$ and $\omega_{I2}^+$.
When this happens, the Type $\mathrm{I}^\downarrow$ diagram drops out of the lower half plane and a second Type VII region appears in the upper half plane touching the vertical ($C=0$) axis.
This occurs at $\gamma_3\simeq 4.51\times10^{-2}$ and is shown as the solid pink lines moving to the dashed pink lines on Fig.~\ref{fig:trans}.

The final transition occurs at the previous mentioned critical value $\gamma_c$, which is associated with the complete disappearance of inflection points from the $\omega_D^\pm(p)$ for all $r$.
This transition is more complicated than the previous three: as $\omega_{I1}^\pm$ and $\omega_{I2}^\pm$ combine to produce $\omega_S^\pm$, the crossings of $\omega_{I1}^+$ and $\omega_{I2}^+$ with $\omega_\lr^+$ simultaneously vanish, leading to the removal from the plane of all remaining diagrams that contain 4 modes concurrently propagating on a single branch $\omega_D^\pm(p)$.
The parameter space for $\gamma>\gamma_c$ is shown on the right panel of Fig.~\ref{fig:trans}.

\section{Resonance conditions} \label{app:res}

\begin{figure*} 
\centering
\includegraphics[width=.9\linewidth]{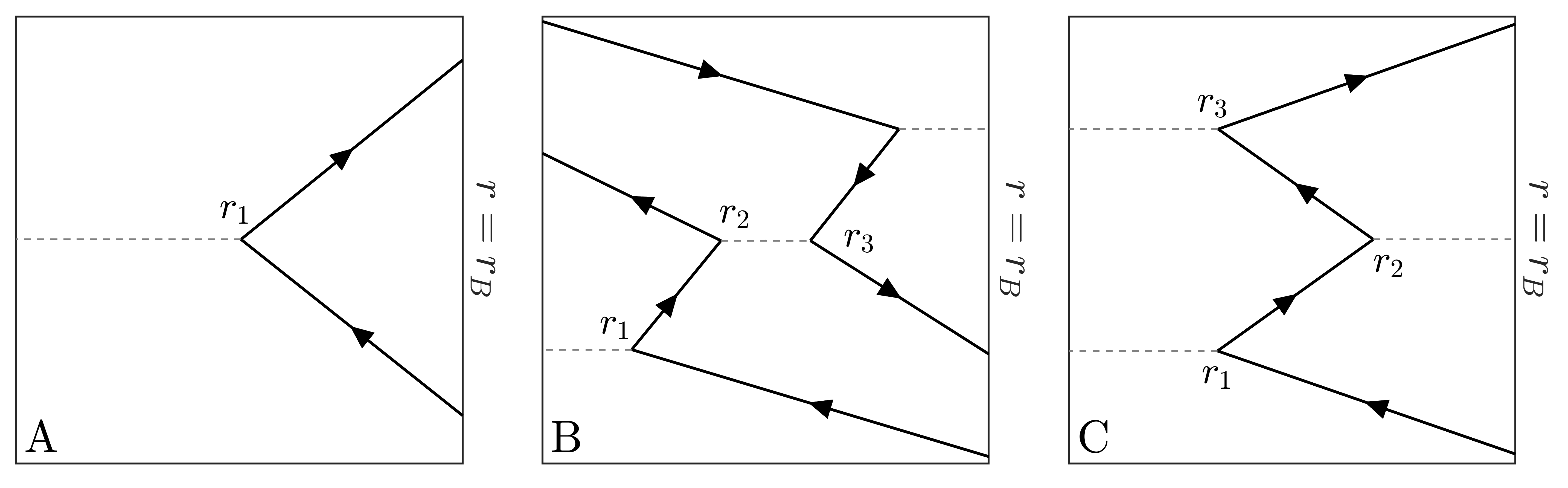}
\caption{Within our formalism, there are three different arrangements of branches on the diagrams in Fig.~\ref{fig:diagrams} which can lead to bound states between the vortex and the hard wall at $r_B$. We schematically illustrate them here and call them bound state A B and C respectively (coinciding with the panel labels in the bottom left corner).} \label{fig:BS}
\end{figure*}

There are three types of bound state we consider in this work (see Fig.~\ref{fig:BS}).
The first (bound state A) arises in the shallow water flows of Section~\ref{sec:shallow}, as well as for capillary-gravity waves which undergo Type II, III, V, VII or IX scattering (see Fig.~\ref{fig:diagrams}).
The other two (bound states B and C) arises for capillary-gravity waves which respectively undergo Type IV/VI and Type XII/XIII scattering.
Below, we derive the resonance formulae for each case separately.
Collectively, the resonance conditions can be grouped under the single formula,
\begin{equation} \label{general_res}
    e^{i\mathrm{Re}[S_\mathrm{path}]} e^{-\nu\mathrm{Re}[\Sigma_\mathrm{path}]}\prod_a X_a = 1,
\end{equation}
where $S_\mathrm{path}$ and $\nu\Sigma_\mathrm{path}$ are respectively the total phase and damping factors accumulated along the path of the bound state and the $X_a$ are scattering coefficients picked up at each turning point.
Note that $S_\mathrm{path}$ and $\Sigma_\mathrm{path}$ are evaluated first for real frequencies, hence, taking the real part amounts to e.g. $S^r\equiv\mathrm{Re}[S_\mathrm{path}] = \int_\mathrm{path}\mathrm{Re}[p] dr$. The extension to complex frequencies $\omega^\mathbb{C}=\omega+i\Gamma$ with $|\omega|\gg|\Gamma|$ is then $S^r(\omega^\mathbb{C})=S^r(\omega)+i\Gamma S^r(\omega)$.
We then find that the leading contributions to the real ($\omega_n$) and imaginary ($\Gamma_n$) parts of the resonant frequency satisfy,
\begin{equation}
\begin{split}
    & \mathrm{Re}[S_\mathrm{path}(\omega_n)] = \pi\left(2n+\frac{1}{2}\right), \\ & \Gamma_n = \frac{\sum_a\log|X_a|-\nu\mathrm{Re}[\Sigma_\mathrm{path}]}{T_\mathrm{path}}\Bigg|_{\omega=\omega_n},
\end{split}
\end{equation}
where $T_\mathrm{path}=\partial_\omega \mathrm{Re}[S_\mathrm{path}]$ is the classical time for the ray to orbit the trapping region.
To arrive at these formulae, one could directly impose the boundary conditions in \eqref{boundaryconditions} on \eqref{scatterM} using the full form of the transfer matrix, e.g. the one in \eqref{M1} for Type VI scattering.
However, it is simpler to transport the amplitudes around the bound state loops in Fig.~\ref{fig:BS} starting from the in-going mode at the boundary.
This will be the approach followed here.

\subsection{Bound state A}

The first case (Fig.~\ref{fig:BS} panel A) involves only two modes and a single reflection at the turning point labelled $r_1$.
We use the mode labels $j=\pm$, although in Type VII scattering on Fig.~\ref{fig:diagrams} the modes involved are labelled $j=t,-$.
Using shift factors \eqref{shift2}, we have the relations $A^\pm_1=\mathcal{F}^\pm_{1B}A^\pm_B$ and the reflection coefficient at $r_1$ is $\mathcal{R}_1 = A^+_1/A^-_1$.
Using \eqref{boundaryconditions}, the boundary condition at $r=r_B$ is $A^+_B=A^-_B$, which leads to the condition,
\begin{equation}
    \mathcal{R}_1 \mathcal{F}^-_{1B}/\mathcal{F}^+_{1B} = 1.
\end{equation}
The square root factors in \eqref{shift2} are equal for the two modes at $r_1$ and therefore cancel in the expression above. The same is approximately true at the outer boundary when $r_B$ is large.
We are left with,
\begin{equation}
    \mathcal{R}_1 e^{i(S^+_{1B}-S^-_{1B})}e^{-\nu(\Sigma^+_{1B}-\Sigma^-_{1B})} = 1,
\end{equation}
where we have defined the integrals,
\begin{equation} \label{phase_int}
    S^j_{ab} = \int^{r_b}_{r_a}p^j dr, \qquad \Sigma^j_{ab} = \int^{r_b}_{r_a}\frac{{k^j}^2}{v^j_{g,r}} dr.
\end{equation}
Defining $S_\mathrm{path} = S^+_{1B}-S^-_{1B}$, which is the accumulated phase difference between the start $(r_B,p^-_B)$ and end $(r_B,p^+_B)$ of the path through phase space (and similarly for $\Sigma_\mathrm{path}$).
We can write this in the form \eqref{general_res}.
In all cases considered, the reflection coefficient has the form $\mathcal{R}_1=-i|\mathcal{R}_1|$ to leading order (see \eqref{R_nu0}).
Hence, we can write this as,
\begin{equation}
    i\left(S_\mathrm{path}+\frac{\pi}{2}\right) -\nu\Sigma_\mathrm{path} = i\pi(2n+1)-\log|\mathcal{R}_1|.
\end{equation}
To solve for the resonant frequency, we assume that the condition is satisfied by $\omega_n^\mathbb{C}=\omega_n+i\Gamma_n$, where $\omega_n\gg\Gamma_n$ are both real quantities.
Perturbing the expression above, under the assumption that $\nu$ and $\log|\mathcal{R}_1|$ are of comparable order to $\Gamma$, we obtain,
\begin{equation}
\begin{split}
    S_\mathrm{path}(\omega_n) = & \ \pi\left(2n+\frac{1}{2}\right), \\ \Gamma_n = & \ \frac{\log|\mathcal{R}_1|-\nu\Sigma_\mathrm{path}}{\partial_\omega S_\mathrm{path}}\Bigg|_{\omega=\omega_n}.
\end{split}
\end{equation}

\subsection{Bound state B}

The second case (Fig.~\ref{fig:BS} panel B) involves a reflection $r_1$ and a transmission between $r_2$ and $r_3$.
We can build the resonance condition by transporting amplitudes around the loop,
\begin{equation} \label{BS_calc}
\begin{split}
    A^-_1 = & \ \mathcal{F}^-_{1B}A^-_B, \\
    A^+_1 = & \ \mathcal{R}_1\mathcal{F}^-_{1B}A^-_B, \\
    \mathcal{F}^+_{12}A^+_2 = & \ \mathcal{R}_1\mathcal{F}^-_{1B}A^-_B, \\
    \mathcal{F}^+_{12}A^+_3 = & \ \mathcal{R}_1\tau_{23}\mathcal{F}^-_{1B}A^-_B, \\
    \mathcal{F}^+_{12}\mathcal{F}^+_{3B}A^+_B = & \ \mathcal{R}_1\tau_{23}\mathcal{F}^-_{1B}A^-_B,
\end{split}
\end{equation}
where we defined the ratio of amplitudes $\tau_{23}=A^+_3/A^+_2$.
We calculate this using \eqref{refl_formula},
\begin{equation}
\begin{split}
    \tau_{23} = & \ \frac{{\mathcal{F}^\downarrow_{23}}^{-1}}{1+f^2_{23}/4}, \\ = & \ \left|\frac{f^j_a\partial_p\mathcal{H}^j_b}{f^j_b\partial_p\mathcal{H}^j_a}\right|^\frac{1}{2}e^{i\mathrm{Re}[S^\downarrow_{23}]}e^{-\nu\mathrm{Re}[\Sigma^\downarrow_{23}]}\mathcal{T}_{23},
\end{split}
\end{equation}
where $f_{23}$ is given by \eqref{f12} and $S^\downarrow_{23},\Sigma^\downarrow_{23}$ by \eqref{phase_int}. 
The transmission coefficient $\mathcal{T}_{23}$ is defined as the ratio of the norm currents \eqref{WKB_curr},
\begin{equation}
    \mathcal{T}_{23} = \frac{f_{23}}{1+f_{23}^2/4},
\end{equation}
and is adiabatically conserved for $\nu=0$, i.e. it is unaffected by the square root factor in \eqref{amp2} which scales the amplitudes.
We now implement the boundary condition $A^+_B=A^-_B$ on \eqref{BS_calc} and cancel the square root factors to write,
\begin{equation}
    \mathcal{R}_1\mathcal{T}_{23}e^{i\mathrm{Re}[S_\mathrm{path}]}e^{-\nu\mathrm{Re}[\Sigma_\mathrm{path}]}=1,
\end{equation}
where $S_\mathrm{path}=S^+_{12}+S^\downarrow_{23}+S^+_{3B}-S^-_{1B}$ and similarly for $\Sigma_\mathrm{path}$.
The argument in \eqref{R_nu0} implies that $\mathcal{R}_1$ and $\mathcal{T}_{23}$ can be evaluated for $\nu=0$, with the reflection coefficient satisfying $\mathcal{R}_1=-i|\mathcal{R}_1|$.
Applying the same argument used for bound state A then leads to,
\begin{equation}
\begin{split}
    & \mathrm{Re}[S_\mathrm{path}(\omega_n)] = \pi\left(2n+\frac{1}{2}\right), \\ & \Gamma_n = \frac{\log|\mathcal{R}_1\mathcal{T}_{23}|-\nu\Sigma_\mathrm{path}}{\partial_\omega \mathrm{Re}[S_\mathrm{path}]}\Bigg|_{\omega=\omega_n}.
\end{split}
\end{equation}

\subsection{Bound state C}

The final case (Fig.~\ref{fig:BS} panel C) involves three reflections at the turning points $r_{1,2,3}$.
Transporting the amplitudes round the loop, we find,
\begin{equation} \label{BS_calc2}
\begin{split}
    A^-_1 = & \ \mathcal{F}^-_{1B}A^-_B, \\
    A^+_1 = & \ \mathcal{R}_1\mathcal{F}^-_{1B}A^-_B, \\
    \mathcal{F}^+_{12}A^+_2 = & \ \mathcal{R}_1\mathcal{F}^-_{1B}A^-_B, \\
    \mathcal{F}^+_{12}A^u_2 = & \ \mathcal{R}_1\mathcal{R}_2\mathcal{F}^-_{1B}A^-_B, \\
    \mathcal{F}^+_{12}A^u_3 = & \ \mathcal{R}_1\mathcal{R}_2\mathcal{F}^u_{32}\mathcal{F}^-_{1B}A^-_B, \\
    \mathcal{F}^+_{12}A^t_3 = & \ \mathcal{R}_1\mathcal{R}_2\mathcal{R}_3 \mathcal{F}^u_{32}\mathcal{F}^-_{1B}A^-_B, \\
    \mathcal{F}^+_{12}\mathcal{F}^t_{3B}A^t_B = & \ \mathcal{R}_1\mathcal{R}_2\mathcal{R}_3 \mathcal{F}^u_{32}\mathcal{F}^-_{1B}A^-_B, \\
\end{split}
\end{equation}
The reflection coefficient $\mathcal{R}_2=A^u_2/A^+_2$ can be calculated using \eqref{tp1} to write,
\begin{equation}
    \begin{pmatrix}
        A^u_2 \\ A^+_2
    \end{pmatrix} = e^{\frac{i\pi}{4}}\begin{pmatrix}
        1 & -\frac{i}{2} \\ -i & \frac{1}{2}
    \end{pmatrix} \begin{pmatrix}
        \mathcal{F}^\downarrow_{2B}A^\downarrow_B \\ \mathcal{F}^\uparrow_{2B}A^\uparrow_B
    \end{pmatrix}.
\end{equation}
From Appendix~\ref{app:BCs} we have $A^\uparrow_B=A^\downarrow_B$, which leads to the expression,
\begin{equation}
    \mathcal{R}_2 = -i\frac{1+if_{2B}^2/2}{1-if_{2B}^2/2}.
\end{equation}
Applying the same argument as \eqref{R_nu0}, we can evaluate $\mathcal{R}_2$ for $\nu=0$, in which case, using $|f_{2B}|\ll 1$, we find $\mathcal{R}_2\simeq e^{i(|f_{2B}|^2-\pi/2)}$ which has unit modulus.

The resonance condition is found from \eqref{BS_calc2} by applying the boundary condition $A^t_B=A^-_B$,
\begin{equation}
    \mathcal{R}_1\mathcal{R}_2\mathcal{R}_3 e^{iS_\mathrm{path}}e^{-\nu\Sigma_\mathrm{path}}=1,
\end{equation}
where $S_\mathrm{path}=S^t_{3B}-S^u_{32}+S^+_{12}-S^-_{1B}$ and similarly for $\Sigma_\mathrm{path}$.
Repeating the steps used for bound state A gives,
\begin{equation}
\begin{split}
    & S_\mathrm{path}(\omega_n) = \pi\left(2n+\frac{1}{2}\right), \\ 
    & \Gamma_n = \frac{\log|\mathcal{R}_1\mathcal{R}_{3}|-\nu\Sigma_\mathrm{path}}{\partial_\omega S_\mathrm{path}}\Bigg|_{\omega=\omega_n}.
\end{split}
\end{equation}
The real part of the frequency also receives a correction $\delta\omega = |f_{2B}(\omega_n)|^2/\partial_\omega S_\mathrm{path}$ in this case (which enters at the same order as $\Gamma$) due to the extra phase shift in $\mathcal{R}_2$.

\end{document}